\documentclass[preprint2]{aastex}
\usepackage{amsmath, amsthm, amssymb}
\usepackage{natbib}
\usepackage{epsfig}
\usepackage[update,prepend]{epstopdf}

\shorttitle{A Fan Beam Model for Radio Pulsars. I. Observational Evidence}
\shortauthors{Wang et al.}

\begin{document}

\title{A Fan Beam Model for Radio Pulsars. I. Observational Evidence}

\author{H. G. Wang\altaffilmark{1,2},  F. P. Pi\altaffilmark{1}, X. P. Zheng\altaffilmark{3}, C. L. Deng\altaffilmark{1},
            S. Q. Wen\altaffilmark{1}, F. Ye\altaffilmark{1}, K.Y. Guan\altaffilmark{1}, Y. Liu\altaffilmark{1}, L. Q. Xu\altaffilmark{1}}

\altaffiltext{1}{ School of Physics and Electronic Engineering, Guangzhou University, 510006, Guangzhou, China}
\email{hgwang.gz@gmail.com}
\altaffiltext{2}{ CSIRO Astronomy and Space Science, PO Box 76, Epping NSW 1710, Australia}
\altaffiltext{3}{ College of Physical Science and Technology, Huazhong Normal University, 430079, Wuhan, China}

\begin{abstract}

We propose a novel beam model for radio pulsars based on the scenario that the broadband and coherent emission from secondary relativistic particles, as they move along a flux tube in a dipolar magnetic field, forms a radially extended sub-beam with unique properties. The whole radio beam may consist of several sub-beams, forming a fan-shaped pattern. When only one or a few flux tubes are active, the fan beam becomes very patchy. This model differs essentially from the conal beam models in the respects of beam structure and predictions on the relationship between pulse width and impact angle $\beta$ (the angle between line of sight and magnetic pole) and the relationship between emission intensity and beam angular radius. The evidence for this model comes from the observed patchy beams of precessional binary pulsars and three statistical relationships found for a sample of 64 pulsars, of which $\beta$ were mostly constrained by fitting polarization position angle data with the Rotation Vector Model. With appropriate assumptions, the fan beam model can reproduce the relationship between 10\% peak pulse width and $|\beta|$, the anticorrelation between the emission intensity and $|\beta|$, and the upper boundary line in the scatter plot of $|\beta|$ versus pulsar distance. An extremely patchy beam model with the assumption of narrowband emission from one or a few flux tubes is studied and found unlikely to be a general model. The implications of the fan beam model to the studies on radio and gamma-ray pulsar populations and radio polarization are discussed.

\end{abstract}



\section{Introduction}

Limited by the fixed orientation of the fixed line of sight (LOS) with respect to the spin axis, it is hard to observe the 2-D structure of radio emission beam for pulsars. This has aroused extensive researches and long-term debates on the radio beam pattern of pulsars.

The conal and patchy beam models are hitherto two general kinds of radio beam models. Based on the empirical classification for the observational properties of radio pulsars, Rankin (1983, 1990, 1993) proposed that the radio beam normally consists of two nested cones and a quasi-axial core. Comparing with the early hollow-cone model (Komesaroff 1970, Backer 1976, Oster \& Sieber 1976), the double-cone-core model has the advantage in explaining a variety of pulse morphology in terms of different LOS trajectories across the beam. For instance, a LOS close to the beam center may sweep across the core and double cones and result in a pulse profile with five components. Following this idea, Gangadhara \& Gupta (2001) identified 9 components for the high-quality pulse profiles of PSR B0329+54 and claimed that this pulsar has 4 distinct cones and a core. Applying a statistical approach to a sample of multi-frequency profiles of conal triple and multiple pulsars, Mitra \& Deshpande (1999) determined the locations of conal components and concluded that a typical radio beam should consist of at least three nested cones, although only one or more of them may be active in individual pulsars. Unlike the regular beam structure in conal beam models, the patchy beam model (Lyne \& Manchester 1988, hereafter LM88) suggested that the intensity distribution within a beam is patchy and usually only parts of the beam are visible. The beam pattern was explained as the product of a pulse window, usually a circular beam common to pulsars, and a source function, which is randomly distributed and unique to each pulsar (Manchester 1995). In spite of the essential difference in intensity distribution, the conal and patchy beam models have a consensus that the radio beam has a boundary, which is presumably circular or elliptical.

Hybrid radio beam models with patchy and conal features were also proposed (hereafter called patchy conal beam models). In one of the series of work on the conal beam model, Rankin (1993) presented a cartoon of spotty inner and outer cones. The luminous spots were presumed to originate from some particular magnetic field lines which have more copious secondary pairs and hence more emitters than the others. In an alternative patchy conal beam model proposed by Karastergiou \& Johnston (2007), the conal structure is supposed to originate from the emission in a wide range of altitudes along the outer field lines close to the polar cap boundary, but the locations of active emission are discrete, forming a patchy pattern within the conal ring. With this model, the authors tried to explain why young pulsars tend to have simpler pulse profiles than old pulsars by assuming distinct altitude ranges for these two types of pulsars. The patchy beam pattern was also introduced by Beskin \& Philippov (2012) into a propagation model where the outer and inner cones, formed by the O-mode and X-mode emission respectively due to different propagation paths, contain discrete bright patches.

There have been many endeavors to test radio beam models. As stated by Gil \& Krawczyk (1996) and Kijak \& Gil (2002), the conal beam model predicts a weak frequency dependence for the relative pulse phase between subpulses and profile components, while the patchy beam model predicts no dependence. The authors found that the observed subpulse behavior favors the conal beam model. Further evidence was suggested to support the multi conal beam model (Kijak \& Gil 2002), including the binomial distribution of the opening angles of beams or profile components (Rankin 1993, Gil et al. 1993, Kramer et al. 1994) and the tendency of large impact angles for the pulsars with single- and double-peak profiles. In order to derive the 2-D image of the mean radio beam, Han \& Manchester (2001, hereafter HM01) collected a sample of 87 normal pulsars with multicomponent pulse profiles and created an intensity stripe along the LOS within the normalized circular beam for each pulsar, and then integrated all the stripes to form an averaged intensity map of the beam. Except for an enhancement near the center, they found only a few mild enhancements in other parts of the beam. This pattern was regarded as the evidence for the patchy beam model. However, Kijak \& Gil (2002) argued that HM01's work is inconclusive, because the frequency dependence of pulse profiles, possible period dependence of beam radius and the diversity in conal and core component positions were not excluded in their data reduction. In the recent series of work, Maciesiak et al. (2011), Maciesiak \& Gil (2011, hereafter MG11) and Maciesiak et al. (2012, hereafter MGM12) suggested that the relationship $W_{50}\sim P^{-1/2}$ for the lower boundary line in the scatter plot of 50\% peak pulse width versus pulsar period can be interpreted by incorporating the spark model and the conal beam model.

Thanks to the discovery of relativistic spin precession in binary pulsars, it provides a unique approach to probe the beam topography directly (see Kramer 2012 for a review). The precession of pulsar spin axis changes the orientation of our LOS and enable us to scan different parts of radio beam. Up to now the tomography has been carried out for 6 pulsars, PSR B1913$+$16, PSR B1534$+$12, PSR J1141$-$6545, PSR J1906$+$0746, PSR J0737$-$3039A and B. To account for the profile changes of PSR B1913$+$16 over two decades, it was suggested that the beam should be elongated in the latitudinal direction to be somewhat hourglass-shaped rather than to be circular (Weisberg \& Taylor 2005, Clifton \& Weisberg 2008). The pulse profile of PSR B1534$+$12 is found to be gradually broadening as out line of sight moves further from the magnetic pole (Arzoumanian 1995, Stairs et al. 2004, Fonseca et al. 2014), which is contrary to the expectation if the beam is a circular cone. The radio beam patterns were reconstructed by the data of pulse profile and flux density for PSR J1141$-$6545 (Manchester et al. 2010) and PSR J1906$+$0746 (Desvignes et al. 2013). PSR J1906$+$0746 has interpulse emission, so the beam intensity maps from both poles were derived. For both pulsars, the scanned parts of their beams are partially illuminated, without any signature of conal rings. These results present unambiguous evidence for the patchy beam model. Meanwhile, the observations revealed a striking limb-darkening phenomenon that the intensity tends to decrease towards the edge of beam in both radial and transverse directions, which has not been explained yet. The profile of PSR J0737$-$3039A is long-term stable, which is interpreted as a result of very small misalignment between the spin axis and the orbital angular momentum. A circular cone model was used to interpret the profile (Manchester et al. 2005, Ferdman et al. 2013, Perera et al. 2014). The radio beam of PSR J0737$-$3039B was found to be partially-filled horse-shoe shape (e.g. Perera et al. 2012, Lomiashvili \& Lyutikov 2013). Unlike the former other double pulsars, the magnetosphere and emission of PSR J0737$-$3039B is strongly influenced by the wind from its companion PSR J0737$-$3039A, therefore, it may not be an ideal case to test the intrinsic beam pattern from an undisturbed pulsar magnetosphere.

The new discoveries from PSR J1141$-$6545 and PSR J1906$+$0746 pose several questions: could the patchy beam be a general pattern for radio pulsars? How is the limb-darkening patchy beam formed? Unfortunately, as far as we know, there is no physical models in literature to account for the formation of a patchy beam. In contrast, the origin of conal beam has been extensively explored. A brief review below will be helpful to understand the current status of the research on this respect.

It has been proposed that the conal and core structure can be generated through the curvature radiation (Sturrock 1971, Ruderman \& Sutherland 1975, hereafter RS75, Gil \& Snakowski 1990, Gangadhara 2004, Wang et al. 2012), or through the curvature maser in the relativistic plasma along curved magnetic field lines (Beskin et al. 1988), or through the inverse Compton scattering (ICS) of the low frequency electromagnetic wave by secondary relativistic particles (Qiao 1988, Qiao \& Lin 1998, Xu et al. 2000, Qiao et al. 2001, Qiao et al. 2004, Lee et al. 2009, Lv et al. 2011). Despite the difference in radiation mechanism and geometry, these models have three common ingredients. (1) The emission is coherent. The demand of coherence for pulsar radio emission has been demonstrated by a number of works both theoretically (Ginzburg et al. 1969, Sturrock 1971, Cheng \& Ruderman 1977, Luo \& Melrose 1992, Kunzl et al. 1998, Qiao \& Lin 1998, Melikidze et al. 2000, Gil et al. 2004, Dyks et al. 2007) and observationally (e.g. Hankins et al. 2003,  Mitra et al. 2009, Jessner et al. 2010). (2) The radio emission is narrow band, i.e. the emission at a particular frequency should come from a fixed or a narrow range of altitude. (3) A cone in the beam is mapped to a bundle of open field lines of which the cross section is annular on the polar cap surface. In the updated version of spark model (originally proposed by RS75), the pair production areas (sparks) above the polar cap may form multi annuli, leading to multi cones (Gil \& Sendyk 2000, hereafter GS00). In the ICS model, emissions at the same frequency can be generated at more than one altitude, thus multi cones can be formed from one annulus of field lines (Qiao \& Lin 1998). Among the three ingredients, the later two points are vital to form a conal beam structure.

The narrowband assumption is thought to be favored by the fact that the average pulse profiles of many pulsars broaden with decreasing frequency, which is usually attributed to the radius to frequency mapping (RFM), i.e. a lower frequency emission comes from a higher altitude (Komesaroff 1970, RS75, Cordes 1978). However, counterexamples are often seen in multi-frequency observations (e.g. Hankins \& Rickett 1986, Johnston et al. 2008). Moreover, it is well known that a number of short-timescale features, which are related to localized emission processes, e.g. single pulses (Bartel \& Sieber 1978, Kramer et al. 2003), microstructures (see Hankins 1996 for a review), giant pulses (Sallmen et al. 1999, Popov et al. 2006, Jessner et al. 2010), nulling (Bartel et al. 1981, Bhat et al. 2007), polarization properties in single pulses (Karastergiou et al. 2002, Mitra et al. 2007), mode changing (Bartel et al. 1982, Chen et al. 2011) and subpulse drifting (Smits et al. 2005), are mostly of a broadband nature. It was postulated that broadband emission may occur near the pair production fronts near polar caps or at higher altitudes cooperating with narrowband emission mechanism (e.g. Melrose 1996), or that a broadband subpulse or microstructure may be an ensemble of a number of narrowband or broadband single emitters (Cordes 1979). The cooperation of narrowband and broadband mechanisms seems possible, because there is evidence that the giant pulses from the Crab pulsar, which can be seen in a very wide frequency range, sometimes contain a number of nanoshots with narrower bandwidths from tens to hundreds of megahertz (Eilek \& Hankins 2006).

Models on broadband emission for pulsars are much fewer. Buschauer \& Benford (1980) studied the properties of both narrowband and broadband coherent curvature radiation and compared them with observations. They suggested that the broadband model could accommodate a wider range of phenomena than the narrowband model. Barnard \& Arons (1986) investigated the effects of refraction on pulse profile, spectrum and polarization in both the narrowband and broadband scenarios. They found that the low-frequency pulse broadening phenomenon, used to be interpreted by RFM, can be alternatively explained by a model that the broadband emission occurring in a narrow range of altitude undergoes more refraction at a lower frequency due to transvers plasma density gradient and hence broadens the low-frequency pulses. Recent simulations on the pair production in the inner vacuum gap (hereafter IVG, Timokhin 2010) and the polar gap due to space charge limited flow (hereafter SCLF-gap, Harding \& Muslimov 2011, Timokhin \& Arons 2013) revealed that the secondary eletrons/positrons are not quasi monotonic in momentum, as many narrowband models had assumed, but can have momentum spectra with flat or moderate slope rates in many situations. Such broad momentum spectra are probably favorable to generate broadband radio emission.

In the past four decades, efforts have been focused on developing empirical and physical models based on the idea of narrowband and coherent emission to explore the origin of multi conal (and core) beam structure. These models can successfully account for parts of the observational properties. Apart from the difficulties in explaining the non-RFM type of frequency dependence of pulse profiles, the discoveries of more pre-/postcursors (Mitra \& Rankin 2011), off-pulse emission (Basu et al. 2011) and patchy beams of binary pulsars present growing challenges against the conal beam model.

This paper is the first one of a series of work on an alternative model for pulsar radio beam in terms of the assumption of broadband and coherent emission. This model predicts a new type of beam pattern, which looks like a fan or part of a fan, and thereby is called the fan beam model. It has different predictions from the conal beam model on the relationships between pulse width and impact angle and between other pairs of quantities. This paper, mainly focusing on the observational evidence, is organized as follows. The beam structure and relationships of the fan beam model are derived in Section 2. The evidence, including the observed beams of binary pulsars and three statistical relationships base on a sample of 64 pulsars with well constrained impact angles are presented in Section 3. The observational tests for the canonical and patchy conal beam models are presented in Section 4. A patchy beam model based on the assumption of narrowband emission from one or a few flux tubes is investigated and finally excluded in Section 5. Section 6 are conclusions and discussions. The implications of this model to the studies on pulsar population and radio polarization are also discussed.

\section{The fan beam model}
\label{section:2}
\subsection{The scenario}

Rather than developing a purely geometric beam model, we are interested in exploring how the emission mechanisms/propagation effects, secondary plasma flow and magnetic field geometry shape the emission beam. The central question is what the beam geometry will be when the secondary relativistic particles produce broadband and coherent emission as they flow along a dipolar magnetic flux tube. Since there is no consensus on pulsar emission mechanisms and propagation effects, we prefer to use a phenomenological parameter $q$ to describe the total effect (will be defined in the assumption (3) below), which is adjustable according to specific
emission and propagation mechanisms. The following basic assumptions are made to simplify the problem.
\begin{itemize}
  \item  The global magnetic field is dipolar.
  \item  Assumptions on the particle flow. (1) Large amount of secondary electrons and positrons are generated in polar gaps (IVG or SCLF-gap) with the multiplicity factor $M>>1$, i.e., the total number density of electrons and positrons is approximately $M$ times of the local Gouldreich-Julian number density $n_{\rm gj}$. (2) Being quasi-neutral, the plasma flows freely along the magnetic flux tubes.
  \item  Assumptions on the emission mechanism and propagation effect. (1) The emission is coherent in an elementary volume $\lambda^3$ ($\lambda$ is the wavelength of emission), where the emission power is proportional to the square of the number of particles therein. The emissions from separated elementary volumes are incoherent. (2) The emission from $\lambda^3$ is broadband. The shape of the averaged radiation spectrum of a single particle is the same everywhere in the emission region. (3) The averaged emission power of a single particle is a power law function of the emission altitude, viz. $p_{\rm e}\propto r^q$, where the altitude $r$ is measured from the stellar ceter in this paper.
\end{itemize}

Some of the assumptions need more words. Firstly, the term ``broadband''  means that either the emission from an elementary coherent volume is intrinsically broadband or it is a collective effect due to assembling of many narrowband emissions, which span a wide range of frequencies. Secondly, the ``averaged power'' and ``averaged spectrum'' for a single particle means that the quantities are averaged over a timescale for obtaining the mean pulse profile, therefore they are stable and useful to model the mean structure of radio beam.

Although the assumption on the emission power and spectrum is phenomenological, it can be adapted to a variety of emission mechanisms and propagation effects by adjusting the power law index $q$. For example, an electron or a positron with constant kinetic energy will emit less power at higher altitudes through the ICS process, roughly in a power law with $q\sim -2$\footnote{In the ICS model, the emission power of a single particle is proportional to $\gamma^2u$, where $\gamma$ is the Lorentz factor of the particle and $u$ is the energy density of low frequency electromagnetic waves (Qiao \& Lin 1998). Since $u\propto r^{-2}$, the emission power is also proportional to $r^{-2}$.}. Considering the curvature radiation instead, the index will be $q\sim -0.5$\footnote{The emission power is proportional to $\gamma^4/\varrho$, where $\varrho \sim r^{0.5}/3$ is the curvature of radius of a field line at the emission altitude $r$ well within the light cylinder (RS75).}. When a particular absorption effect correlated with the plasma density is considered, the plasma may be more transparent to the emission at higher altitudes, therefore the relationship becomes flatter and may even have a positive index. In the case that the real relationship is not a power-law, e.g. an exponential or some other functions, the index $q$ itself needs to be adjusted as a function of the coordinates of emission location to mimic the real relationship. Anyway, it is a practical treatment for the emission and propagation mechanisms to simplify our model. In this paper, $q$ is assumed to be a constant for a single pulsar. The statistical value will be constrained with a sample of pulsars in Section 3.2.2.

\begin{figure*}
\centering
\resizebox{11cm}{7.5cm}
{
\includegraphics{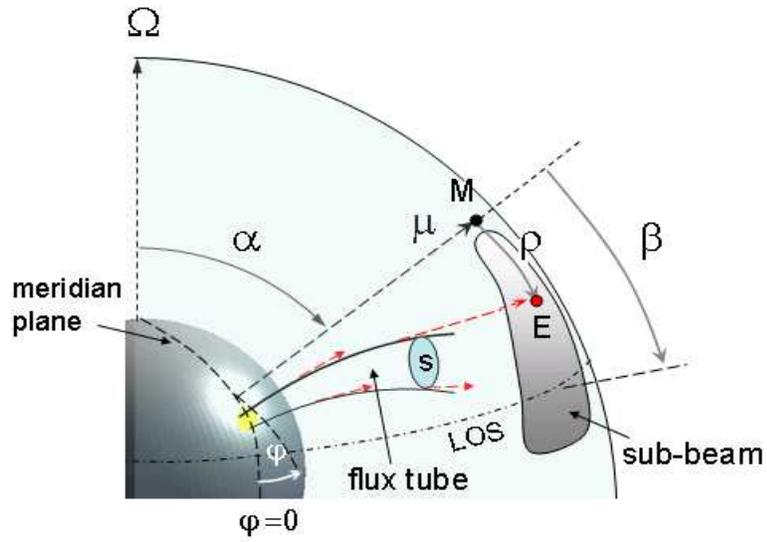}
}

\caption{The schematic diagram for the formation of limb-darkening sub-beam from a flux tube extending from the polar cap. $\alpha$ is the inclination angle between the rotation ($\Omega$) and magnetic ($\mu$) axes. $\beta$ is the impact angle between line of sight (LOS) and $\mu$ axis. It is assumed that the secondary pair plasma streams along the flux tube and generate broadband and coherent emission. Emissions at a single frequency can be generated at all the altitudes (e.g. the red dashed arrows), and hence forming a sub-beam, which fades outwards from the magnetic pole. The beam radius $\rho$ stands for the angular distance between the projection of magnetic pole ``M'' and the projection of an emission direction ``E'' on the celestial plane. }

\label{figure:schematic}
\end{figure*}

We first describe the general picture of the emission beam qualitatively with the above assumptions. As the secondary particles flow out along a flux tube, the number density keeps decreasing due to the divergent nature of dipolar field. This normally leads to decreasing emission intensity, except if too large a positive index $q$ is assumed ($q>3$ according to Section 2.2). Since the emission at a higher altitude points further from the magnetic pole, the intensity will decreases with increasing beam radius when $q<3$, or on the contrary when $q>3$, where the beam radius $\rho$ is defined as the angle between the emission direction and the magnetic pole (see Fig. \ref{figure:schematic}). In this paper, the first case is called the radial limb-darkening intensity distribution. Fig. \ref{figure:schematic} shows the schematic diagram for the formation of such a sub-beam generated from a flux tube.

The azimuthal (transverse) intensity distribution depends on further assumptions about the particles distribution across the polar cap. In the simplest case, if the secondary particles are uniformly distributed, the intensity will be constant in any circular ring around the magnetic pole. Such a model can only account for single pulse profiles. More practically, the secondary particles may be generated in a number of separated flux tubes, and each flux tube will form a wedge-shaped sub beam, which widens with increasing radius due to the divergence of flux tube. There is an intensity valley between a pair of sub-beams due to lack of particles in the region between two neighboring flux tubes. Then, the whole beam looks like an axial fan with a set of wedge-shaped sub-beams. This kind of beam is called the fan beam. When only a few flux tubes are assumed to be active in emission, the fan beam will show a very patchy pattern. Therefore, this model may give a reasonable explanation for the origin of patchy beam.

The above qualitative analysis has revealed a new type of emission beam pattern strikingly different from the conal beam structure. Unlike the conal beam which has a circular or an elliptical boundary, the fan beam has no boundary. This is because we choose the broadband assumption while the conal beam model choose the narrowband assumptions. In the following subsections, we will derive the radial intensity-radius relationship, the transverse intensity-azimuth relationship and an important prediction $-$ the relationship between pulse width $W$ and impact angle $\beta$. The beam structure and resultant profiles at various viewing geometries are simulated.

\subsection{Radial intensity distribution}

In this paper, the intensity is defined as the emission power over a unit solid angle around an emission direction. Since the emission from a secondary relativistic particle is beamed into a very narrow solid angle with a half opening angle of $\sim1/\gamma$, where the Lorentz factor $\gamma$ is typically larger than $100$, we assume that the emission direction is aligned with the tangent of field line for the convenience of calculation. In order to derive the intensity, we first select an arbitrary subregion in a flux tube and calculate its solid angle formed by the tangents of the boundary field lines. Then, we calculate the volume, the particle number density and the total emission power, and finally the emission intensity from the region.

\begin{figure*}
\centering
\resizebox{12cm}{5cm}
{
\includegraphics{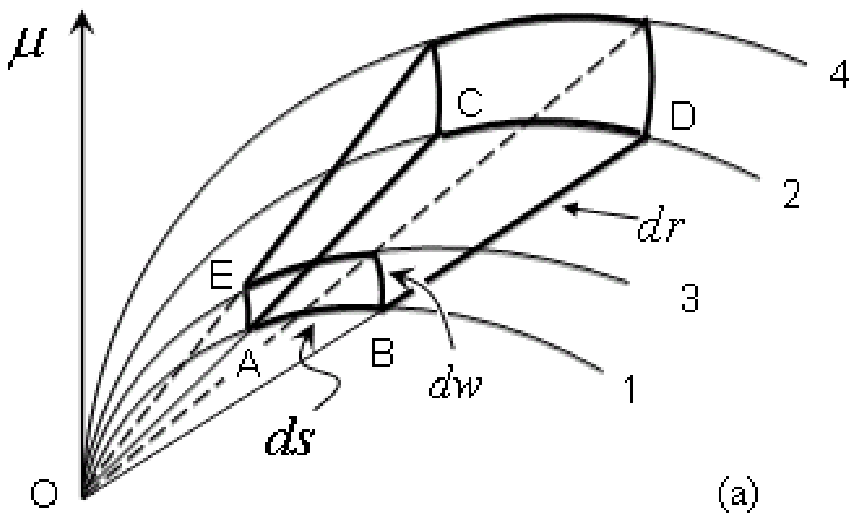}
\includegraphics{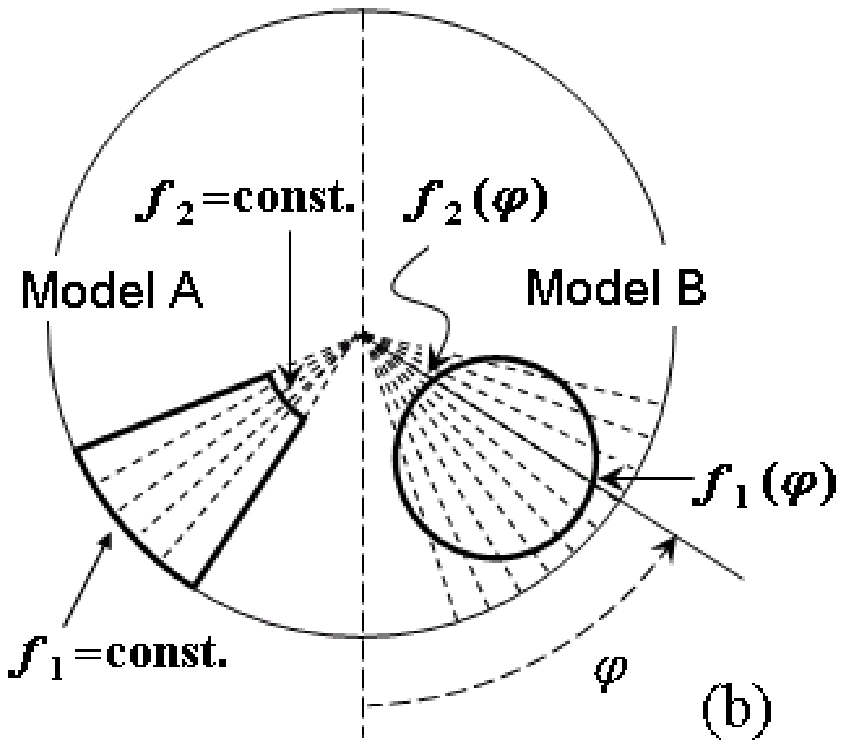}
}
\resizebox{12cm}{4.5cm}
{
\includegraphics{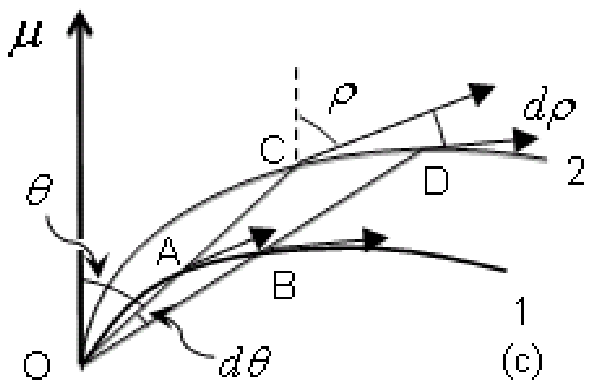}
\includegraphics{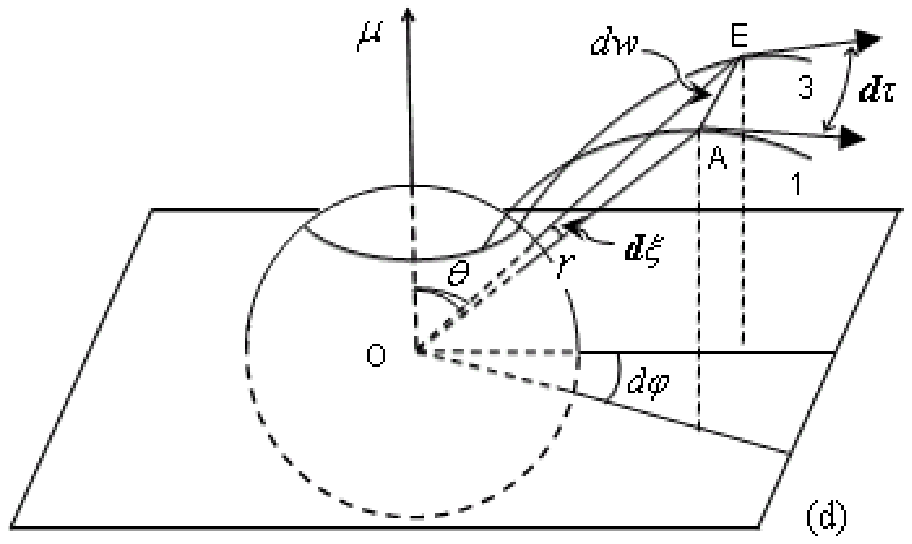}
}

\caption{(a) The 3-dimensional structure of a sub emission region in a flux tube, as enclosed by the thick lines and arcs within two groups of coplanar open field lines ``1, 2'' and ``3, 4''. $O$ is the stellar center. (b) The cross section for flux tubes on the polar cap. The thick solid curves stand for the boundary of flux tubes in Models A and B, and the dashed lines represent the division for slices of sub tubes. Each slice has approximately the shape confined by the field lines ``1, 2, 3, 4'' in panel (a). (c) The poloidal cross section of the subregion enclosed by ``ABCD'', where $\theta$ is the polar angle and $\rho$ is the opening angle of field line tangent with respect to the $\mathbf{\mu}$ axis at an emission point. (d) The transverse scale ${\rm d}w$ of the subregion. The vertices A and E are separated azimuthally by ${\rm d}\varphi$ and open an angle of ${\rm d}\xi$ with respect to the stellar center. }

\label{figure:region_geometry}
\end{figure*}

Regarding the axial symmetry of magnetic dipole fields, it is convenient to describe an emission point with the polar angle $\theta$, the azimuth angle $\varphi$ and the altitude $r$ (counted from the stellar center) in the spherical coordinate system where the polar axis is the magnetic pole (Figs. \ref{figure:schematic} and \ref{figure:region_geometry}). The open field lines can be distinguished by two parameters: the azimuth $\varphi$ between a field line plane and the meridian plane containing the rotation and magnetic axes, which is counted anticlockwise from the equatorial side of meridian plane, and the parameter $f\equiv R_{\rm e}/R_{\rm c}$, where $R_{\rm e}$ is the maximum altitude for an open field line and $R_{\rm c}=Pc/(2\pi)$ is the light cylinder radius, with $P$ the pulsar period. The parameter $f$ tells how close the field line is with respect to the magnetic pole. For the last open field lines, $R_{\rm e}$ varies from $R_{\rm c}$ to $\sim 3R_{\rm c}$ depending on inclination and azimuth angles. For simplicity, we neglect this difference and assume $R_{\rm e}=R_{\rm c}$ for all the last open field lines. Therefore, we have $f=1$ for last open field lines and $f>1$ for inner open field lines.

Fig. \ref{figure:region_geometry}(a) shows a typical subregion in a slice of a flux tube. The slice confined by the field lines ``1, 2, 3, 4'' is divided into a number of this kind of subregions, started from the polar cap surface to high altitudes. The field lines marked with ``1'' and ``2'' (``3'' and ``4'') stand for the outer and inner (with respect to the magnetic pole) boundary field lines of the slice. Each pair of boundary lines have the same azimuth angle but different $f$ parameters, where $f_2>f_1$ and $f_4>f_3$, meanwhile, $f_1\simeq f_3$ and $f_2\simeq f_4$. The whole flux tube can be divided azimuthally into a number of slices. The slices may have different boundary parameters $f_1$ and $f_2$, depending on the shape of the cross section of flux tube. We will consider two kinds of flux tube geometry in Section 2.4, one has a sector-shaped cross section (Model A) and the other has a circular shape (Model B). In Model A, all the slices have the same pair of $f_1$ and $f_2$, while in Model B, $f_1$ and $f_2$ depends on the azimuth $\varphi$ of each subregion. The cross sections of flux tube and the corresponding slices are illustrated by Fig. \ref{figure:region_geometry}(b) for these two models.

For clarity, the poloidal and toroidal cross sections of the subregion are presented by Fig. \ref{figure:region_geometry}(c) and (d), respectively. In panel (c), the two straight lines with the polar angles $\theta$ and $\theta+{\rm d}\theta$ stand for the lower and upper boundaries (with respect to the polar cap) of the subregion. At a pair of points where a boundary field line intersect the straight lines, e.g. A and B, or C and D, the tangents form an small angle ${\rm d}\rho$, where the beam radius $\rho$ is the anger between the tangent of a field line at $r$ and the magnetic pole. In panel (d), the two azimuthal boundary field lines, marked with ``1'' and ``3'', are separated from each other by a small azimuth angle ${\rm d}\varphi$. The tangents of field lines at two boundary points A and E open an angle ${\rm d}\tau$. ${\rm d}\rho$ and ${\rm d}\tau$ should be much larger than the $1/\gamma$ beaming angle of a single particle so that the angular power distribution in the single-particle emission beam can be ignored. Whenever $\gamma$ is higher than hundreds, ${\rm d}\rho\sim 1^{\rm o}$ and ${\rm d}\tau \sim 1^{\rm o}$ would be large enough to ensure that. In this circumstance, the particle number density varies little poloidally between A and B (or C and D) as well as toroidally between A and E, enabling the following derivation to be valid.

\begin{figure}
\centering
\resizebox{7cm}{5.7cm}
{
\includegraphics{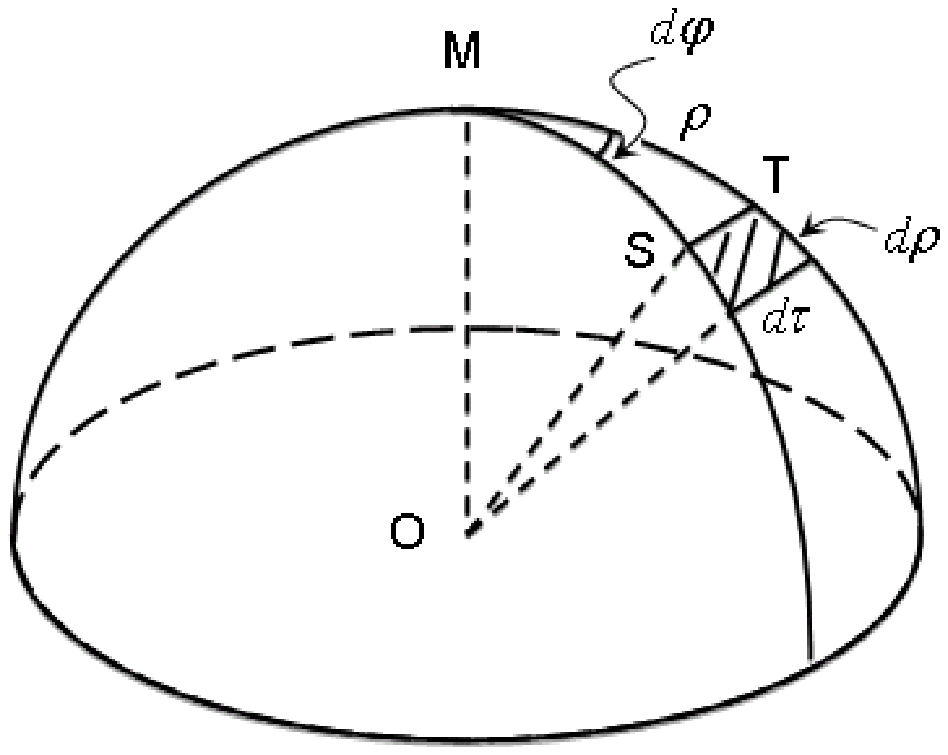}
}

\caption{The solid angle (shadowed region) formed by the subregion shown in Fig. \ref{figure:region_geometry}(a). $M$ stands for the projection of the magnetic pole. The emission directions from $A$ and $E$ are projected to $S$ and $T$ on the celestial sphere.}
\label{figure:celestial}
\end{figure}

The tangents at eight vertices of the subregion form a solid angle ${\rm d}\Omega$, as projected in the celestial sphere centered on the star (Fig. \ref{figure:celestial}). It is easy to find
${\rm d}\Omega={\rm d}\rho{\rm d}\tau$, where ${\rm d}\tau\simeq\sin\rho{\rm d}\varphi$, which can be derived by using the law of cosines in the spherical triangle $\triangle$MST.
In the dipole field geometry, one has $\rho\simeq 3/2\theta$ when $\theta$ is not so large (see Appendix A). Then, the solid angle reads
\begin{equation}
{\rm d}\Omega\simeq \frac{9}{4}\theta{\rm d}\theta{\rm d}\varphi.
\label{eq:sd}
\end{equation}
%

When deriving the emission power ${\rm d}P$ for this subregion, it should be noted that the arc length ${\rm d}s$ of the inner boundary line would be significantly larger than that of the outer boundary line when
$f_2>>f_1$, and the particle number density would vary remarkably with $f$ as well. Therefore, one needs to divide the whole subregion within $f_1$ and $f_2$ into a number of smaller ones with an interval ${\rm d}f$.
The volume of such an elementary region is (see Appendix B)
\begin{equation}
{\rm d}V_{\rm f}\simeq R_{\rm c}^3\theta^7f^2{\rm d}f{\rm d}\theta{\rm d}\varphi.
\label{eq:dv}
\end{equation}
The next step is to find the particle number density in ${\rm d}V_{\rm f}$ and calculate the emission power ${\rm d}P_{\rm f}$, where the subscription ``f'' indicates that these quantities are for the elementary region within $f$ and $f+{\rm d}f$. Then, the total emission power ${\rm d}P$ is the integration of ${\rm d}P_{\rm f}$ over $f$.

The altitude dependence of particle number density can be derived using the dipole field geometry and the assumption of free flow. According to the assumption, the initial total number density of
secondary electrons and positrons is
\begin{equation}
n_0=Mn_{\rm gj0},
\label{eq:n0}
\end{equation}
with the Gouldreich-Julian number density
\begin{equation}
n_{\rm gj0}\simeq |\mathbf{\Omega}\cdot\mathbf{B_{\rm s}}|/(2\pi ce)\simeq B_{\rm s}\cos\alpha/Pce,
\label{eq:ngj0}
\end{equation}
where $\mathbf{\Omega}$ is the rotation angular velocity, $B_{\rm s}$ is the surface magnetic field, $\alpha$ is the inclination angle, $c$ and $e$ are the light velocity and the electron charge,
respectively. In the case of free flow, when neglecting the trivial energy loss due to radiation, the particle number density follows the flux conservation law $nS=n_0S_0$, where $S$ and $S_0$
are the cross section areas of a flux tube at altitudes $r$ and $R$ (stellar radius), respectively. $S$ and $S_0$ are related to each other by the law of magnetic flux conservation, i.e. $BS=B_0S_0$.
Combing these relations with the dipole magnetic field strength $B\simeq B_{\rm s}(R/r)^3$, one has
\begin{equation}
n\simeq n_{\rm 0}(R/r)^3,
\label{eq:n}
\end{equation}

Under the basic assumptions of emission mechanism and propagation effect, the emission power from a coherent volume $\lambda^3$ is
\begin{equation}
\delta P_{\rm c}= (n\lambda^3)^2i_{\rm e0}(r/R)^{ q},
\label{eq:deltapc}
\end{equation}
where $i_{\rm e0}$, the averaged emission power of a single particle, is modified by a power-law term $(r/R)^{q}$ to represent the possible dependence on altitude. Note that $i_{\rm e0}$ may be a function of emission frequency, but in the following calculation we ignore such dependence. Its possible influence to frequency dependence of pulse profiles will be studied in a subsequent paper.

In a larger volume ${\rm d}V_{\rm f}$ that consists of many coherent volumes, the emissions are incoherently superposed, thus the
emission power is
\begin{equation}
{\rm d}P_{\rm f} = \frac{{\rm d}V_{\rm f}}{\lambda^3}\delta P_{\rm c}={\rm d}V_{\rm f} n^2{\lambda^3}i_{\rm e0}(r/R)^{q}.
\label{eq:dPf}
\end{equation}
Integrating over all the elementary regions from $f_1$ to $f_2$, one has the
total emission power from the subregion
\begin{equation}
{\rm d}P = \int_{f_1}^{f_2}{\rm d}P_{\rm f}
\label{eq:dP}
\end{equation}
Eventually, the emission intensity reads
\begin{equation}
I = \frac{{\rm d}P}{{\rm d}\Omega}
\label{eq:Lomg1}
\end{equation}

Substituting Eqs. (\ref{eq:sd})$-$(\ref{eq:dP}) into Eq. (\ref{eq:Lomg1}), and using the dipole field relations
$B_{\rm s}\simeq 3.2\times 10^{19}(P\dot{P})^{1/2}({\rm Gauss})$ and $r\simeq {4/9} R_{\rm e}\rho^2$ (for $r<<R_{\rm e}$),
we have the emission intensity (see Appendix B for details)
\begin{equation}
I=AP^{q-4}\dot{P}\cos^2\alpha \left(f_{1}^{q-3}-f_{2}^{q-3}\right)\rho^{2q-6},
\label{eq:Lomg2}
\end{equation}
where the coefficient is
\begin{equation}
A={1\over{3-q}}\left({{6.4\times10^{19}} \over {3ce}}\right)^2 \left( {{2c}\over{9\pi}}\right)^{q-3} R^{6-q} \lambda^3  M^2 i_{\rm e0}.
\label{eq:A}
\end{equation}

The last term in the right hand, $\rho^{2q-6}$, indicates a radius dependence of emission intensity resulting from the assumptions of coherent emission, plasma free flow and the index $q$, which have been qualitatively discussed before. When $q<3$, it predicts a limb-darkening relationship.

Beside this, the term $\left(f_{1}^{q-3}-f_{2}^{q-3}\right)$ also affects the
$I-\rho$ relationship. It may cause different trends in the central and outer parts of emission beam. To see this, without losing generality, we consider a flux tube located between the last open field lines ($f_{\rm om}=1$) and a layer of innermost open field lines with a fixed $f_{\rm im}$, where the initial particle density $n_0$ is constant on the polar cap and the coherent emission starts from the polar cap (relaxation of these assumptions will be discussed soon). In order to see the emission from all the field lines between $f_{\rm om}$ and $f_{\rm im}$, there must be $\rho\geq \rho_{\rm t}$, where the transition radius $\rho_{\rm t}$ equals the opening angle of the polar cap boundary $\rho_{\rm pc}\simeq(3/2)(R/R_{\rm c})^{1/2}$. In this case, namely, for the outer beam with $\rho\geq \rho_{\rm t}$, the term $\left(f_{\rm om}^{q-3}-f_{\rm im}^{q-3}\right)$ is a constant, and the intensity follows  $I\propto\rho^{2q-6}$. However, in the inner beam with $\rho\leq \rho_{\rm t}$, for any emission direction with $\rho$, the outermost visible field line is determined where the emission direction from its foot point on the polar cap is aligned with the LOS, which leads to $f_1=(9/4)(R/R_{\rm c})\rho^{-2}$. Therefore, the visible emission region is confined between $f_1$ and $f_2=f_{\rm im}$, which obviously shrinks, because $f_1>1$. Then,
Eq. (\ref{eq:Lomg2}) reduces to
\begin{eqnarray}
I_{\rm inner} =  AP^{-1}\dot{P}\cos^2\alpha \nonumber \\
\times \left[1-\left(\frac{9\pi R}{2c}\right)^{3-q} f_{\rm im}^{q-3}P^{q-3}\rho^{2q-6}\right].
\label{eq:Lomg3}
\end{eqnarray}
When $q<3$, it shows an opposite trend that the intensity increases with increasing radius in the inner beam.

In the above analysis, the polar cap boundary plays the role of transition location for intensity distributions, because we assume a uniform $n_0$, set the lowest coherent emission altitude to be $R$ and select the last open field lines as the outer boundary of flux tube. Any deviation from these three assumptions will lead to different transition location. For example, if a layer of inner open field lines is selected as the outer boundary of a flux tube, namely $f_{\rm om}>1$, or $n_0$ is peaked at an inner open field line within the flux tube, the transition location will move inwards to the magnetic pole, thereby $\rho_{\rm t}<\rho_{\rm pc}$. The later case can be seen in Figs. \ref{figure:discharge} and \ref{figure:model_B_u_a10} (Models B1 and B2). On the contrary, if the lowest coherent emission altitude is high above the polar cap, i.e. $r_{\rm L}>R$, while the other two assumptions remain, the transition location will have $\rho_{\rm t}\simeq\sqrt{r_{\rm L}/R}\rho_{\rm pc}>\rho_{\rm pc}$.

To summarize, the radial intensity distribution in the sub-beam formed by a flux tube is twofold: starting from the magnetic pole (or a place near the magnetic pole), the intensity increases first, reaching its maximum at $\rho=\rho_{\rm t}$ and then fades with increasing $\rho$ in a form of power law. This feature can be seen in the upper panels of Figs. \ref{figure:model_A_u_a10}-\ref{figure:model_B_nu_a80}.

\subsection{Transverse intensity distribution}

The transverse (azimuthal) intensity distribution in the beam depends on the azimuthal distribution of number density of secondary particles. Being enlightened by the idea of sparks in RS75 and GS00, it is assumed that there are some discharging flux tubes on the polar cap, but their transverse sizes are not necessary to be the same as those of sparks. The number density of secondary particles, or equivalently the multiplicity factor $M$, is probably a function of $f$ and $\varphi$ within a flux tube. Then, Eqs. (\ref{eq:Lomg2})-(\ref{eq:Lomg3}), which only apply to the homogeneous distribution of $M$, must be revised. For simplicity, the lowest coherent emission height is set as $r_{\rm L}=R$ in the following derivation. The effect of a higher $r_{\rm L}$ will be discussed by the end of this section.

We start from an elementary region within $f-f+{\rm d}f$ at a given $\varphi$ in a flux tube. Its intensity should be
${\rm d}I\propto (3-q)M(f,\varphi)^2f^{q-4}{\rm d}f$ (by replacing $f_1$ and $f_2$ with $f$ and $f+{\rm d}f$ in Eq. (\ref{eq:Lomg2}), respectively).
Then, for the outer part of the sub-beam, namely, $\rho>(3/2)\sqrt{R/(R_{\rm c}f_1)}$, the total intensity from the whole flux tube reads
\begin{eqnarray}
I_{\rm outer}(\varphi)=A_1P^{q-4}\dot{P}\cos^2\alpha\rho^{2q-6} \nonumber \\
 \times \int_{f_1}^{f_2}M(f,\varphi)^2f^{q-4}{\rm d}f,
\label{eq:Lomg_varphi_out}
\end{eqnarray}
where the coefficient is
\begin{equation}
A_1=\left({{6.4\times10^{19}} \over {3ce}}\right)^2 \left( {{2c}\over{9\pi}}\right)^{q-3} R^{6-q}\lambda^3 i_{\rm e0}.
\label{eq:A_1}
\end{equation}
Note that the boundaries $f_1$ and $f_2$ may also be a function of $\varphi$, depending on the shape of the cross section of a flux tube on the polar cap.

For the inner part where $\rho\lesssim (3/2)\sqrt{R/(R_{\rm c}f_1)}$, the outmost visible field line corresponding to a given $\rho$ shrinks inwards, which satisfies $f_{\rm o}=9R/(4R_{\rm c}\rho^{2})$. Thus the lower boundary of integration should be determined by $f_1^{\prime}=\min[f_{\rm o}, f_1]$, where $f_1$ is the real outer boundary of the flux tube at a given azimuth $\varphi$. Then the
the total intensity is
%
\begin{eqnarray}
I_{\rm inner}(\varphi)=A_1P^{q-4}\dot{P}\cos^2\alpha\rho^{2q-6} \nonumber \\
\times \int_{f_1^{\prime}}^{f_2}M(f,\varphi)^2f^{q-4}{\rm d}f.
\label{eq:Lomg_varphi_in}
\end{eqnarray}

To explore how the geometry of flux tubes and the secondary particle distribution affect the beam geometry, we consider two kinds of models below.

In the first case (Model A), $M$ is assumed to be constant along the colatitude dimension but follows a gaussian distribution in the azimuthal dimension, peaked at the central azimuth $\varphi_0$ of the flux tube, i.e. $M=M_0\exp[-(\varphi-\varphi_0)^2/(2\sigma^2)]$.
The flux tube is assumed to be within [$f_1, f_2$] for all the azimuth angles, where $f_1$ and $f_2$ are constant. For the intensity in the outer part, Eq.
\ref{eq:Lomg_varphi_out} is modified as
%
\begin{eqnarray}
I_{\rm outer}^{\rm A}(\varphi)=A^\prime P^{q-4}\dot{P}\cos^2\alpha {\rm e}^{-(\varphi-\varphi_0)^2/\sigma^2}  \nonumber \\
\times \left(f_{1}^{q-3}-f_{2}^{q-3}\right)\rho^{2q-6},
\label{eq:Lout_caseA}
\end{eqnarray}
where the coefficient is
\begin{equation}
A^\prime={1\over{3-q}}\left({{6.4\times10^{19}} \over {3ce}}\right)^2 \left( {{2c}\over{9\pi}}\right)^{q-3} R^{6-q} \lambda^3  M_0^2 i_{\rm e0}.
\label{eq:A_1_caseA}
\end{equation}
For the intensity in the inner part, Eq. (\ref{eq:Lomg_varphi_in}) is modified as
%
\begin{eqnarray}
I_{\rm inner}^{\rm A}(\varphi)=A^\prime P^{-1}\dot{P}\cos^2\alpha {\rm e}^{-(\varphi-\varphi_0)^2/\sigma^2} \nonumber \\
\times \left[1-\left(\frac{9\pi R}{2c}\right)^{3-q}f_{\rm im}^{q-3}P^{q-3}\rho^{2q-6}\right].
\label{eq:Lin_caseA}
\end{eqnarray}

In the second case (Model B), it is assumed that the cross section of a flux tube is circular and $M$ follows a gaussian distribution around the center of the flux tube. For convenience, a dimensionless colatitude $\chi\equiv\theta/\theta_{\rm pc}$ is used below to describe the latitudinal position of the foot point of a field line on the polar cap, where $\theta$
and $\theta_{\rm pc}$ are the polar angles of the foot points of an inner and the last open field lines, respectively. $\chi$ and
 $f$ can be converted to each other by
\begin{equation}
\chi={3\over 2}\left(\frac{R}{fR_{\rm c}}\right)^{1/2}\rho_{\rm pc}^{-1}.
\label{eq:chi}
\end{equation}
Given the position of the center ($\chi_0$, $\varphi_0$) and
the dimensionless radius $\Re_0$ of the flux-tube cross section ($\Re_0\le 1-\chi_0$), the Gaussian distribution of $M$ is $M=M_0\exp{[-\Re^2/(2\sigma^2)]}$, where $\Re$ is the dimensionless angular distance from an arbitrary point to the flux tube center. Then the intensity of the outer part reads
%
\begin{eqnarray}
I_{\rm outer}^{\rm B}(\varphi)=A_2P^{q-4}\dot{P}\cos^2\alpha\rho^{2q-6} \nonumber \\
\times \int_{f_1}^{f_2}{\rm e}^{-\frac{\chi_0^2+\chi^2-2\chi_0\chi\cos(\varphi-\varphi_0)}{\sigma^2}}f^{q-4}{\rm d}f,
\label{eq:Lout_caseB}
\end{eqnarray}
where
\begin{equation}
A_2=\left({{6.4\times10^{19}} \over {3ce}}\right)^2 \left( {{2c}\over{9\pi}}\right)^{q-3} R^{6-q}\lambda^3 i_{\rm e0} M_0^2,
\label{eq:A_2}
\end{equation}
with the boundary
$$f_{1,2}=(9/4)(R/R_{\rm c})[\rho_{\rm pc}\Re_0\sin\vartheta_{1,2}/\sin(\varphi-\varphi_0)]^{-2},$$ where $\vartheta_{1,2}=\arccos[\sin(\varphi-\varphi_0)^2 \chi_0/\Re_0\mp\cos(\varphi-\varphi_0)\sqrt{1-(\chi_0/\Re_0)^2\sin(\varphi-\varphi_0)^2}]$ (see Appendix C for derivation).
In the inner part of the sub-beam, there is
%
\begin{eqnarray}
I_{\rm inner}^{\rm B}(\varphi)=A_2P^{q-4}\dot{P}\cos^2\alpha\rho^{2q-6} \nonumber \\
\times \int_{f_1^\prime}^{f_2}{\rm e}^{-\frac{\chi_0^2+\chi^2-2\chi_0\chi\cos(\varphi-\varphi_0)}{\sigma^2}}f^{q-4}{\rm d}f.
\label{eq:Lin_caseB}
\end{eqnarray}

In fact, Eqs. (\ref{eq:Lout_caseA})-(\ref{eq:Lin_caseA}) and (\ref{eq:Lout_caseB})-(\ref{eq:Lin_caseB}) give the full description for the radial and transverse intensity distribution of a sub-beam in Models A and B, respectively. The beam shape and resultant profiles will be modeled in the following subsection.

In the above derivation it is assumed that the coherent emission starts from the polar cap. If the lowest emission altitude is higher, the transition radius for the outer and inner parts of beam should be $\rho_{\rm t}=(3/2)\sqrt{r_{\rm L}/(R_{\rm c}f_1)}$.

\subsection{Modeling beam patterns and average pulse profiles}
The above formulae only apply to a single flux tube; to form a global beam pattern, one needs additional assumptions on the configuration of discharging flux tubes across the polar cap, including the locations of flux tubes and the multiplicity factors therein (may be inhomogeneous). The following simulations for Models A and B are presented as simple examples. There are a vast variety of configurations of discharging areas and particle density distributions, whose observational consequences differ in details. However, these simple examples are still useful to infer the necessary configurations that can account for observations.

\begin{figure*}
\centering
\resizebox{15cm}{9cm}
{
\includegraphics{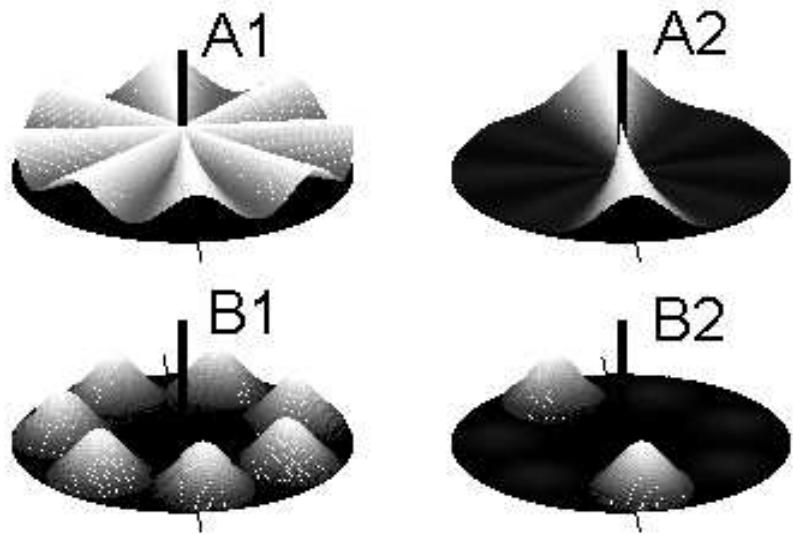}
}

\caption{Four models of density distribution of secondary plasma across the polar cap (the dark ellipse in each panel). The discharging flux tubes are assumed to be 8 in Models A1 and A2, and 7 in Models B1 and B2. In Model A, the cross section of a flux tube is sector-shaped, and the particle density is a Gaussian function that only depends on the magnetic azimuth $\varphi$. In Model B, the cross section is circular, and the particle density is a two dimensional Gaussian function peaked at the center of a flux tube.
The central vertical line represents the magnetic pole. The dashed line stands for the projection of the meridian plane. In Models B1 and B2, the flux tubes are located in the outer part of polar cap. The heights of the bumps are proportional to the local particle densities. In Models A1 and B1, the maximum particle densities are the same for all the flux tubes. In Models A2 and B2, two flux tubes are dominant, with the maximum number density being 10 times the others.}
\label{figure:discharge}
\end{figure*}

The simulations follow two steps. In the first step we designate the geometry of flux tubes. In model A, 8 discharging areas with sector-shaped cross sections are assumed to exist on the polar cap, each occupying an azimuthal range of $45^{\rm o}$ (see Fig. \ref{figure:discharge}). Their central azimuths are located every $45^{\rm o}$ starting from $\varphi=0$. In model B, the discharging areas are assumed to be circular, which are all centered at $\chi_0=0.7$ with a radius $\Re_0=0.3$. The centers are placed every $51.5^{\rm o}$ in azimuth dimension starting from $\varphi=-5^{\rm o}$, so there are totally 7 discharge areas (see Fig. \ref{figure:discharge}).

In the second step we assign the multiplicity distribution across the polar cap. It is of particular interest to see how the inhomogeneity of the multiplicity among flux tubes affects the beam pattern and average pulse profiles. Therefore, we investigate two subclasses for each model: one is the homogeneous case that all the discharging areas have the same multiplicity pattern, which are called models A1 and B1, respectively, the other is the inhomogeneous case that two areas are dominant in pair production over the others, which is called model A2 and B2. The dominant areas are centered at $\varphi=0$ and $180^{\rm o}$ in Model A2 and centered at $\varphi=-5^{\rm o}$ and $149^{\rm o}$ in Model B2. In both of these sub models, the maximum multiplicity $M_0$ in the dominant areas (at central azimuth in Model A2 and at center of the area in Model B2) are 10 times of those in the other areas.

In the following simulation the index $q$ is fixed as 2, close to the values constrained from a sample of pulsars in Section 3.2. The lowest emission altitude is set as $r_{\rm L}=R$ for simplicity.

\begin{figure*}
\centering
\resizebox{16cm}{6cm}
{
\includegraphics{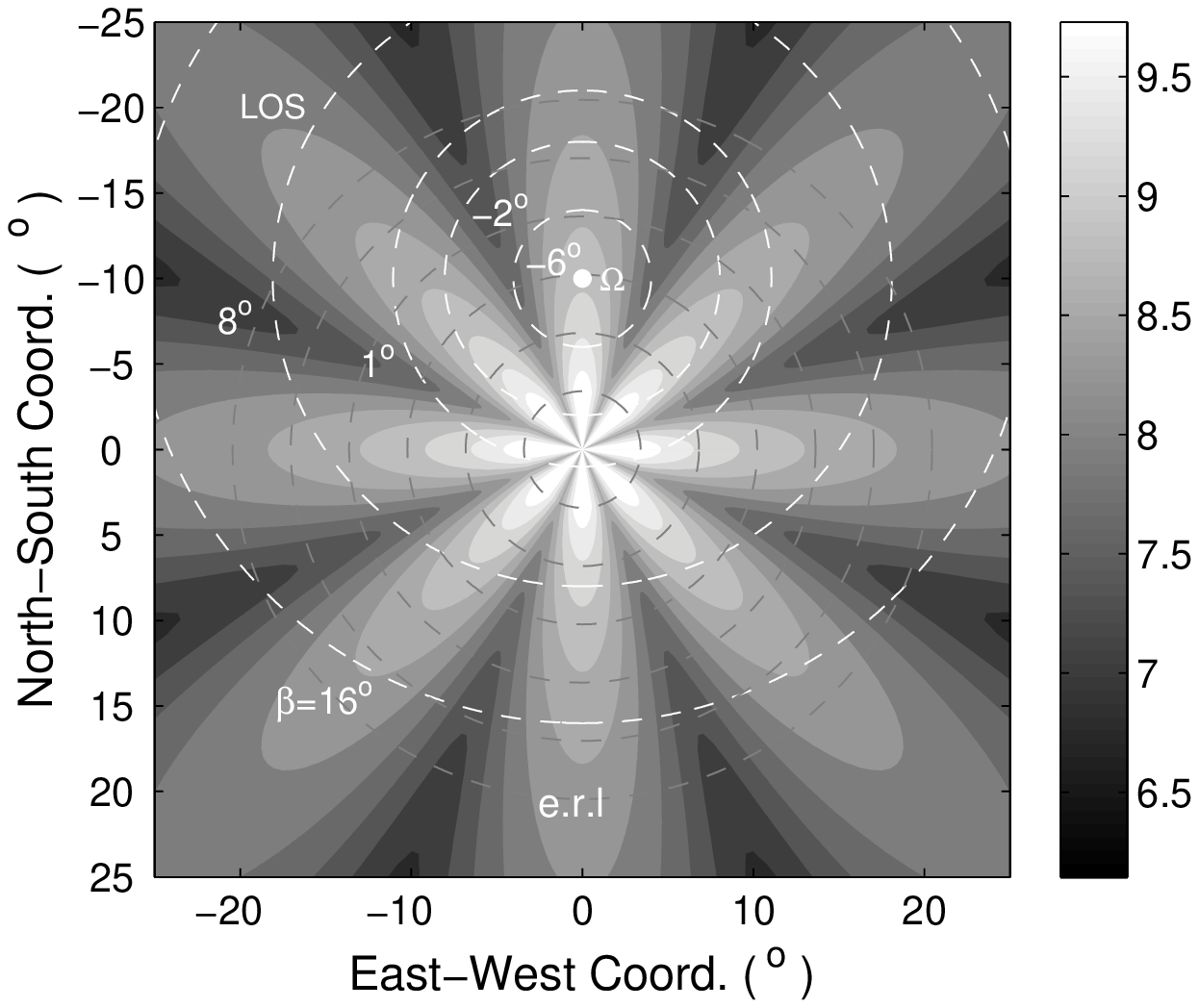}
\includegraphics{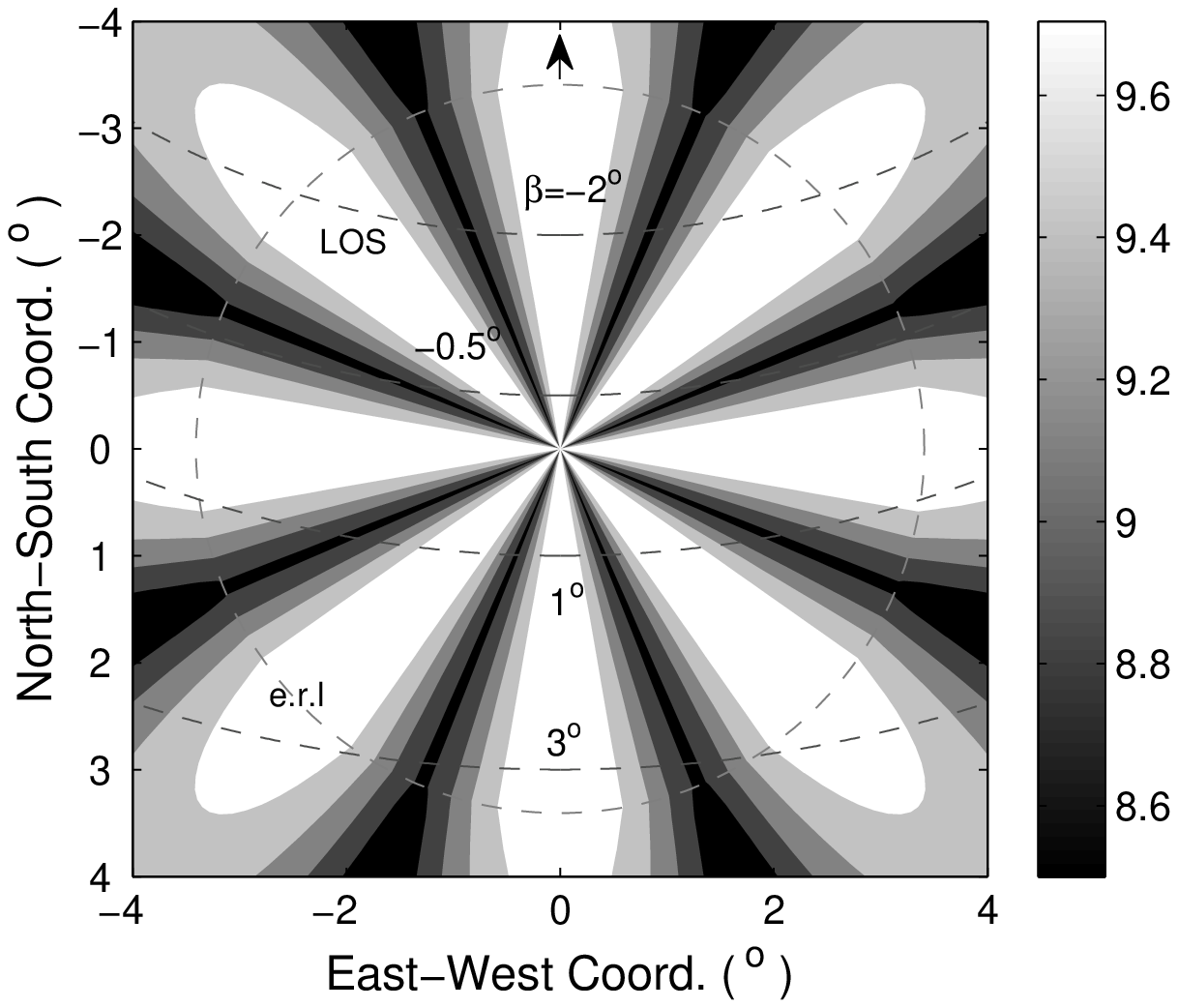}
}
\resizebox{13cm}{8cm}
{
\includegraphics{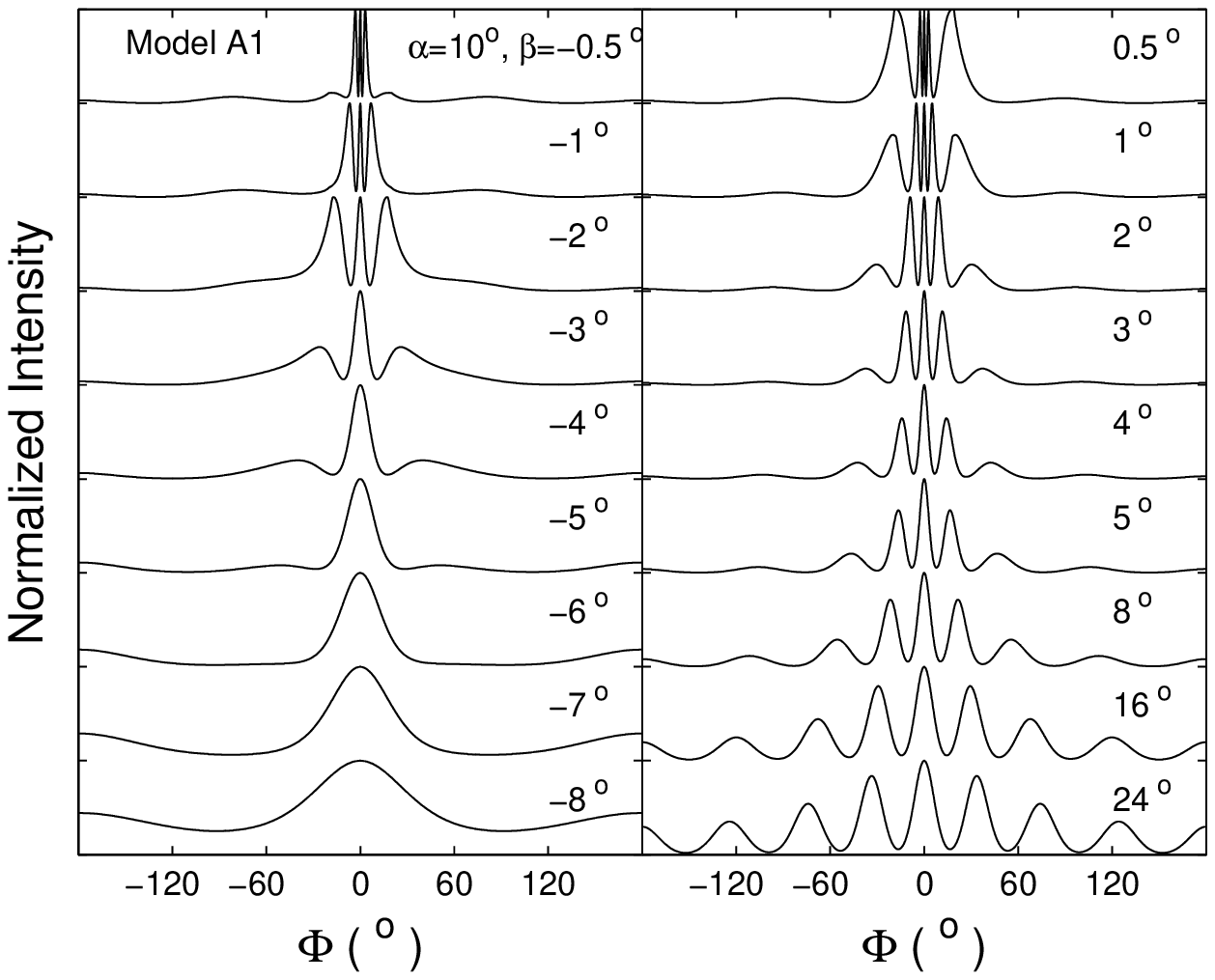}
}

\caption{The modeled beam and pulse profiles for Model A1, where 8 discharging flux tubes with sector-shaped cross section are distributed in the polar cap. The top left panel is the beam projected on the viewing plane when looking down to the magnetic pole. The pole is located at the center (0,0), and the white dot marked with the symbol ``$\Omega$'' stands for the spin axis. In this case, the inclination angle is $\alpha=10^{\rm o}$. The ``North-South'' coordinate means the angular distance from the magnetic pole toward or away from the spin axis, and ``East-West'' coordinate is the angular distance along the direction perpendicular to magnetic pole-spin axis baseline. The white dashed circles around the spin axis, one of which marked with ``LOS'', represent a set of line of sights, with the corresponding impact angle marked for each line. A group of 6 gray dashed circles centered on magnetic pole, the outmost one of which is marked with ``e.r.l'', are the equal-radius lines spanning from $\rho_{\rm pc}$ to $6\rho_{\rm pc}$ with a step of $\rho_{\rm pc}$. The intensity contours are plotted in logarithmic scale in arbitrary units, stepping outwards by a factor of $1/2$. Since all the flux tubes have identical particle distribution, the sub-beams also have the same pattern of intensity distribution. The limb-darkening index and the lowest emission altitude are set as $\delta=-2$ and $r_{\rm L}=R$ in the calculation, respectively. The top right panel is the zoom-in picture of the central part of the top left panel, reflecting the beam structure near and within the polar cap boundary with the radius of $\rho_{\rm pc}$, as presented by a gray circle marked with ``e.r.l''. The arrow indicates the direction toward the spin axis. Four lines of sight from $\beta=-2^{\rm o}$ to $3^{\rm o}$ are shown by darker dashed lines. The lower panel are pulse profiles calculated with $\alpha=10^{\rm o}$ for a number of $\beta$ angles, which are marked beside individual profiles.}
\label{figure:model_A_u_a10}
\end{figure*}

\begin{figure*}
\centering
\resizebox{16cm}{6cm}
{
\includegraphics{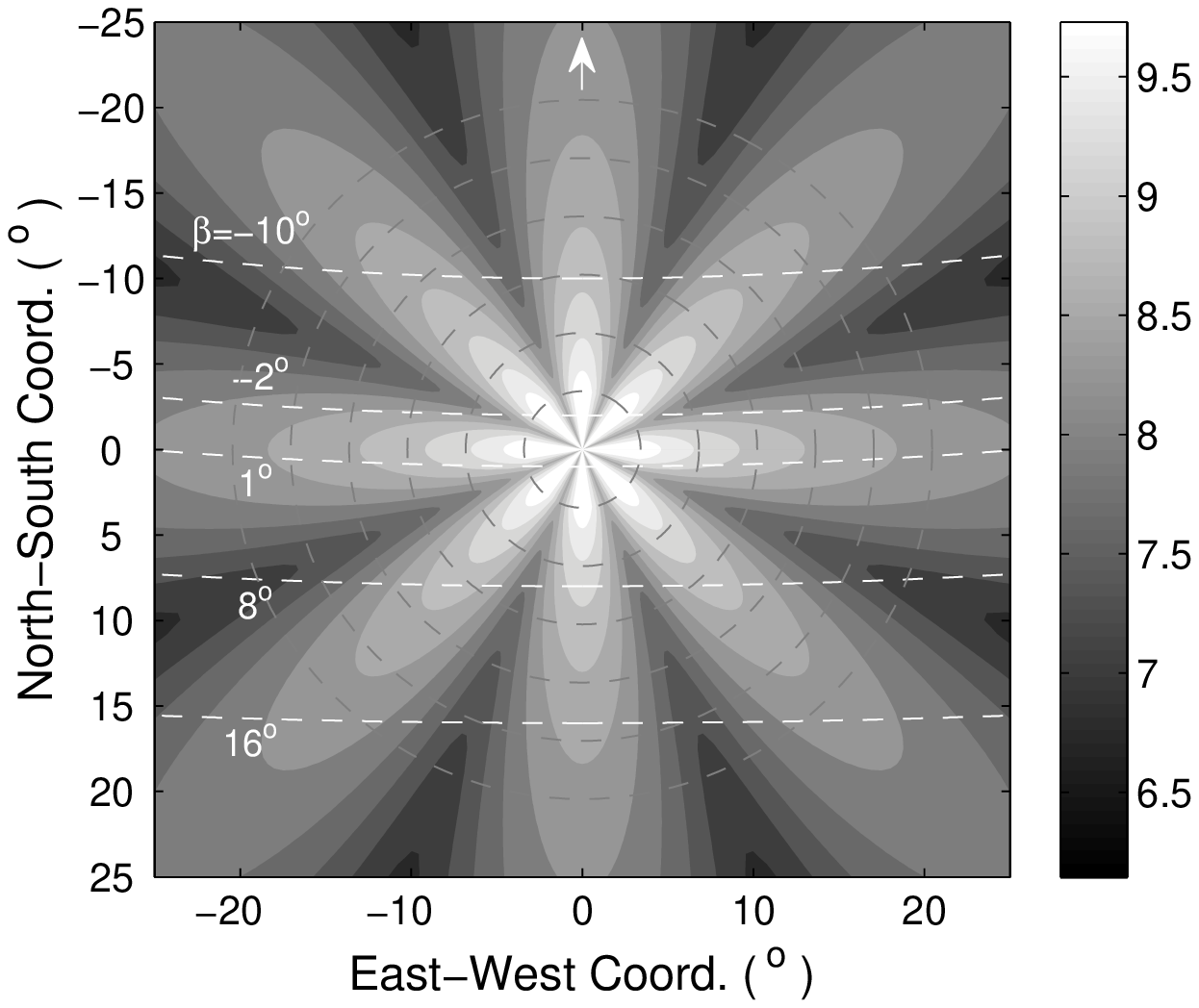}
\includegraphics{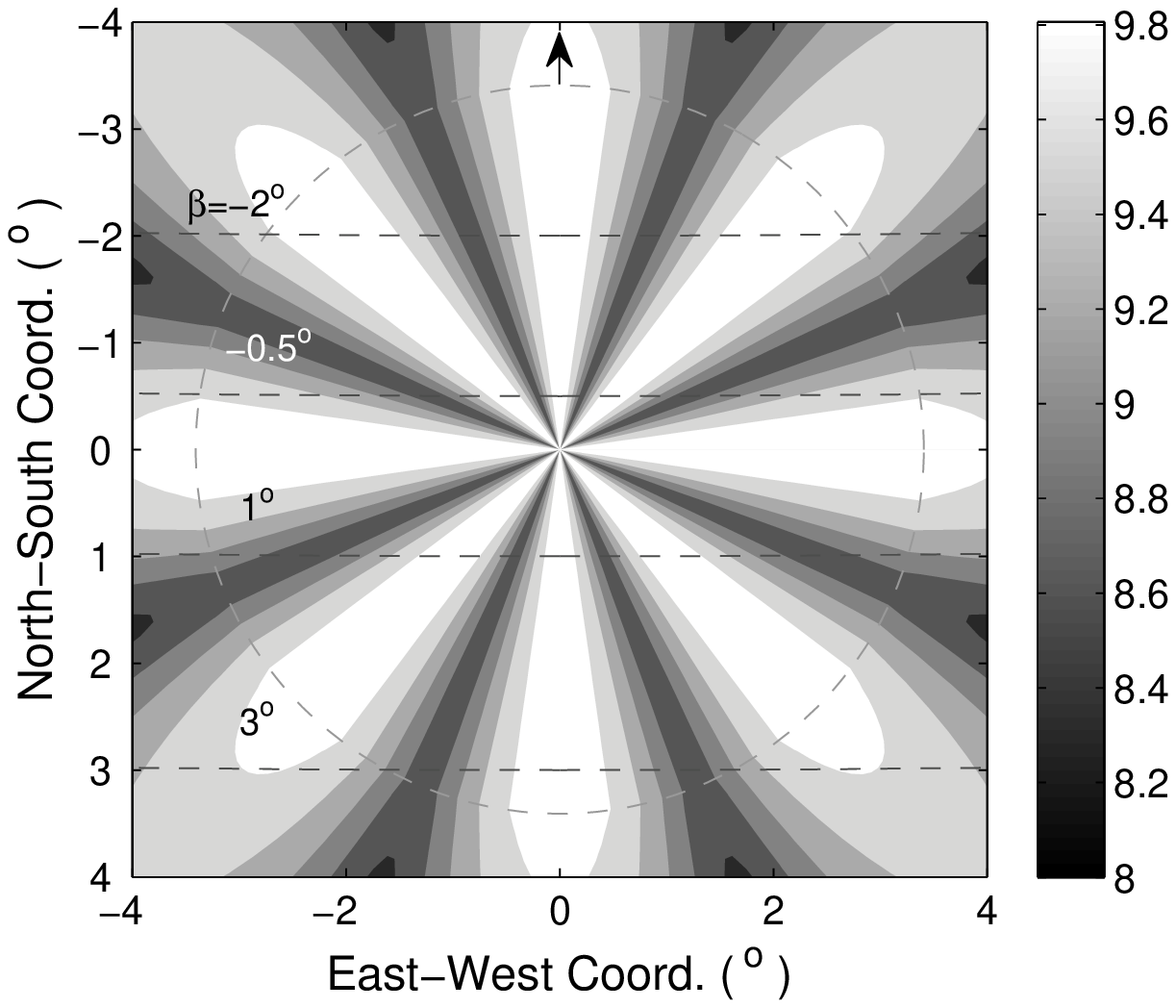}
}
\resizebox{15cm}{11cm}
{
\includegraphics{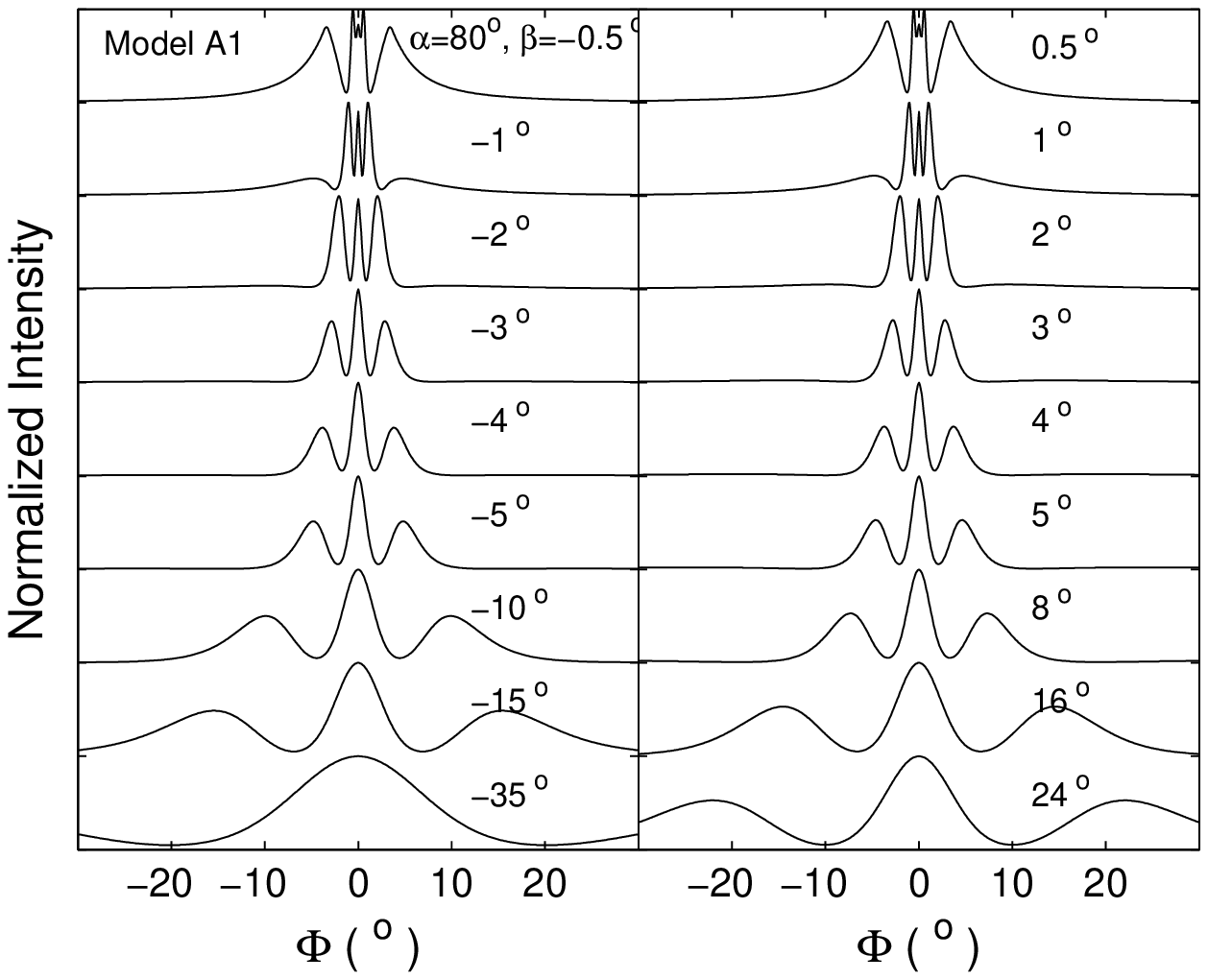}
}

\caption{The simulated beam and pulse profiles for Model A1. As Fig. \ref{figure:model_A_u_a10} except $\alpha=80^{\rm o}$. The arrows indicate the direction toward the spin axis, which is out of the top boundaries of the plots. }
\label{figure:model_A_u_a80}
\end{figure*}

\begin{figure*}
\centering
\resizebox{16cm}{6cm}
{
\includegraphics{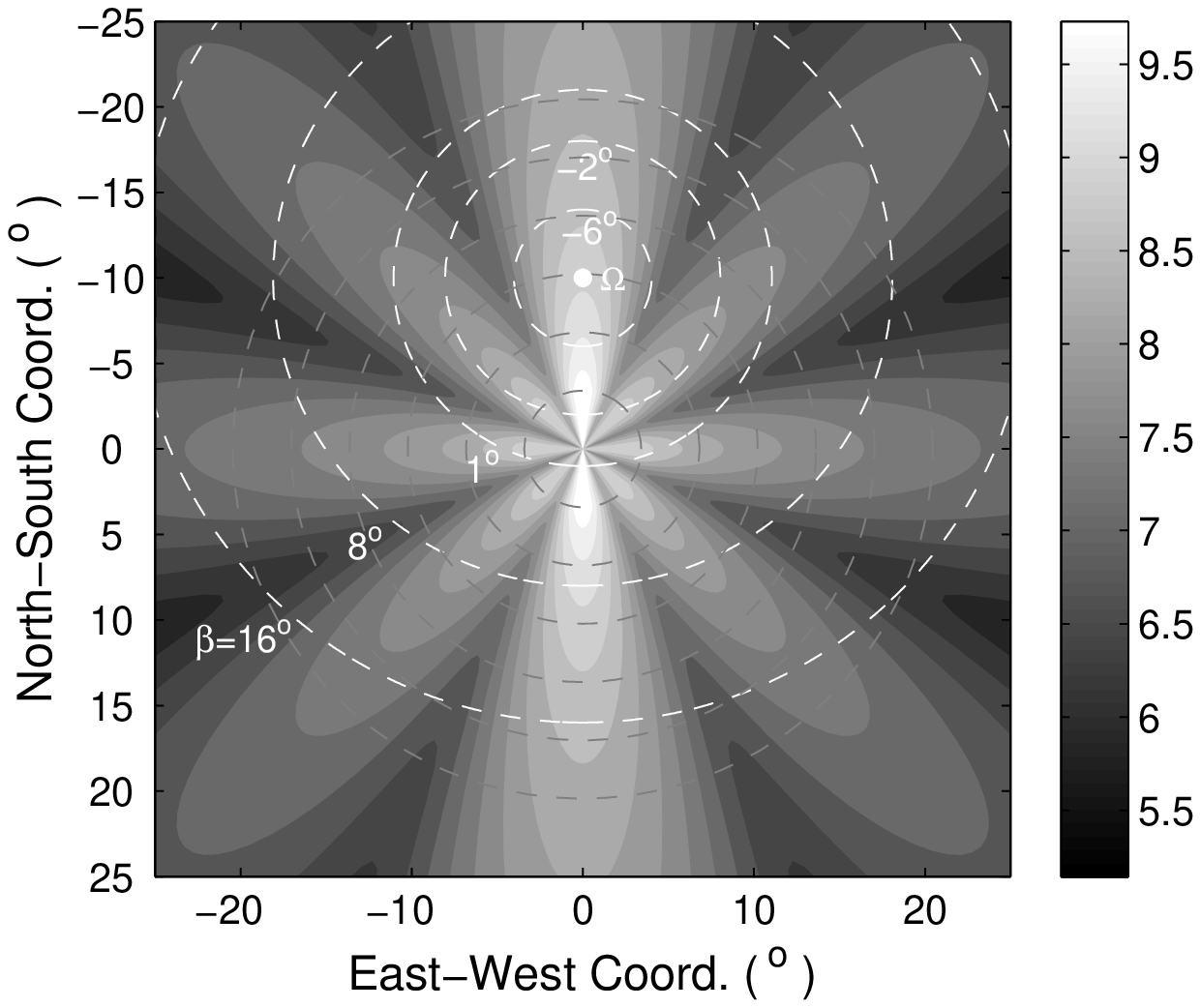}
\includegraphics{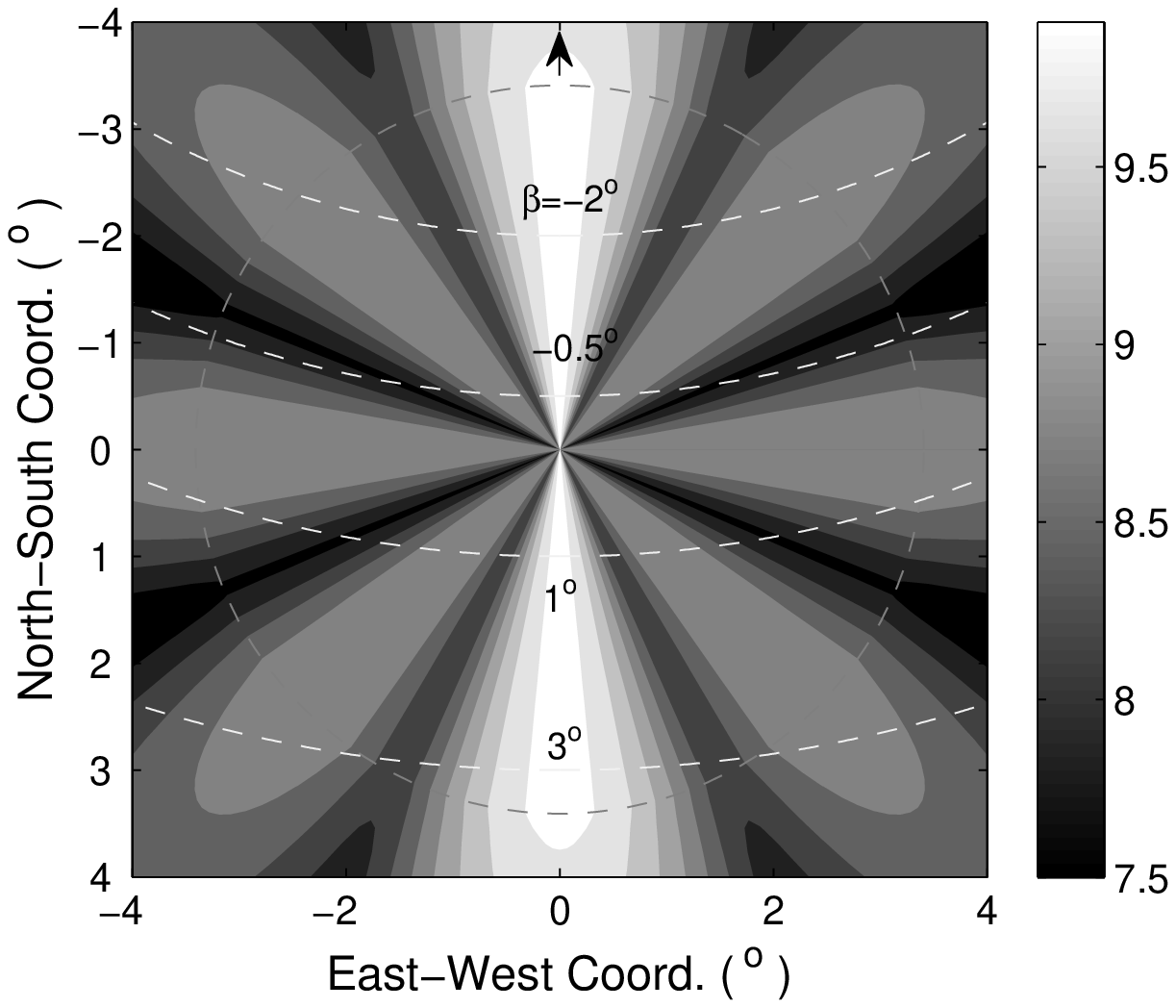}
}
\resizebox{15cm}{11cm}
{
\includegraphics{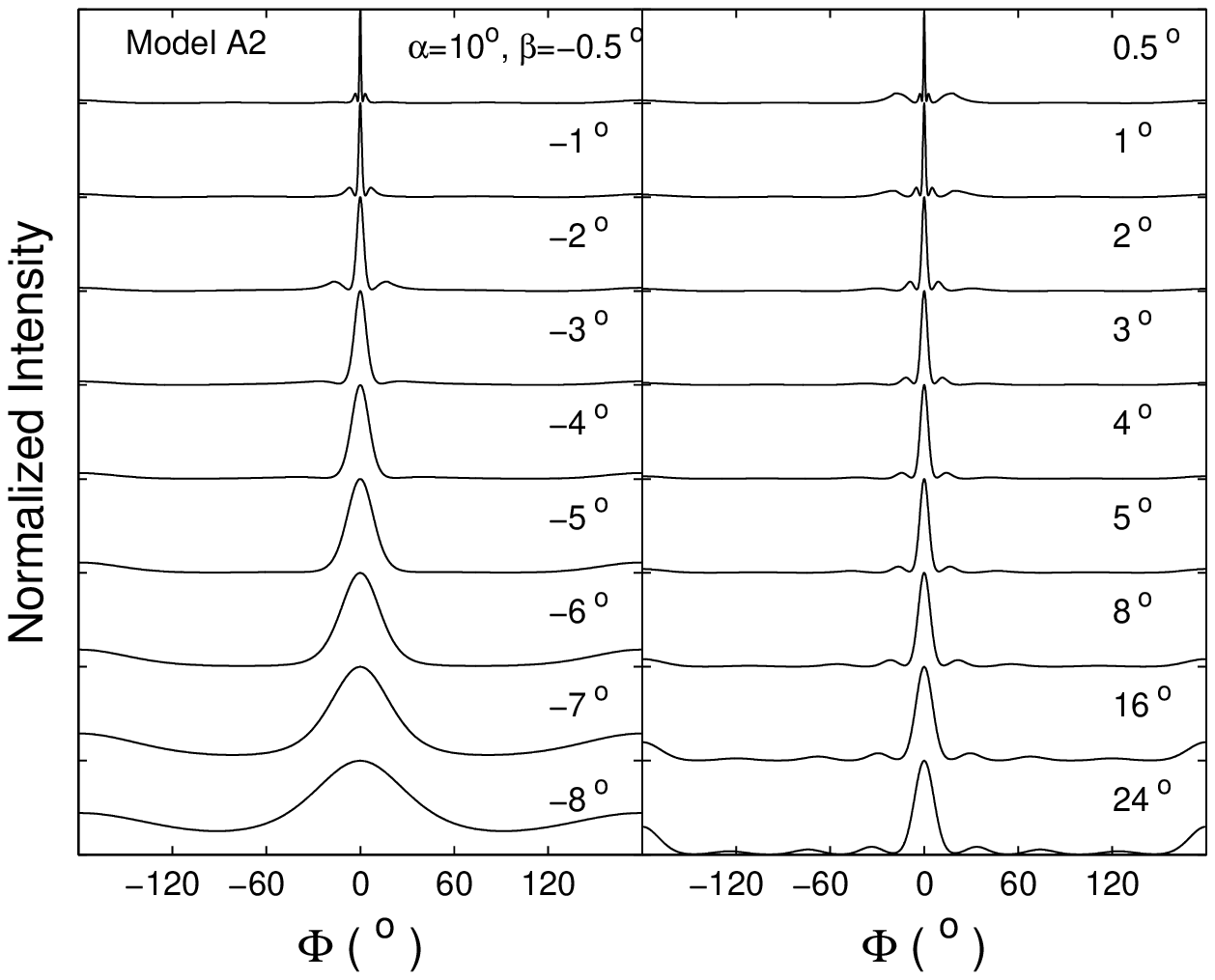}
}

\caption{The beam and pulse profiles for Model A2, where two flux tubes are dominant in particle density. The two sub-beams produced by the dominant flux tubes are more luminous than the others. See Fig. \ref{figure:model_A_u_a10} for details.}
\label{figure:model_A_nu_a10}
\end{figure*}

\begin{figure*}
\centering
\resizebox{16cm}{6cm}
{
\includegraphics{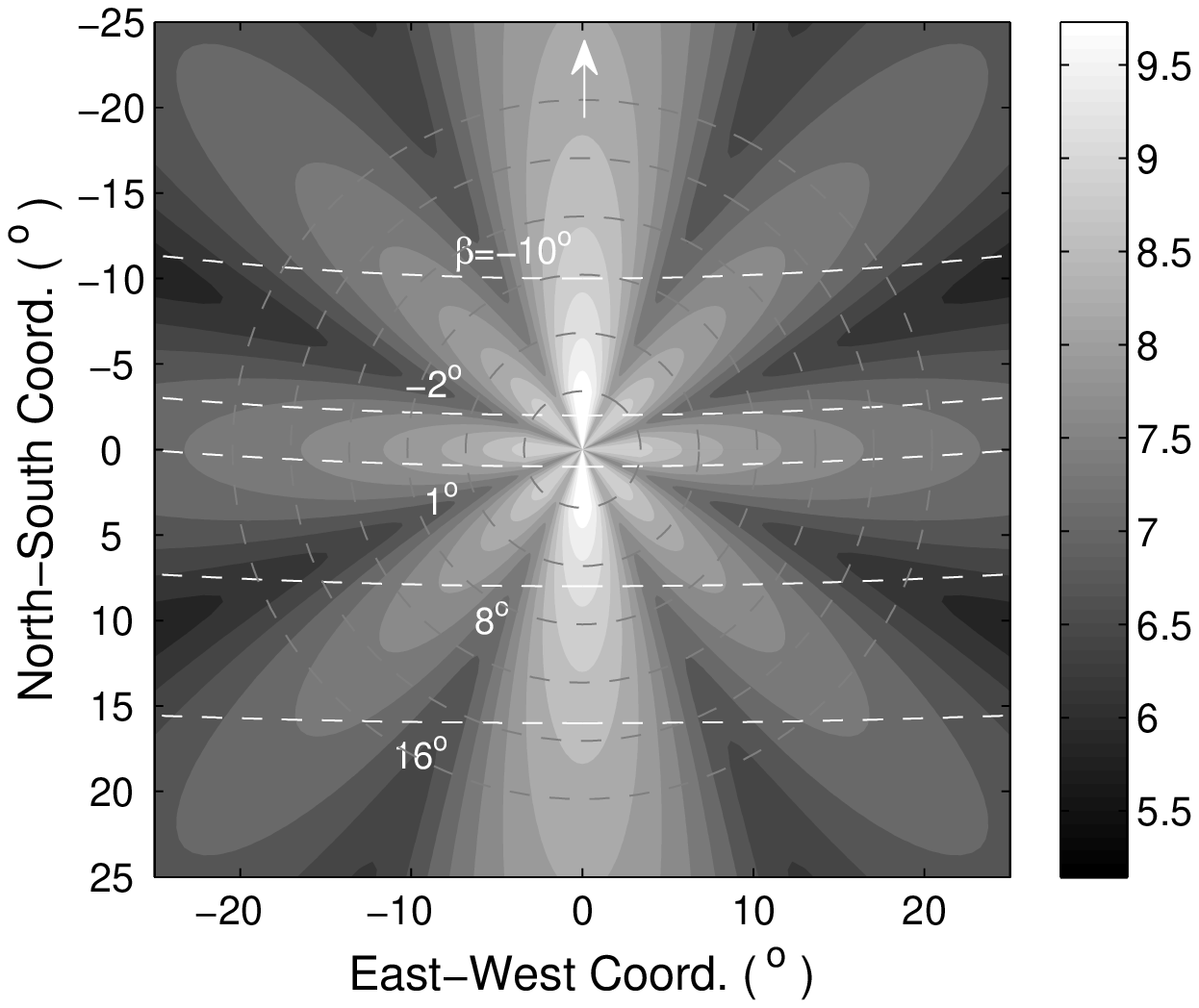}
\includegraphics{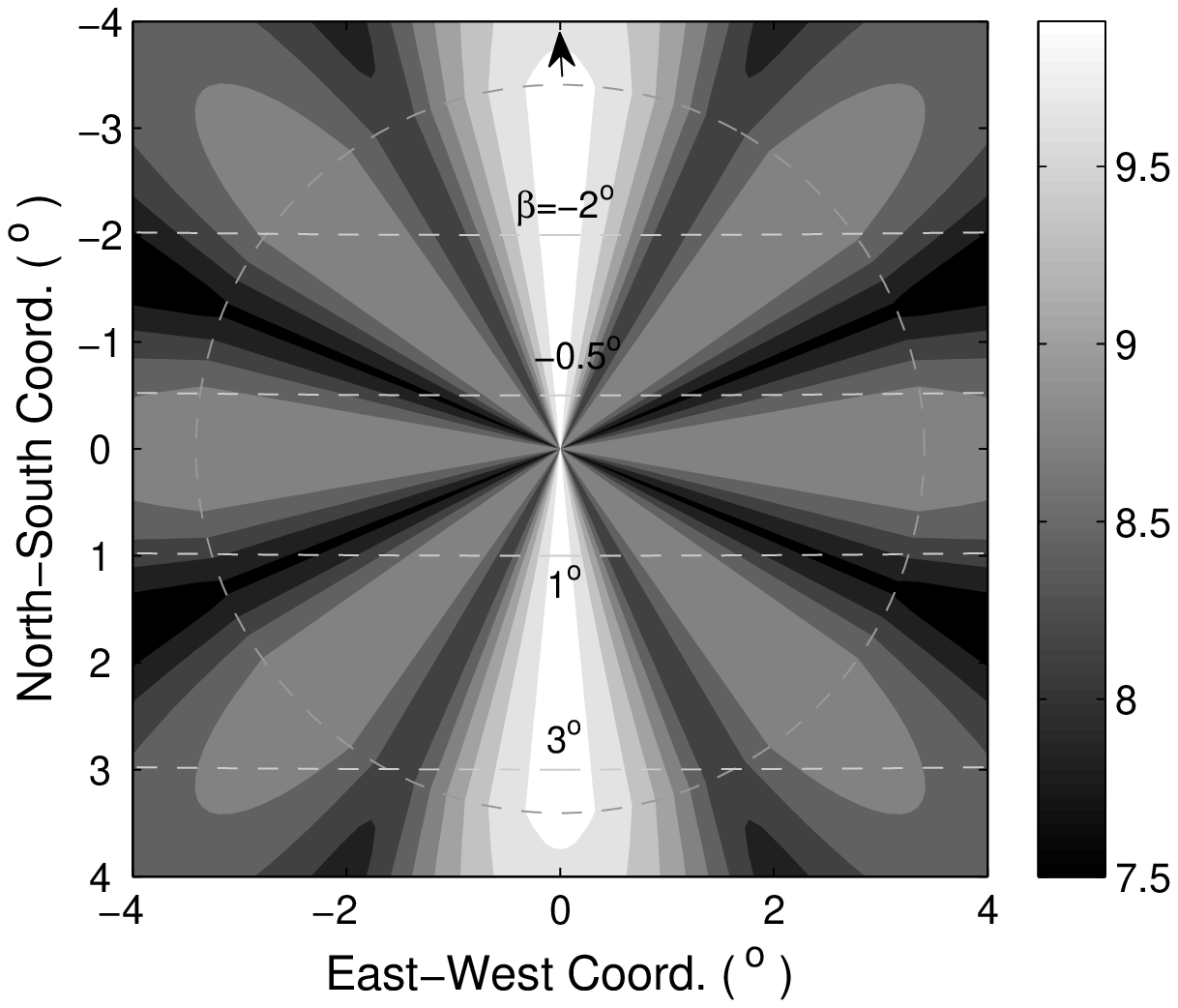}
}
\resizebox{15cm}{11cm}
{
\includegraphics{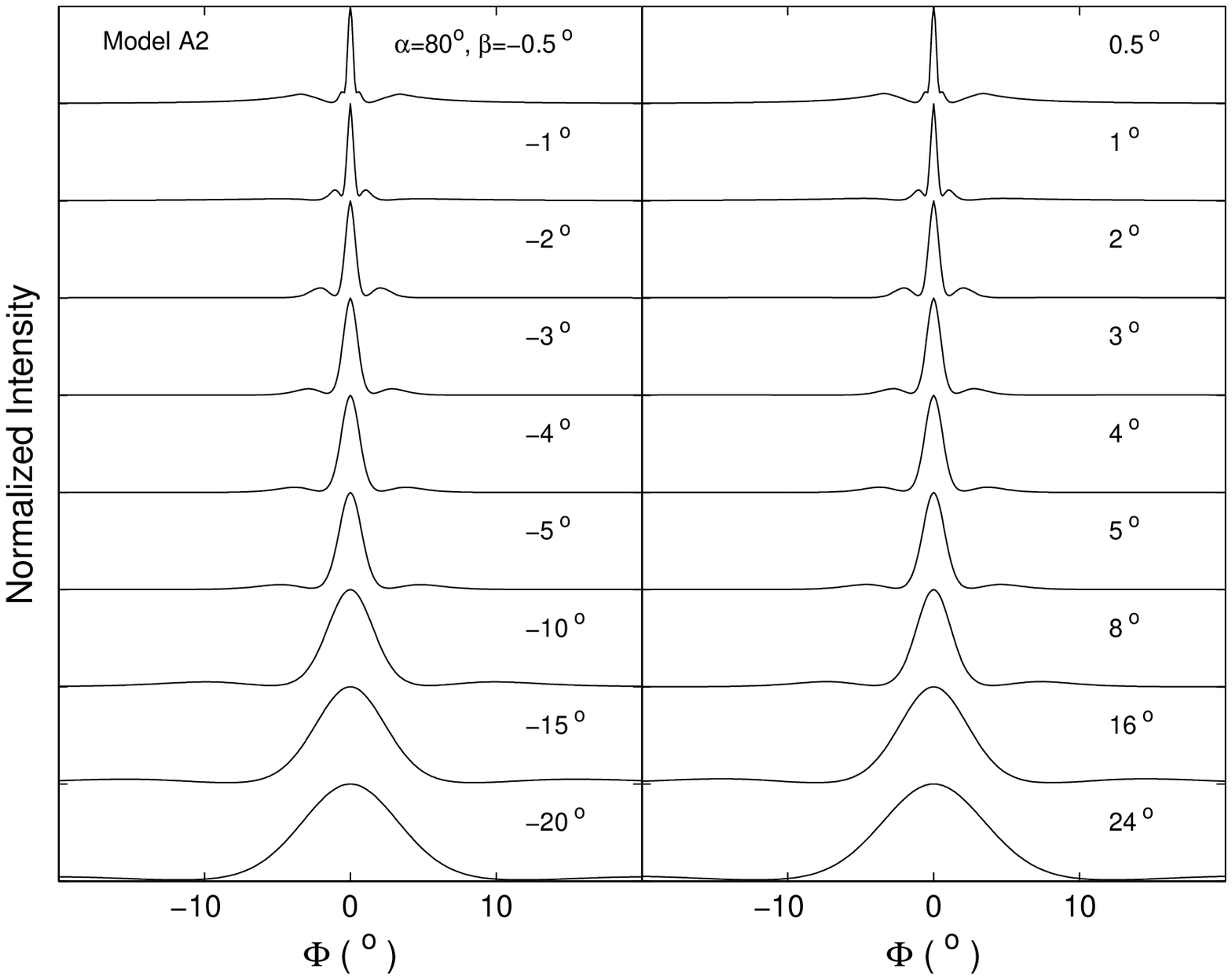}
}

\caption{The modeled beam and pulse profiles for Model A2. As Fig. \ref{figure:model_A_nu_a10} except that $\alpha=80^{\rm o}$. }
\label{figure:model_A_nu_a80}
\end{figure*}

\begin{figure*}
\centering
\resizebox{16cm}{6cm}
{
\includegraphics{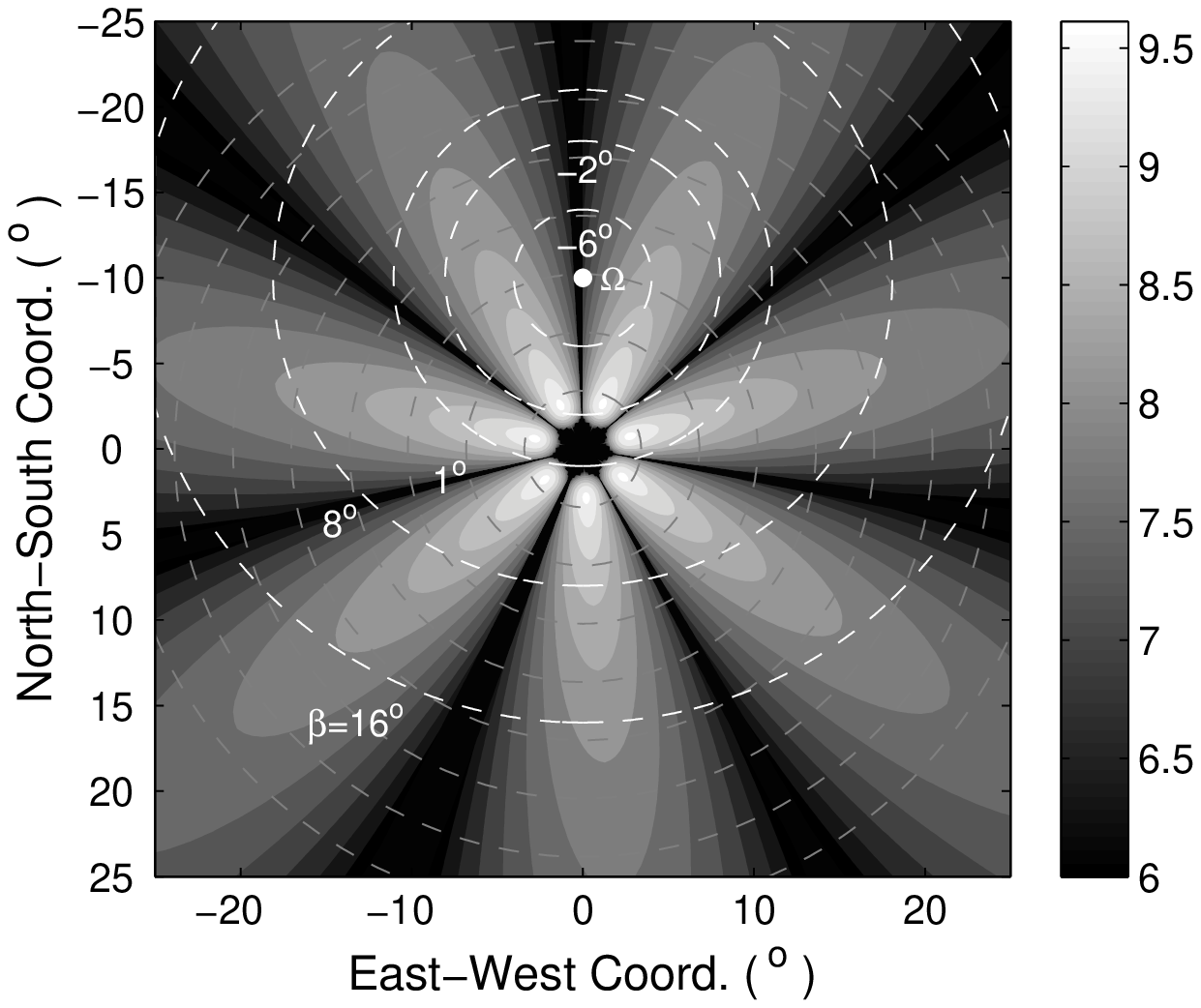}
\includegraphics{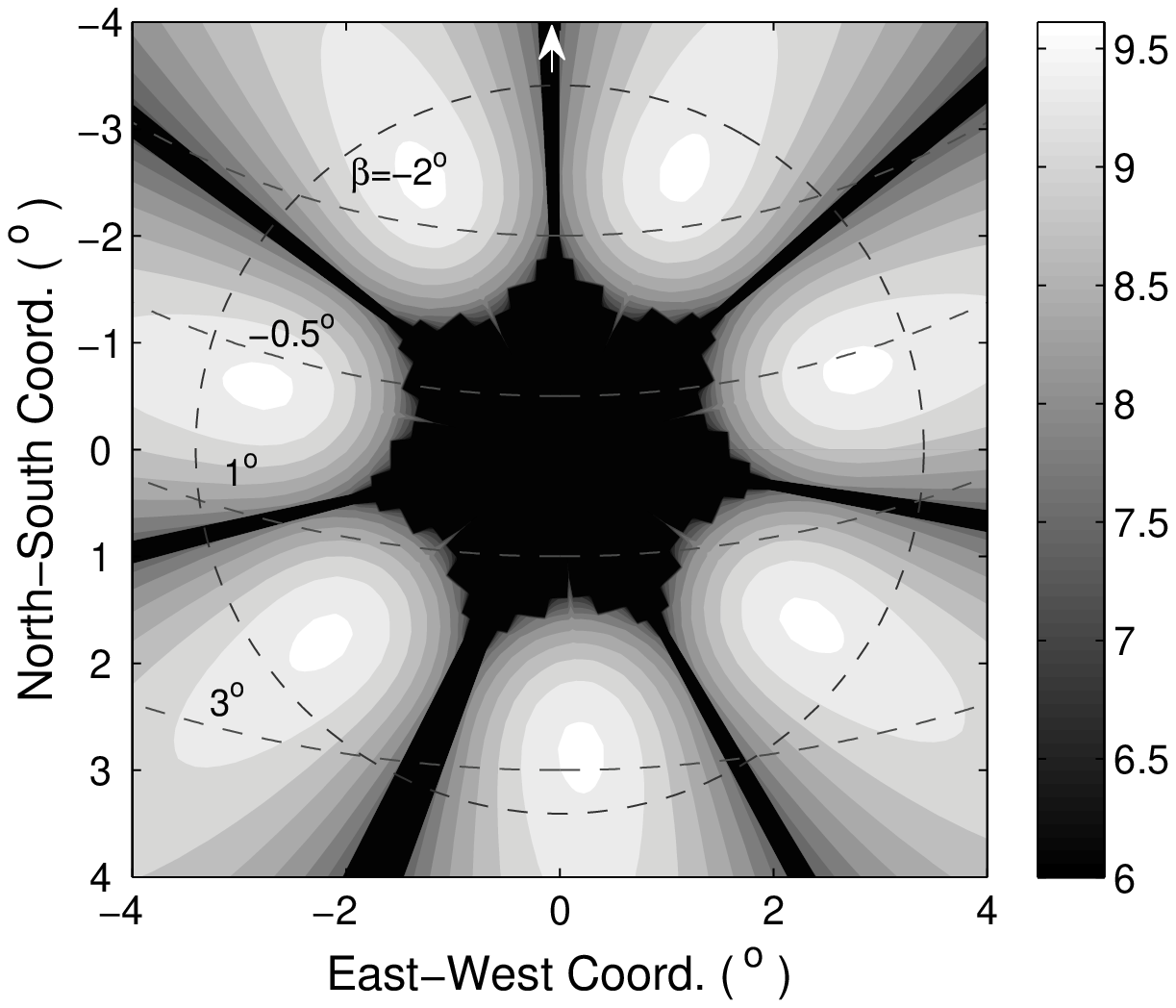}
}
\resizebox{15cm}{11cm}
{
\includegraphics{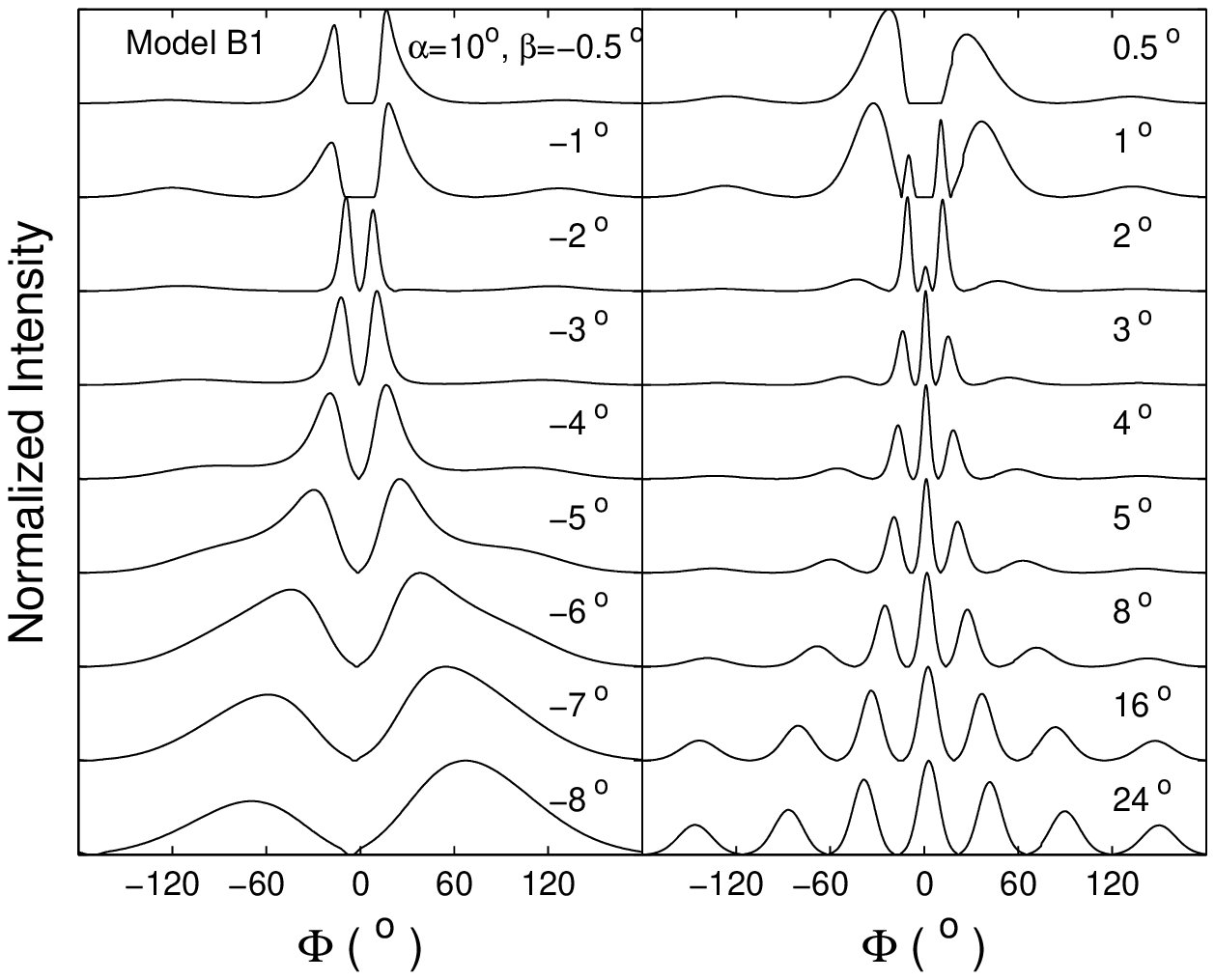}
}

\caption{The beam and pulse profiles for Model B1. 7 identical discharging flux tubes are assumed to be located in the outer part of polar cap, leaving a small dark region around the beam center. The inclination angle is $\alpha=10^{\rm o}$. See Fig. \ref{figure:model_A_u_a10} for other details.}
\label{figure:model_B_u_a10}
\end{figure*}

\begin{figure*}
\centering
\resizebox{16cm}{6cm}
{
\includegraphics{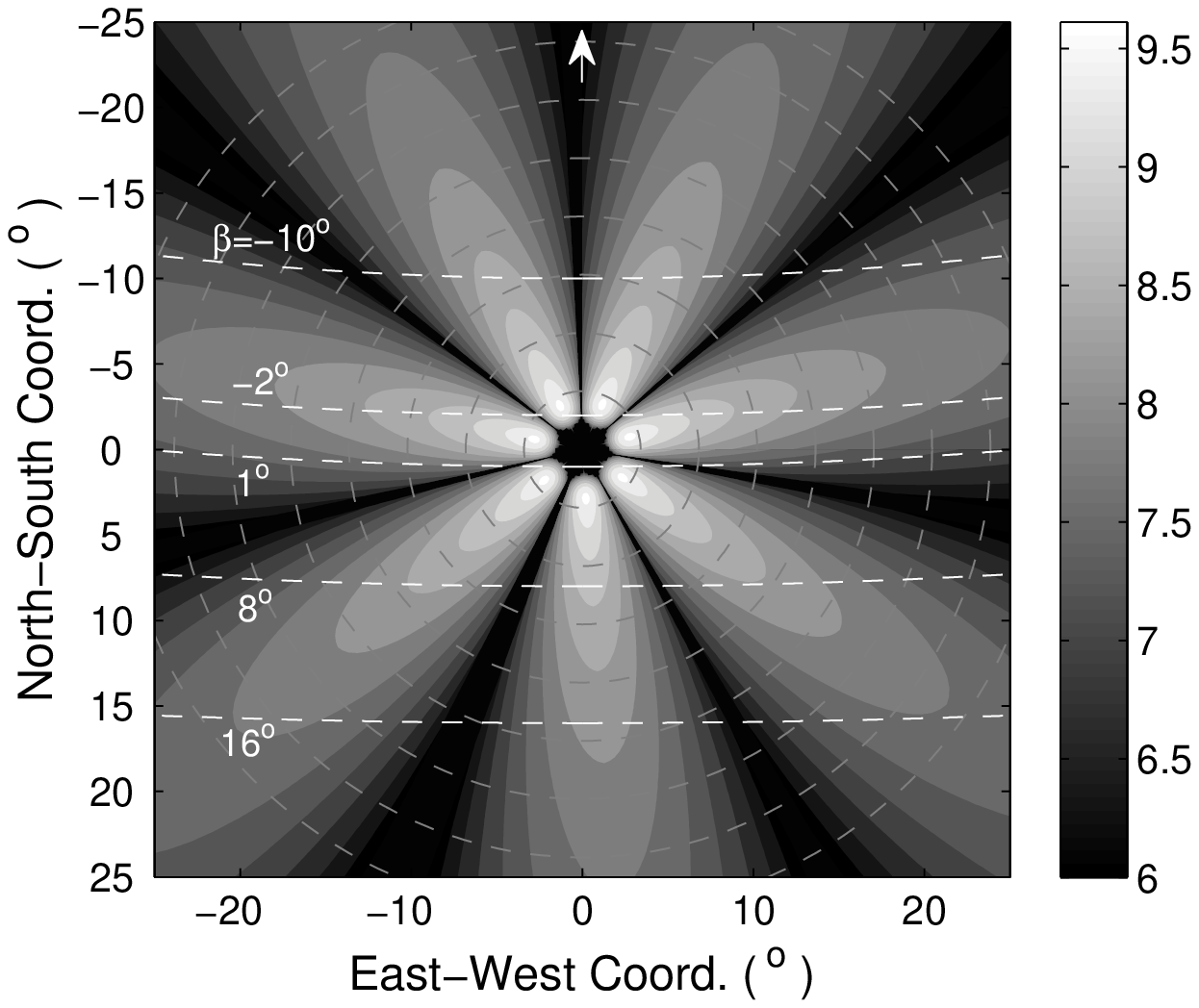}
\includegraphics{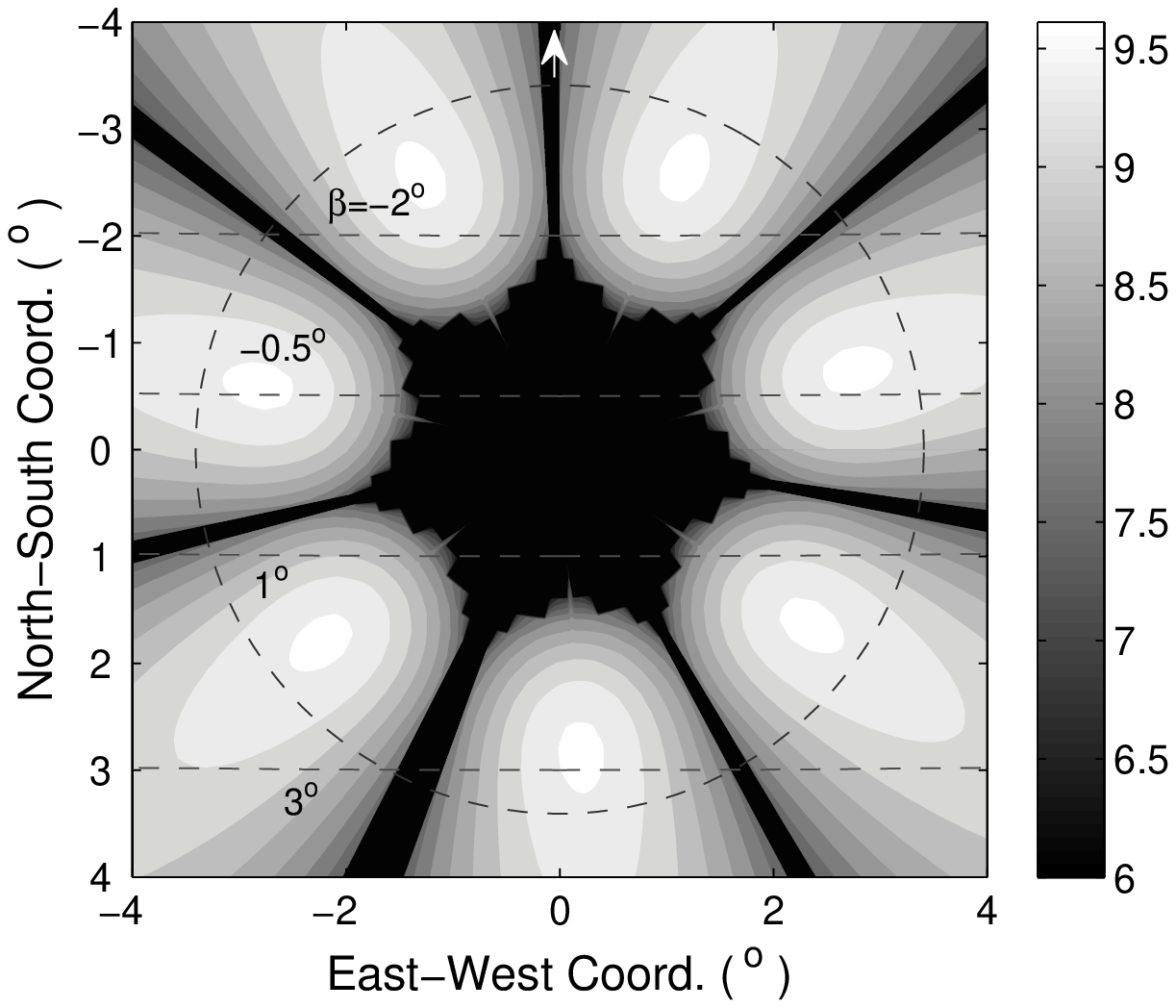}
}
\resizebox{15cm}{11cm}
{
\includegraphics{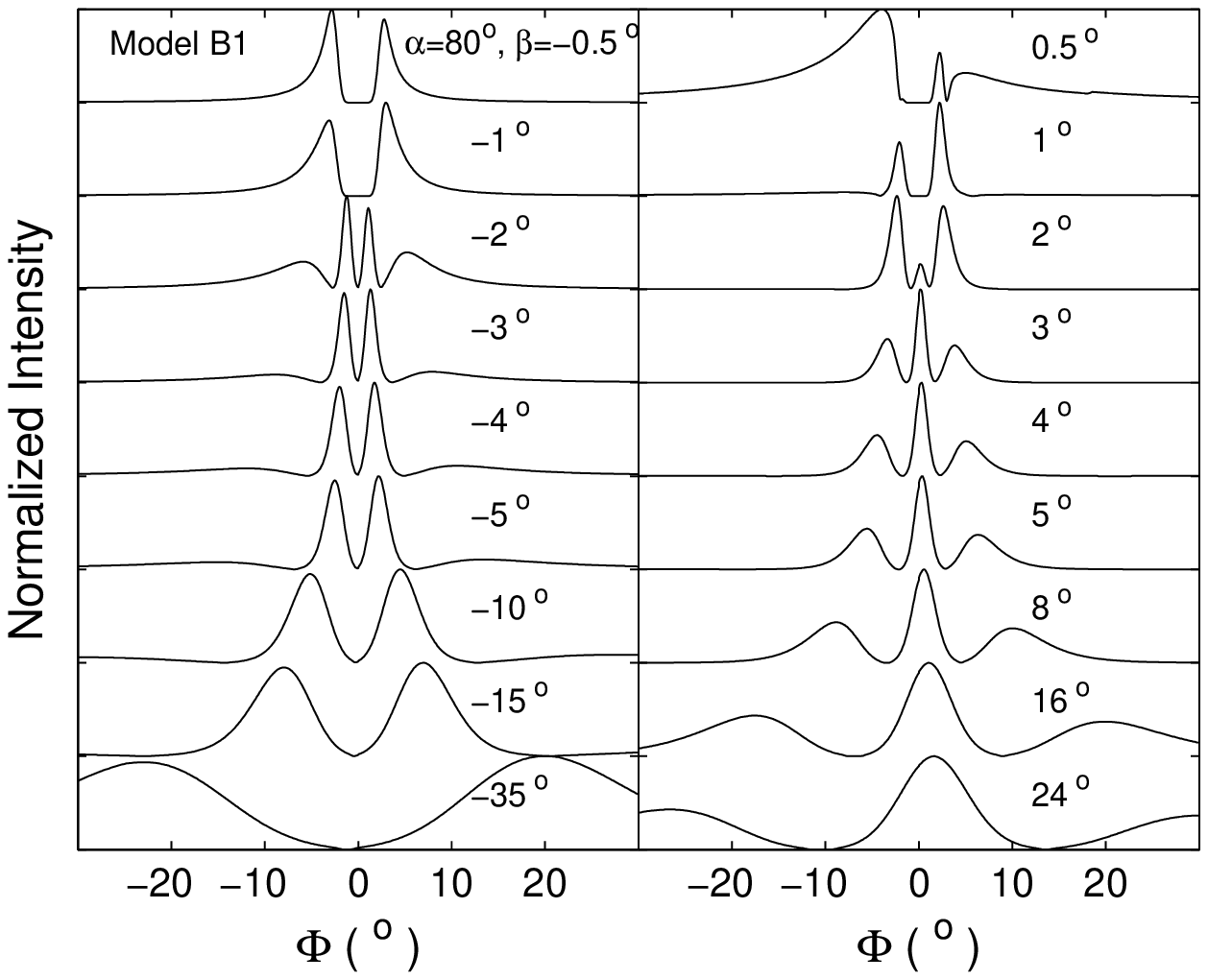}
}

\caption{The modeled beam and pulse profiles for Model B1. As Fig. \ref{figure:model_B_u_a10} except that $\alpha=80^{\rm o}$. }
\label{figure:model_B_u_a80}
\end{figure*}

\begin{figure*}
\centering
\resizebox{16cm}{6cm}
{
\includegraphics{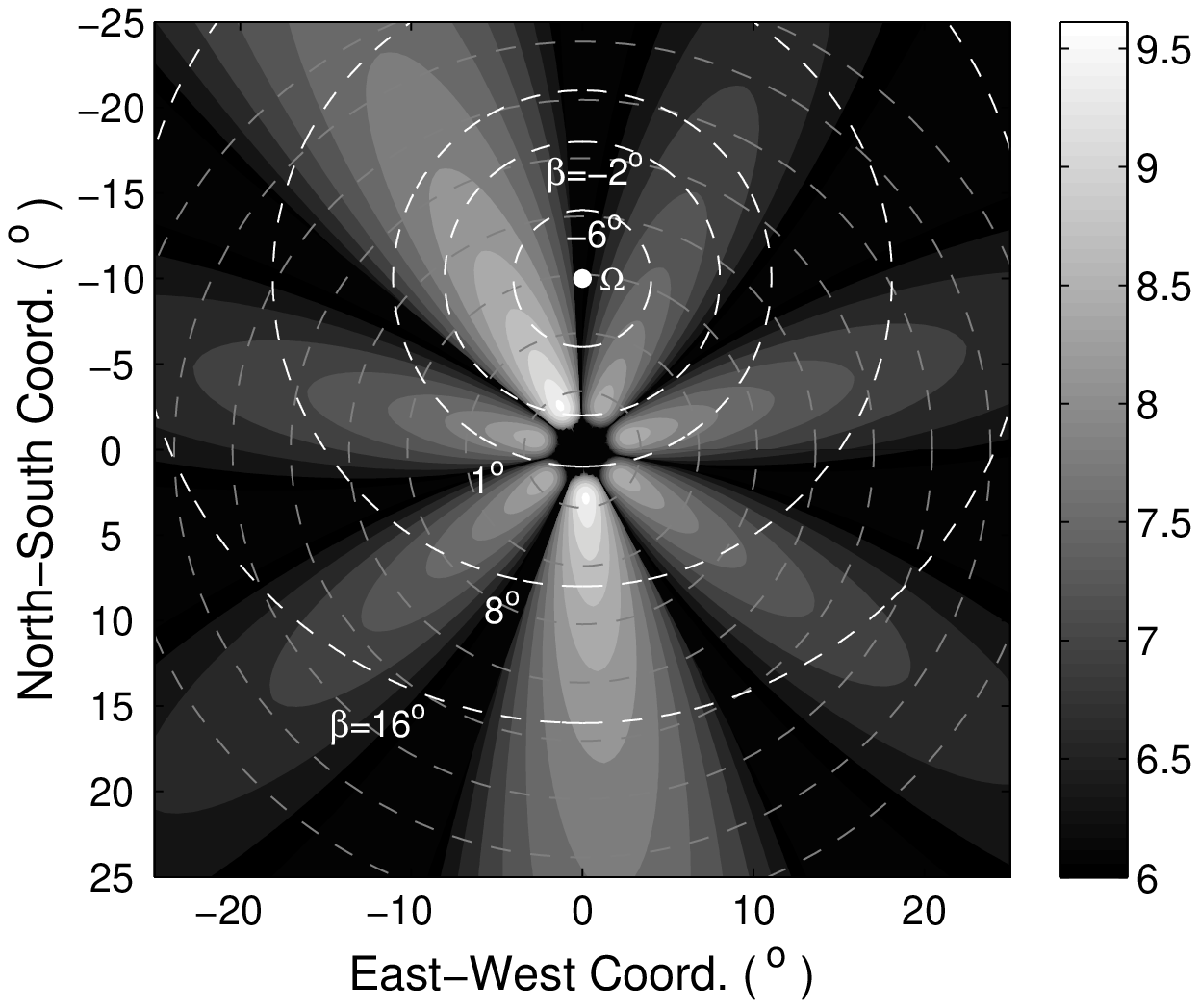}
\includegraphics{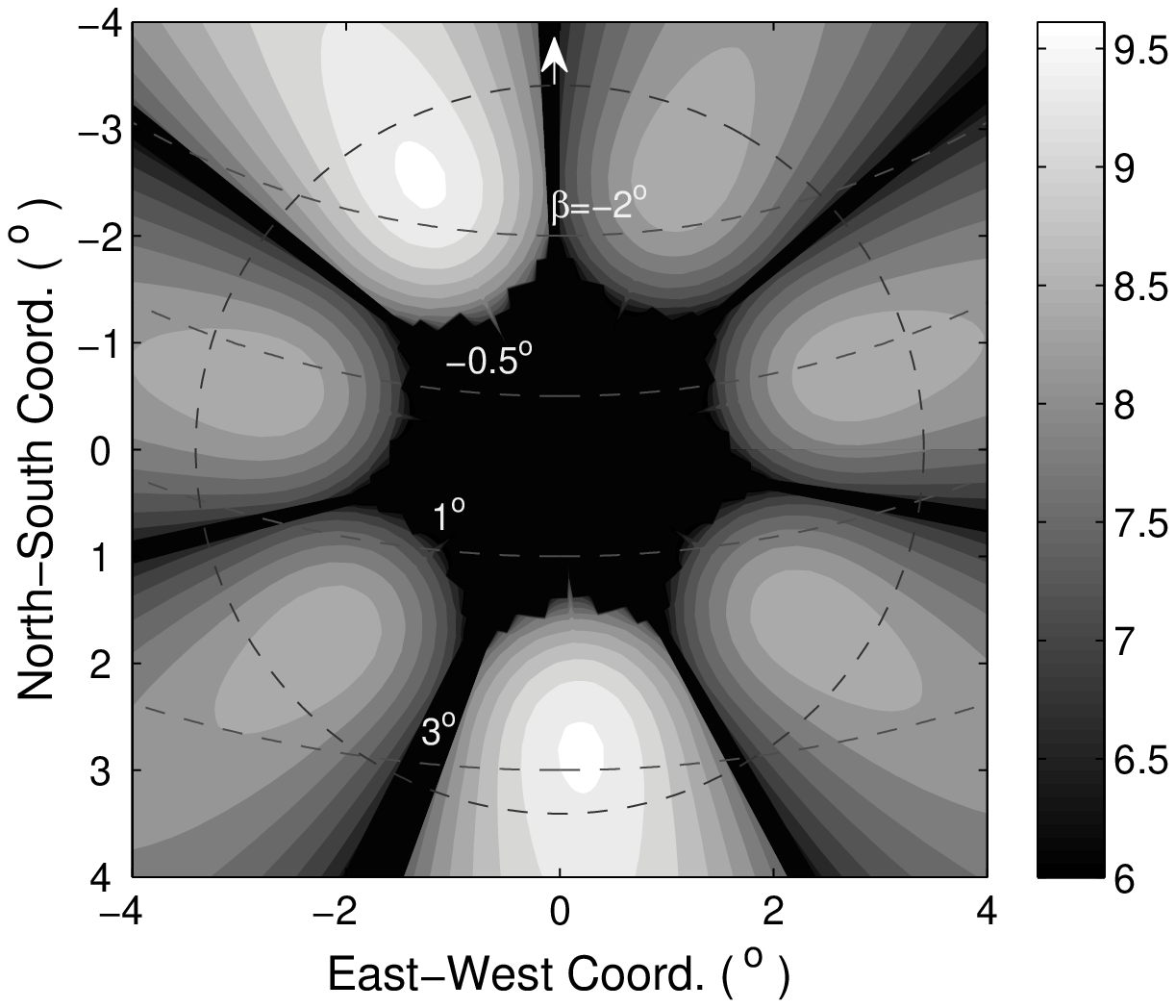}
}
\resizebox{15cm}{11cm}
{
\includegraphics{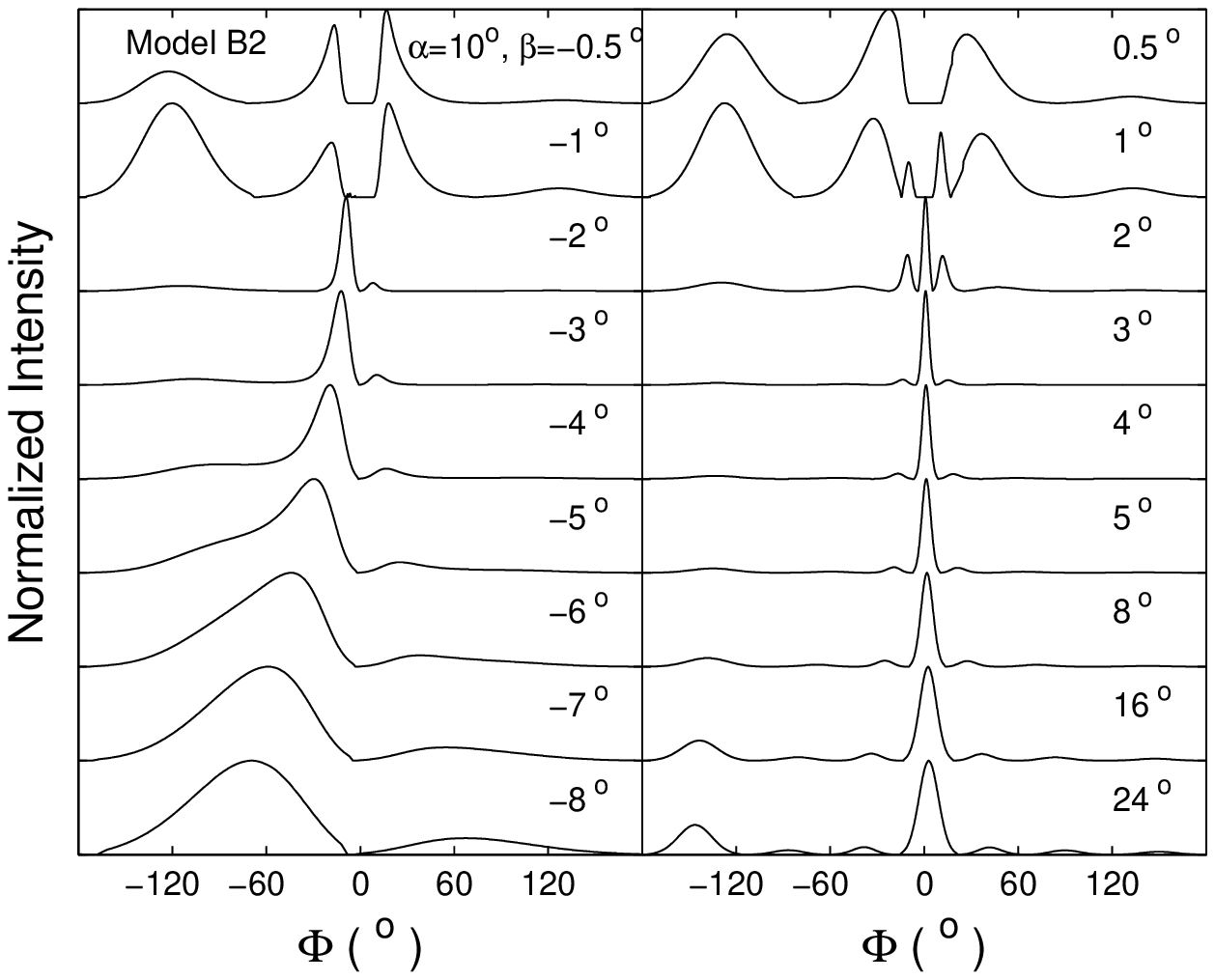}
}

\caption{The beam and pulse profiles for Model B2, where two flux tubes are dominant. The inclination angle is $\alpha=10^{\rm o}$.}
\label{figure:model_B_nu_a10}
\end{figure*}

\begin{figure*}
\centering
\resizebox{16cm}{6cm}
{
\includegraphics{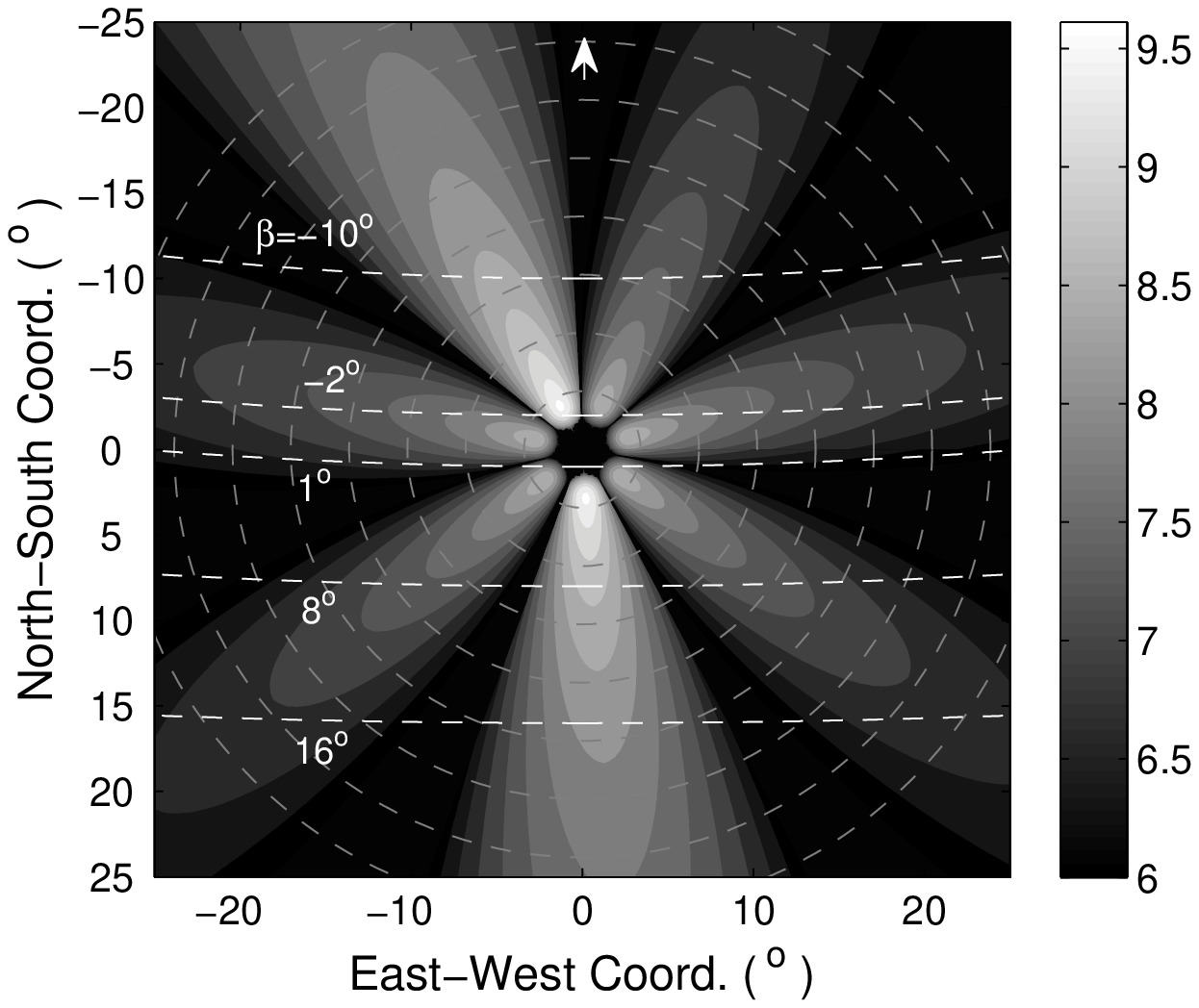}
\includegraphics{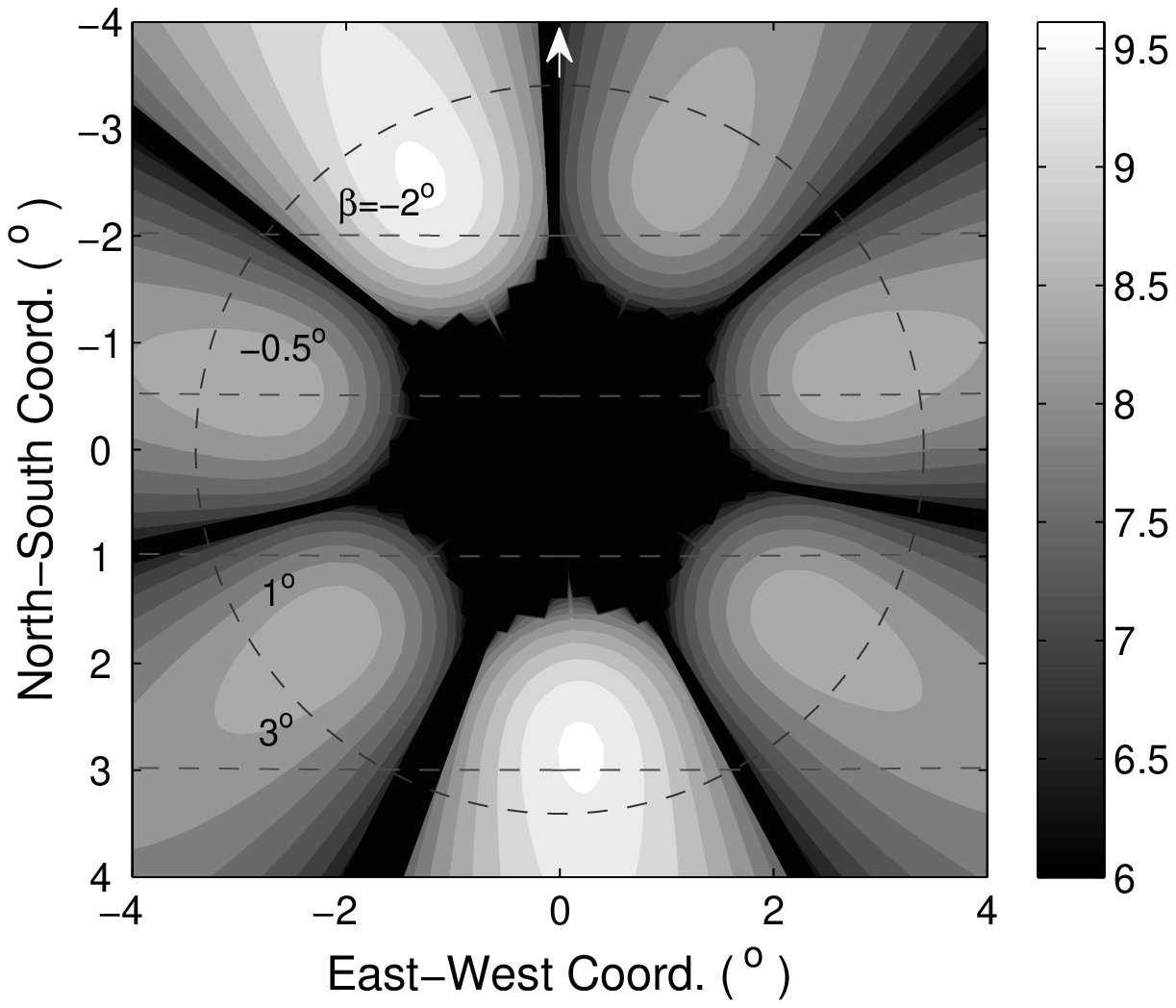}
}
\resizebox{15cm}{11cm}
{
\includegraphics{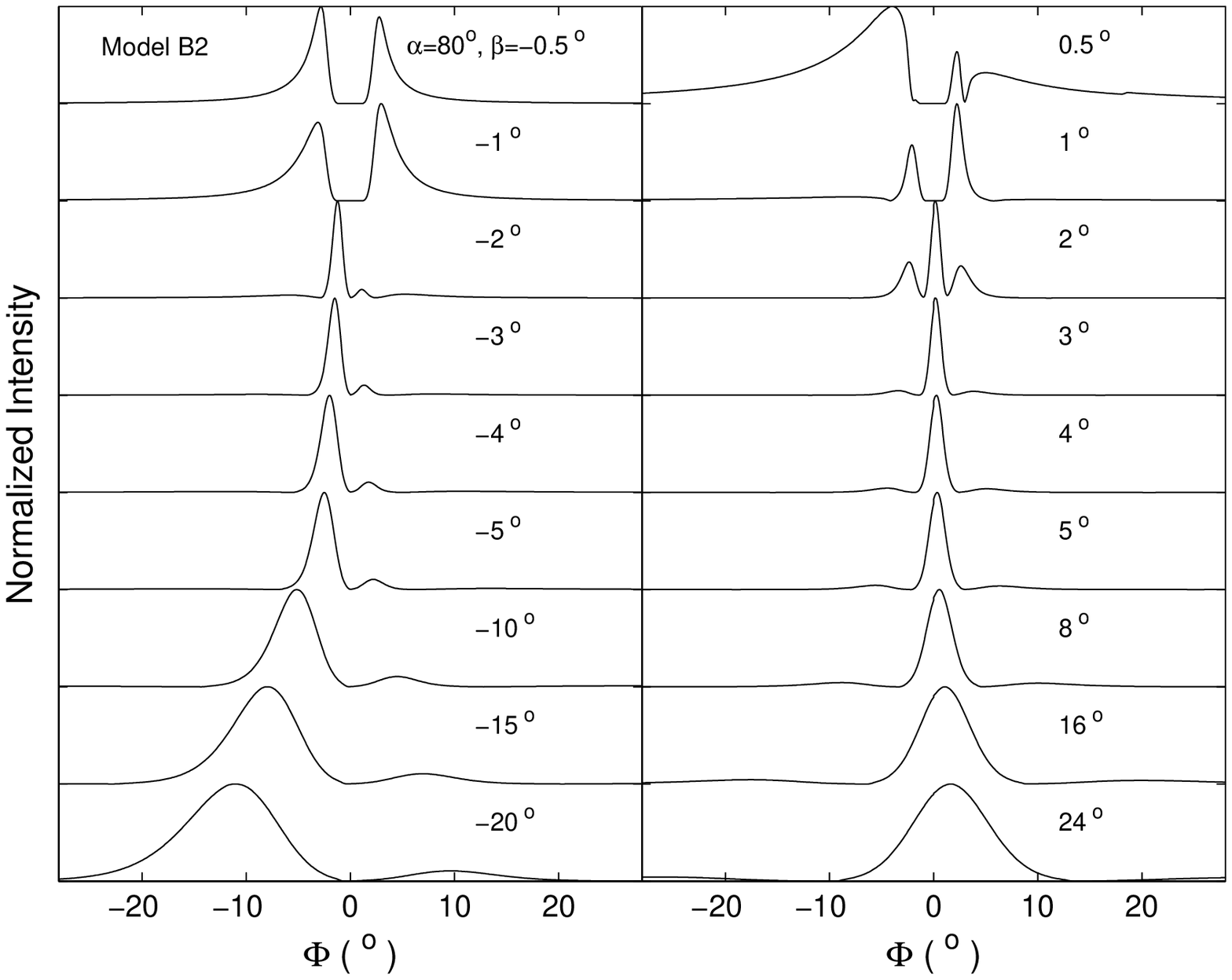}
}

\caption{The and pulse profiles for Model B2. As Fig. \ref{figure:model_B_nu_a10} except that $\alpha=80^{\rm o}$.}

\label{figure:model_B_nu_a80}
\end{figure*}

Figs. \ref{figure:model_A_u_a10} to \ref{figure:model_B_nu_a80} show the beam patterns of Models A1, A2, B1 and B2 for inclination angles of $\alpha=10^{\rm o}$ and $80^{\rm o}$, respectively, together with the average pulse profiles at a number of impact angles for each model. In the contour maps of intensity distribution (the upper-left panels in each figure), we plot the equi-radius circles from 1 to 6 times of $\rho_{\rm pc}$. This is especially helpful to view the intensity structure in the inner beam (the upper-right panels). A set of LOSs are also plotted in the contour maps, marked with their impact angles. We take $\rho_{\rm pc}\simeq 3.3^{\rm o}$ in our calculation, corresponding to a pulsar period $P=0.15$ s.

\begin{figure*}
\centering
\resizebox{12cm}{7cm}
{
\includegraphics{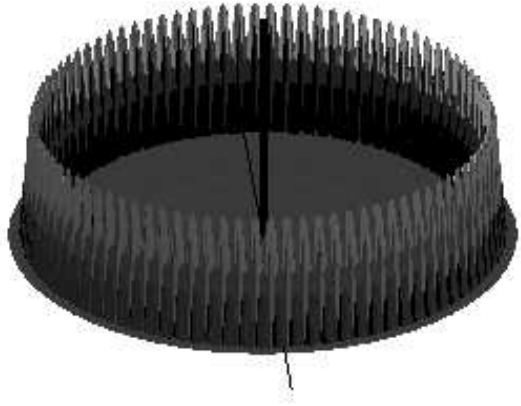}
}

\caption{An example of densely distributed flux tubes in the polar cap (Model B3). 90 evenly spaced flux tubes are located near the polar cap boundary. The particle density follows 2-dimensional Gaussian distribution in each flux tube, and all the flux tubes are identical. }
\label{figure:discharge2}
\end{figure*}

\begin{figure*}
\centering
\resizebox{16cm}{6cm}
{
\includegraphics{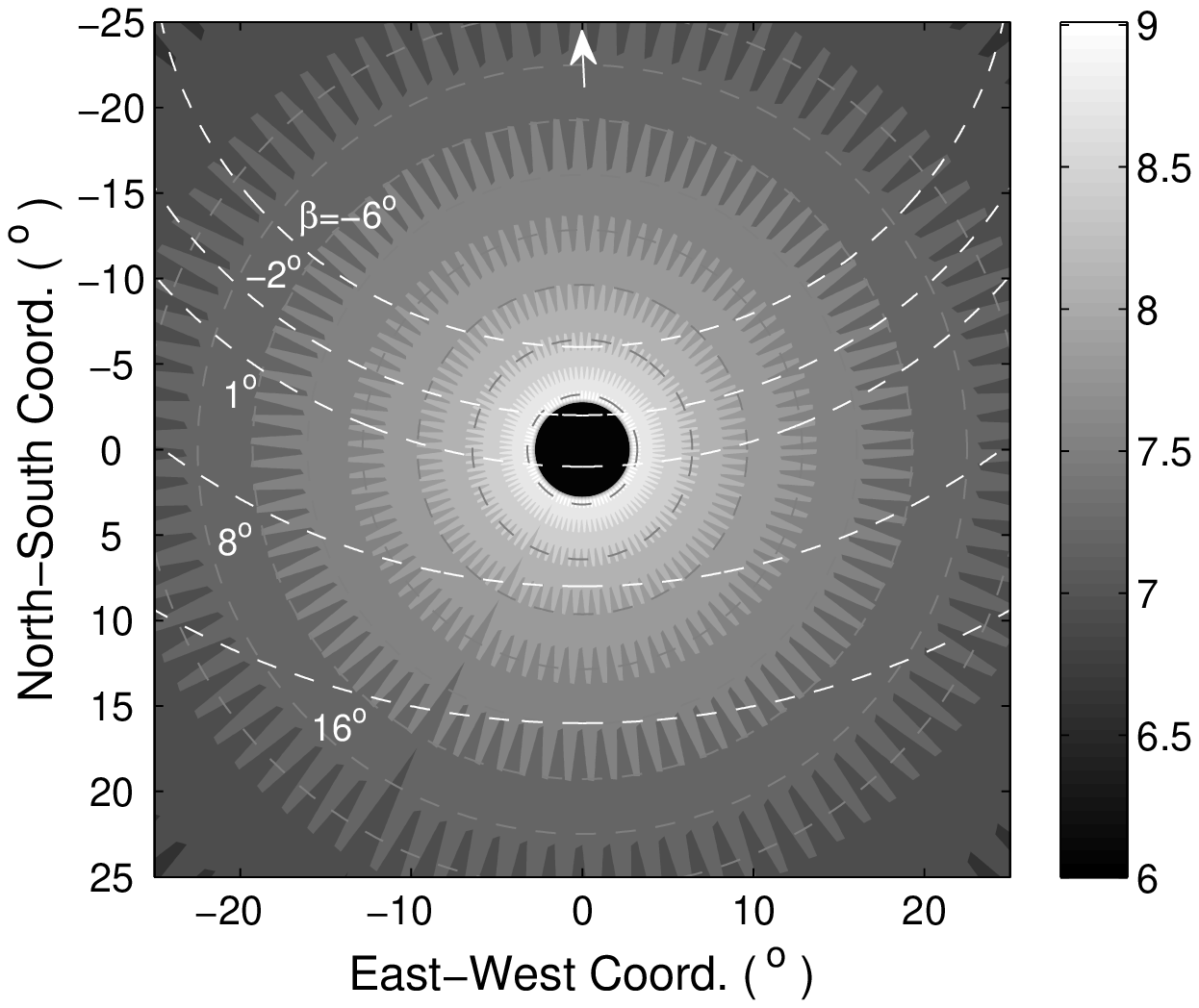}
\includegraphics{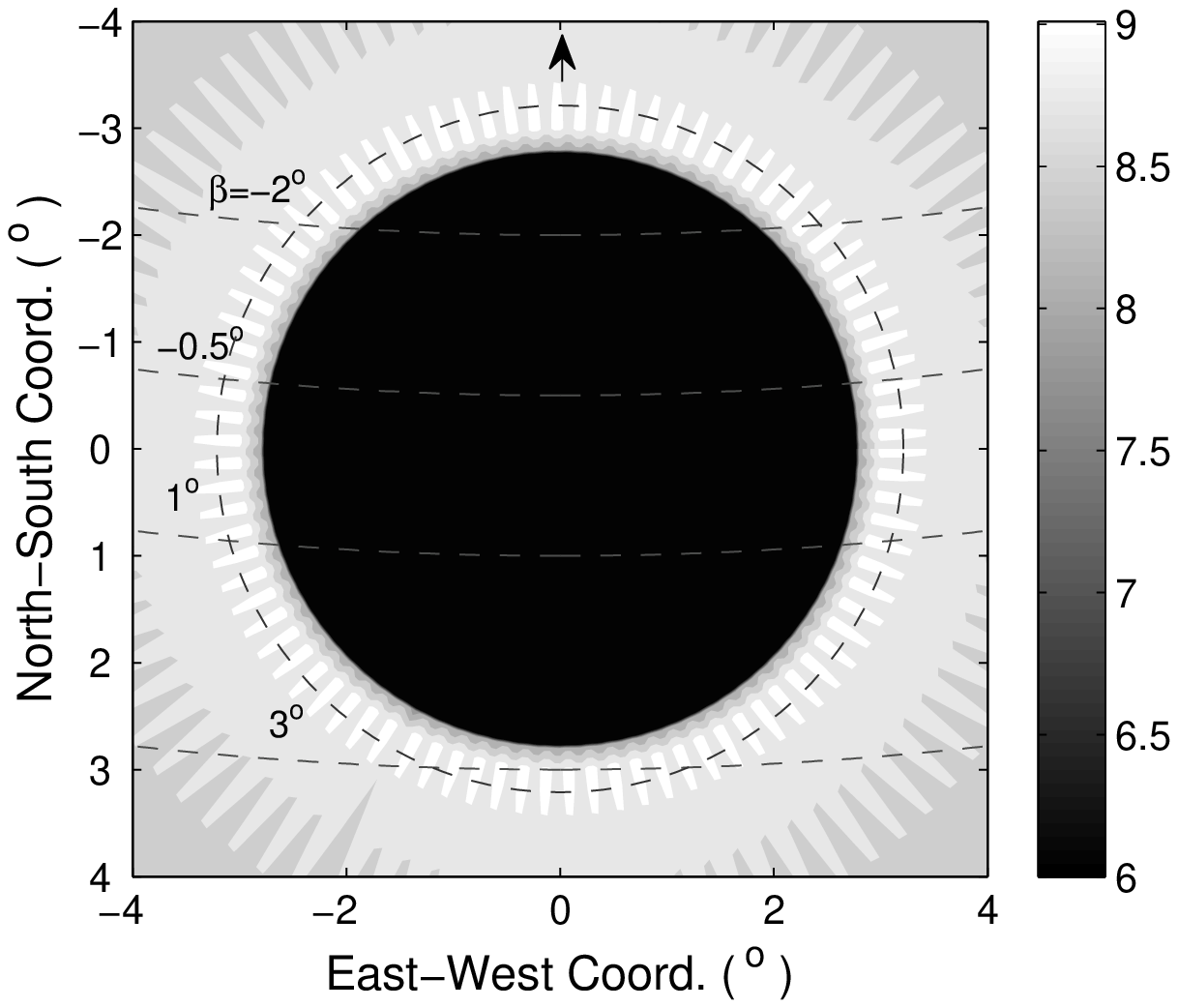}
}
\resizebox{15cm}{11cm}
{
\includegraphics{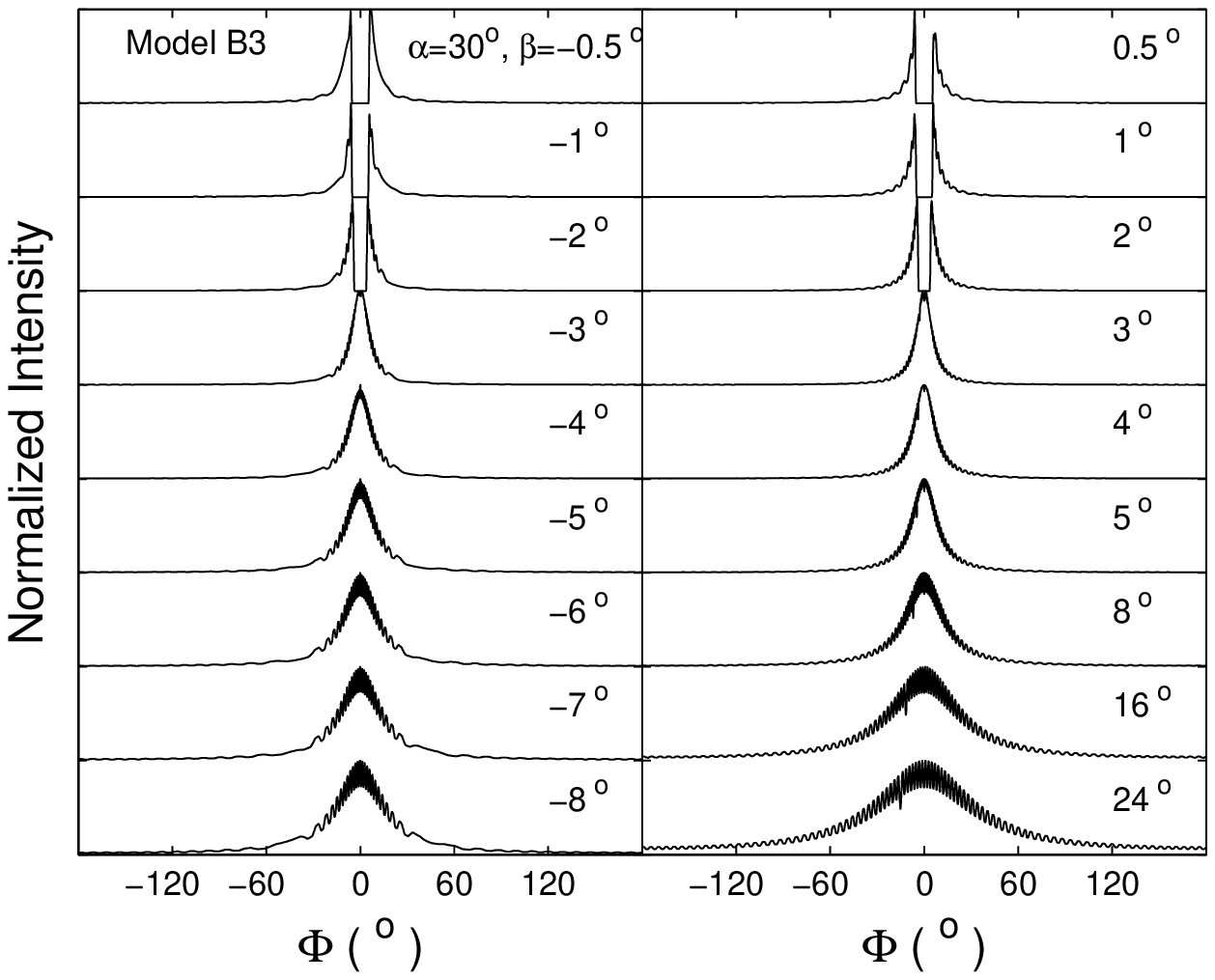}
}

\caption{The beam and pulse profiles for Model B3. The profiles are calculated with $\alpha=30^{\rm o}$.}
\label{figure:model_B_u_a30}
\end{figure*}

The above models assume a few flux tubes in the polar cap. To compare with the case of a large number of flux tubes, we simulate a model B3, where the single flux tube follows the pattern of Model B, but totally 90 flux tubes are evenly spaced along the polar cap boundary and the size of each flux tube is very small. We take the following parameters/assumptions in the simulation: $\chi_0=0.9$, $\Re_0=0.05$ and the same peak multiplicity factors for all the flux tubes. The diagram of the flux tubes and the simulated beam and profiles for $\alpha=30^{\rm o}$ are presented by Figs. \ref{figure:discharge2} and \ref{figure:model_B_u_a30}.

Below we summarize the main features and the inference that can be drawn from the simulated results.

(1) Both kinds of models present limb-darkening feature in both the radial and the azimuthal dimensions, except that the intensity trend undergoes transition near or within the polar cap boundary (mainly because of the assumption $r_{\rm L}=R$). The radial limb-darkening feature in the outer beam follows a power-law $\rho^{2q-6}$, which is a consequence of cooperation of coherent emission, particle free flow and altitude dependence of single-particle emission power. The transition of radial intensity trend in the inner beam is caused by the shrinkage of emission region for very small $\rho$ and the attenuation of secondary particle density towards the magnetic pole (if it exists, e.g. in Model B). The transvers limb-darkening phenomenon is caused by the continuous attenuation of particle density towards the edge of flux tube.

(2) Discarding the detailed intensity distribution in the sub-beams, the shapes of sub beams in Models A and B looks very similar. This is because the shape is determined by the divergence nature of dipolar flux tubes.

(3) In the outer beam, the pulse width broadens with increasing $|\beta|$. This is also a natural consequence of the divergence nature of dipolar flux tube. However, this trend does not hold when the LOS sweeps across the central part of the beam, because the intensity distribution therein is more complex and the LOS may sweep across the bright parts of a few sub beams. This break of pulse-width-impact-angle relationship is indeed seen in observational data, as shown in Section 3.2.

(4) Models A1 and B1 tend to predict more complex and wider profiles than Models A2 and B2. This is because all the flux tubes have the same activity in emission in Models A1 and B1, thus the emission from the flux tubes outside the meridian plane are still strong enough to be observed, even though the radial limb-darkening effect has caused more attenuation in the observed intensity for them. Especially, Models A1 and B1 predicts growing complexity with increasing $|\beta|$ in the cases of small inclination angles, which is not supported by the data (see Table 1 for $\alpha$ and $\beta$ and the number of pulse components $N_{\rm c}$). Since normal pulsars have simple profiles (e.g. Karastergiou \& Johnston 2007), these features strongly suggest that only a limited number of discharging flux tubes should be dominant in pair production and emission activity.

(5) Ignoring the spikes, Model B3 can only produce single-component profiles for relatively large impact angles. In order to predict double- and multiple-component profiles, one has to assume that the peak multiplicity factors in the flux tubes are modulated azimuthally and form a few large-scale structures. General speaking, this kind of model with small-size flux tubes and large-scale multiplicity modulation, is equivalent to the models invoking large-size flux tubes in the capability of predicting various kinds of pulse profiles.

%
%



\subsection{The $W-|\beta|$ relation}
\label{subsect:w-b}

The above simulation shows that the pulse width $W$ increases with increasing impact angle $|\beta|$ in the outer beam. Below we derive the relationship, which will be used in Section 3.2 to test our model.

The above points (2)-(5) suggest that probably only a few flux tubes are active and visible, we simply assume that the effective emission region is confined within an azimuth range $[-\varphi, \varphi]$ around the meridian plane (note that this range may contain more than one flux tubes).
According to the spherical geometry relationships, we have
\begin{equation}
\cos\zeta=\cos\alpha\cos\rho-\sin\alpha\sin\rho\cos\varphi,
\label{eq:zeta-cosine}
\end{equation}
\begin{equation}
\cos\rho=\cos\alpha\cos\zeta+\sin\alpha\sin\zeta\cos\Phi,
\label{eq:rho-cosine}
\end{equation}
and
\begin{equation}
\frac{\sin\rho}{\sin\Phi}=\frac{\sin\zeta}{\sin\varphi},
\label{eq:rho-sine}
\end{equation}
where $\zeta=\alpha+\beta$ is the viewing angle between the LOS and the rotating axis, $\Phi$ is the pulse longitude.

\begin{figure*}
\centering
\resizebox{16cm}{5.7cm}
{
\includegraphics{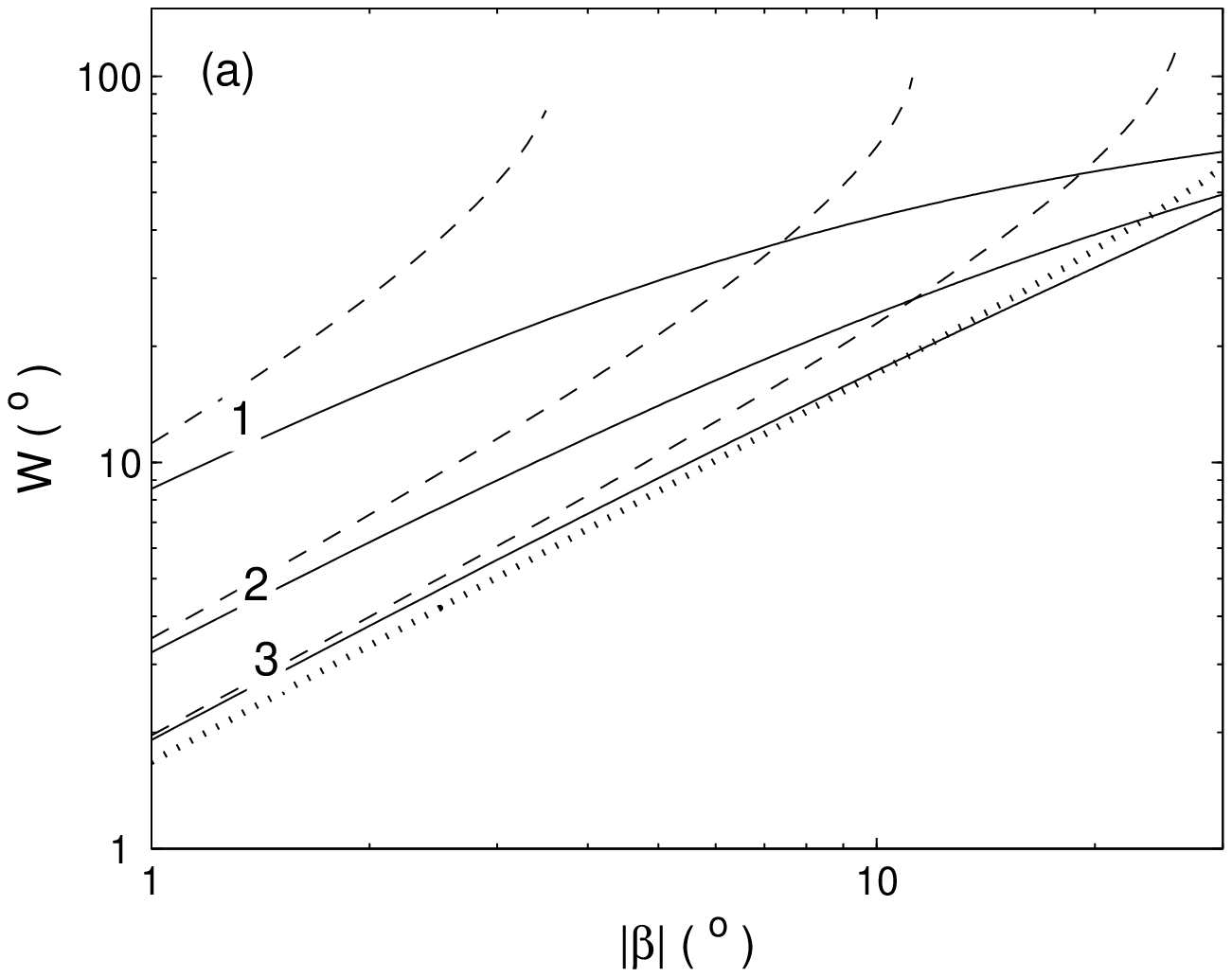}
\includegraphics{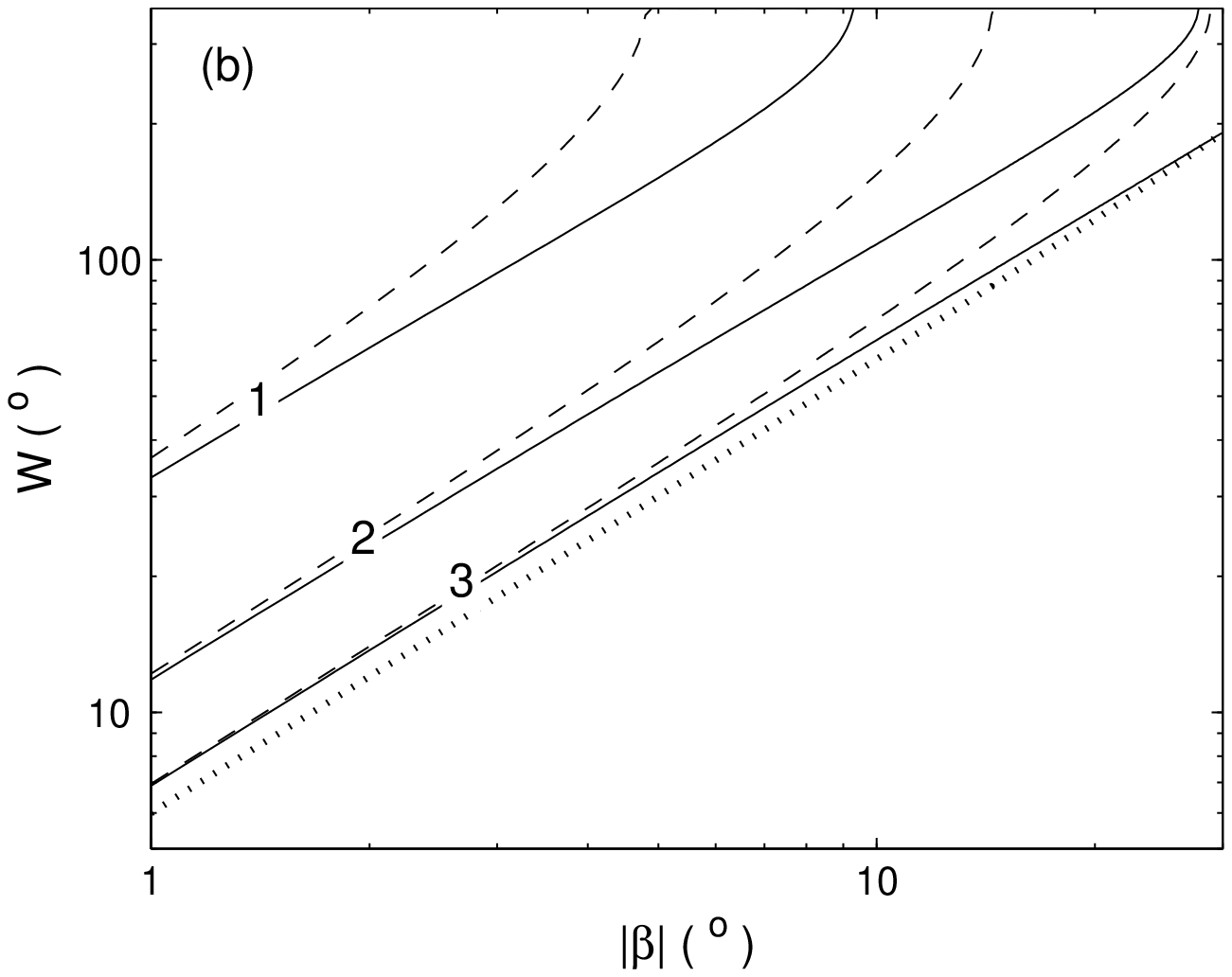}
}

\caption{The modeled relationship between pulse width and impact angle, with (a) for Eq. (\ref{eq:w-b}) and (b) for Eq. (\ref{eq:W-unif}). In panel (a), three group of solid and dashed curves marked with ``1, 2, 3'' are calculated with $\alpha=10^{\rm o}$, $30^{\rm o}$ and $60^{\rm o}$ respectively, where $\varphi=40^{\rm o}$ is fixed. The solid and dashed lines are for $\beta>0$ and $\beta<0$, respectively. When $\alpha=90^{\rm o}$, negative and positive impact angles will produce the same curve, as presented by the the dotted line. In panel (b), the solid, dashed and dotted curves have the same meaning as in panel (a), except that the parameter $\varphi=40^{\rm o}$ is no longer useful but replaced by $\delta=-2$ and $\eta=0.1$, where $\eta=0.1$ means we always measure the full pulse width at the level of 10\% peak intensity for any impact angle. }
\label{figure:wbmodel}
\end{figure*}

Substituting Eqs. (\ref{eq:rho-cosine}) and (\ref{eq:rho-sine}) into Eq. (\ref{eq:zeta-cosine}), and using $\Phi=W/2$, the formula is simplified as
\begin{equation}
\cos(\frac{W}{2}+C)=\frac{\sin\alpha}{\tan(\alpha+\beta)\left(\cos^2\alpha+\tan^{-2}\varphi\right)^{1/2}},
\label{eq:w-b}
\end{equation}
where $C=\arctan(\sec\alpha/\tan\varphi)$. It can be found that no matter $\beta>0$ or $\beta<0$, $W$ will increase with $|\beta|$, which is shown by Fig. \ref{figure:wbmodel}(a). Note that Eq. (\ref{eq:w-b}) only applies to the outer beam.

To derive the $W-|\beta|$ relationship for the inner beam is difficult because of the reversal intensity distribution. In some cases, the LOS may see emissions from other less active flux tubes far away from the meridian plane (see $\beta=1^{\rm o}$ and $2^{\rm o}$ in Fig. (\ref{figure:model_B_nu_a80}) for examples). Since one tends to see more parts of the polar cap, $\rho_{\rm b}\simeq 1\sim 2\rho_{\rm pc}$ may be a rough approximation for the beam size when $|\beta|\lesssim\rho_{\rm t}$\footnote{The relationship $W\propto P^{-1/2}$ derived from the lower boundary line (hereafter LBL, e.g. MG11) in the diagram of pulse width versus period suggests that this approximation is viable, because the opening angle of polar cap, $\rho_{\rm pc}\propto P^{-1/2}$, does follow this relation. Further studies on the LBL will be presented in a subsequent paper}.

We must point out that the above mentioned pulse width is completely determined by the boundary of the flux tube, irrespective of the intensity distribution in the emission beam, hence it can be called the geometric pulse width. The most often used pulse width, $W_{50}$ or $W_{10}$, which is measured at the 50\% or 10\% level of pulse peak, obviously depends on intensity distribution. It is possible that the intensity drops so abruptly as the LOS sweeps towards the lateral beam edges that $W_{10}$ becomes smaller than the geometric width $W$. Will the increasing trend of $W-|\beta|$ relationship still hold? Below we demonstrate that even in the extreme case such a trend still exists.

We consider an extreme case that the intensity is constant for any circle around the beam center. In the outer part of the beam, the intensity follows a simple limb-darkening relation, $I=I_{0}\rho^{\delta}$ ($\delta<0$). Given a LOS with $\beta$, one will see the maximum intensity
when $\rho$ reaches its minimal value, i.e. $\rho=|\beta|$, therefore, $I_{\rm pk}=I_{0}\beta^{\delta}$. For a level $\eta I_{\rm pk}$ ($\eta<1$), at which the pulse width is defined, the boundary intensity drops to $I_{\rm b}=\eta I_{\rm,pk}$. Using the limb-darkening relation, one finds the boundary radius
$\rho_{\rm b}=\eta^{1/\delta}\beta$. Then the corresponding pulse phase can be figured out via $\cos\Phi_{\rm b}=(\cos\rho_{\rm b}-\cos\alpha\cos\zeta)/(\sin\alpha\sin\zeta)$
(from Eq. (\ref{eq:rho-cosine})), and the pulse width will be
\begin{equation}
W=2\Phi_{\rm b}=2\arccos\left[\frac{\cos(\eta^{1/\delta}\beta)-\cos\alpha\cos\zeta}{\sin\alpha\sin\zeta}\right].
\label{eq:W-unif}
\end{equation}
Again, no matter $\beta>0$ or $\beta<0$, $W$ will increase with $|\beta|$, which is shown by Fig. \ref{figure:wbmodel}(b). This tendency can also be found in the profiles in Fig. \ref{figure:model_B_u_a30}.

In the following section, we will use Eq. (\ref{eq:w-b}) to calculate the pulse width when $|\beta|>\rho_{\rm pc}$ but fix the beam size as $2\rho_{\rm pc}$ when $|\beta|\le\rho_{\rm pc}$.


\section{Tests for the fan beam model}

In this section, we focus on the observational tests for the predictions on pulse width evolution and radial intensity distribution by the fan beam model. In Section 3.1, the observed beam properties of a couple of precessional pulsars are compared with the model. The phenomenon of increasing pulse width with increasing impact angle for PSR B1534$+$12, the radial intensity distributions of PSR J1141$-$6545 and PSR J1906$+$0746 are in general agreement with the main features of fan beam model. Three relationships predicted by the model, i.e. $W-|\beta|$ relationship, the radial limb-darkening relationship for the outer beam ($I\propto\rho^{2q-6}$) and the relation between the upper-limit of impact angle and pulsar distance, are tested statistically based on a sample of 64 pulsars collected from literature, of which the impact angles are known by fitting the linear polarization position angle (hereafter PPA) data with the rotating vector model (hereafter RVM, Radhakrishnan \& Cooke 1969). The relationships derived from the data can be successfully reproduced by the fan beam model. Details are described in Sections 3.2, 3.3 and 3.4, respectively.

\subsection{The observed radio beams of precessional binary pulsars}

It is known that binary pulsars may undergo relativistic spin-precession due to coupling between the spin and orbital angular momenta (Damour \& Ruffini 1974, Barker \& O'Connell 1975). As a result, the spin axis of the pulsar rotates around the total angular momentum vector, changing the viewing and impact angles. This effect enables us to ``scan'' the emission beam. So far, efforts to construct the beam structure have been been made for 6 pulsars, PSR B1913$+$16 (e.g. Weisberg \& Taylor 2002), B1534$+$12 (e.g. Arzoumanian 1995), PSR J1141$-$6545 (Manchester et al. 2010), PSR J1906$+$0746 (Desvignes et al. 2013), PSR J0737$-$3039A (Ferdman et al. 2013, Perera et al. 2014) and J0737-3039B (e.g. Perera et al. 2012, Lomiashvili \& Lyutikov 2013). For half of them, PSR J1141$-$6545, PSR J1906$+$0746 and PSR J0737$-$3039B, the two-dimensional beam structures were constructed with multi-epoch absolute flux density data, therefore they can be used to probe the models via both the evolution of pulse width and the radial distribution of emission intensity. But the magnetosphere of PSR J0737$-$3039B is distorted by the wind from its companion PSR J0737$-$3039A, it is not an ideal case for model test. For the other 3 pulsars, the profile at each epoch is first normalized by a pulse peak and then used to study the profile evolution or construct the beam structure. This method focuses on the evolution of relative intensity of different pulse components rather than the radial distribution of absolute intensity in the pulse beam. Therefore, only the evolution of pulse width is useful to test the fan beam model for these 3 pulsars. Below we first present the clear evidence from PSR J1141$-$6545 and PSR J1906$+$0746, and then discuss the remaining pulsars, which show less prominent evidence.

\begin{figure*}
\centering
\resizebox{14cm}{5cm}
{
\includegraphics{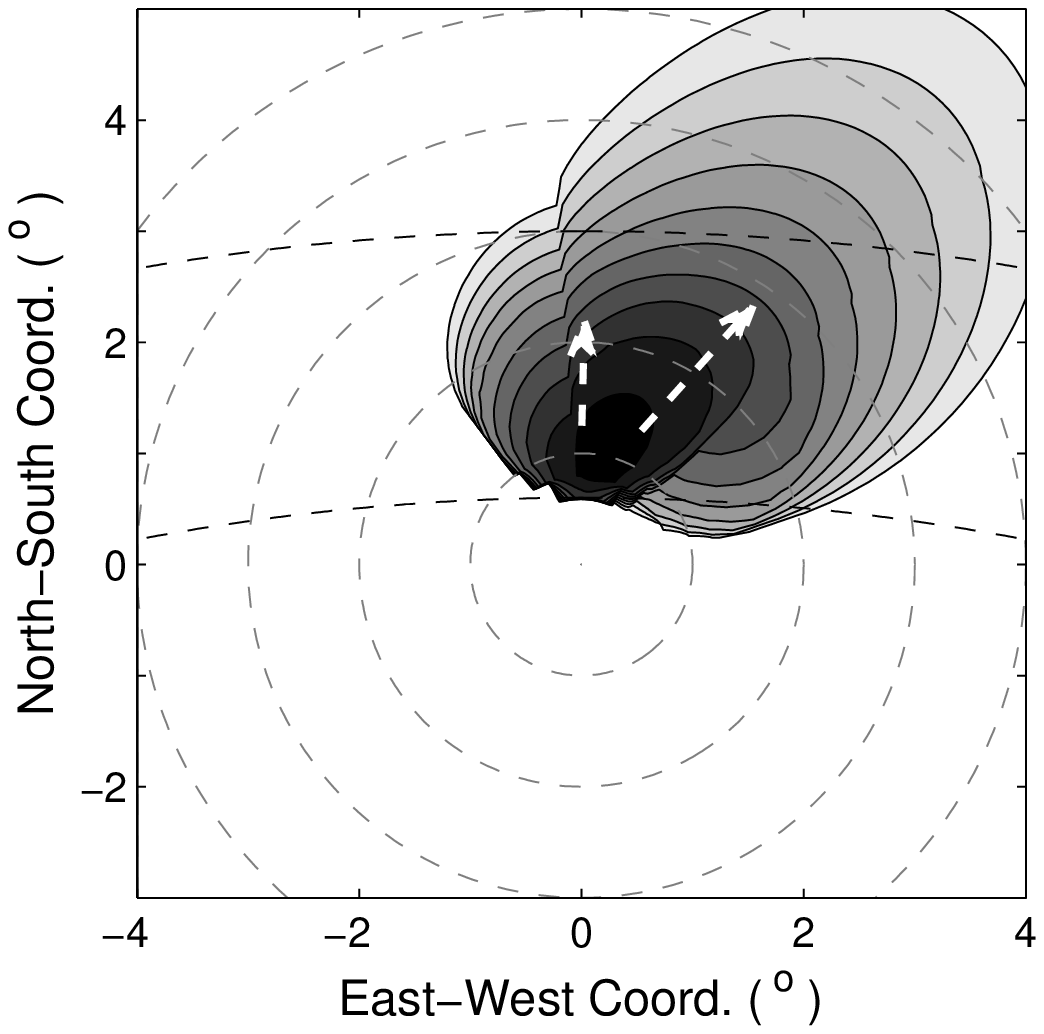}
\includegraphics{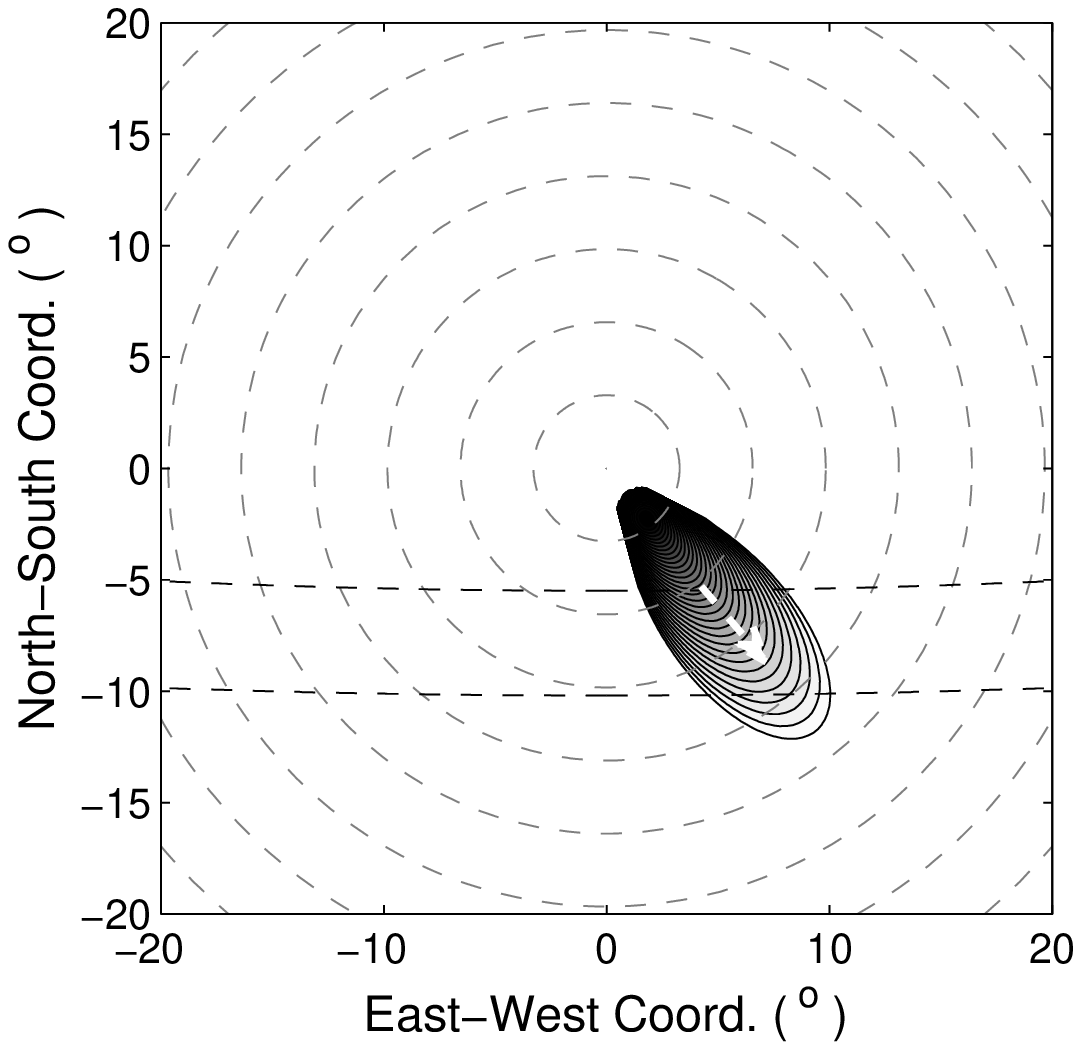}
}
\resizebox{14cm}{5cm}
{
\includegraphics{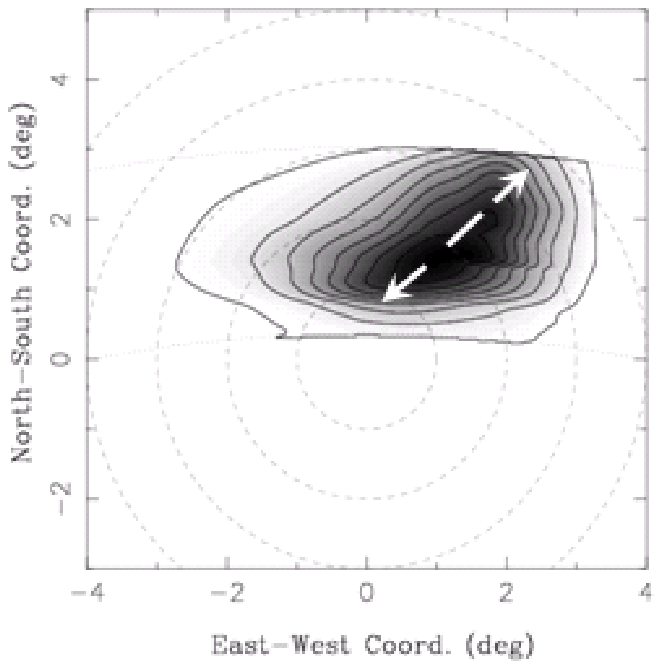}
\includegraphics{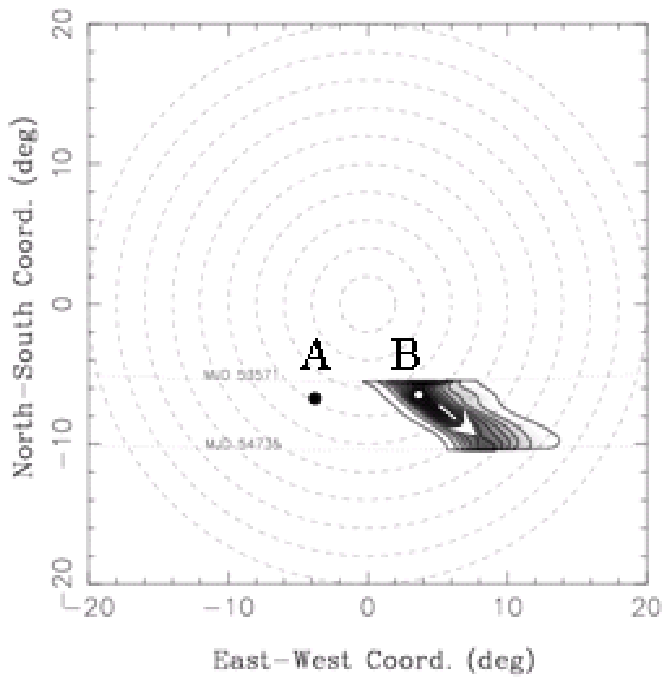}
}

\caption{The simulated and observed beams for PSR J1141$-$6545 (left panels) and the main pulse of PSR J1906$+$0746 (right panels). Two flux tubes are used to model the beam of PSR J1141$-$6545 (left upper), which have the same center coordinates, $\chi_0=0.7$ and $\Re_0=0.3$, and the same peak multiplicity factor, but with different central azimuths, $\varphi_{0,{\rm a}}=0$ and $\varphi_{0,{\rm b}}=35^{\rm o}$ (indicated by a short and a long white arrows, respectively). The radial limb-darkening indices are different too, with $\delta=-4$ for the left flux tube and $\delta=-2$ for the right one. The full range of intensity contours is of a factor of $\sim$16. For PSR J1906$+$0746, only one flux tube with $\chi_0=0.7$, $\Re_0=0.3$, $\varphi_0=40^{\rm o}$ and $\delta=-2$ is used in the simulation, as shown in the upper right panel, where the arrow indicates the central azimuth of the flux tube. The arrow in the observed beam (lower right panel) indicates the radial intensity gradient direction. The full range of intensity contours is of a factor of $\sim$40. The observational beams of PSR J1141$-$6545 and PSR J1906$+$0746 are taken from Manchester et al. (2010) and Desvignes et al. (2013). See Section 3.1 for details. }
\label{figure:obsbeam}
\end{figure*}

PSR J1141$-$6545 is a young binary pulsar with a precession rate 1.4~deg$/$yr. Manchester et al. (2010) found that the average pulse profiles of this pulsar show remarkable variation at 1.4~GHz between 1999 and 2008. In the first a couple of years the profile was dominated by the trailing part. The leading part grew stronger with time and later became comparable with the trailing part, leading to a bump-shaped profile. Despite the profile evolution, the pulse width at a very low intensity level, e.g. 1\% of the peak, is nearly constant. The authors used a precessional beam model to fit the pulse profile and the absolute central PPAs obtained by fitting PPA data with the RVM. The derived impact angles varied from about $-3.7^{\rm o}$ in 1999 to $-0.9^{\rm o}$ in 2007, meaning that the LOS was moving towards the magnetic pole. With the data of flux density, the two-dimensional beam intensity structure was inferred, which shows that the maximum intensity was reached when $\beta\simeq -2^{\rm o}$. Beyond this angle, the intensity decreases as our LOS goes further to the magnetic pole, while below this angle, the intensity decreases as the LOS moves towards the magnetic pole, as indicated by two opposite arrows in the lower left panel of Fig. \ref{figure:obsbeam}. The most striking feature is that the beam is quite asymmetric and is partially filled without any core or conal structure. Based on this point, the authors concluded that the beam is patchy.

The twofold radial intensity distribution is generally consistent with our model. The observed transition radius of intensity distribution, $\rho_{\rm t}\simeq 2^{\rm o}$, is close to the opening angle of the polar cap $\rho_{\rm pc}\simeq 2^{\rm o}$ for this pulsar with a period of 0.394~s. This suggests that the lowest coherent emission altitude should be close to the polar cap. The nearly constant pulse width does not conflict with our model. We have discussed in Section 2.4 that the pulse width near the transition radius may not follow the same increasing trend with impact angle as it does for the outer beam.

PSR J1906$+$0746 is a young binary pulsar with a higher precession rate of 2.2~deg$/$yr. The radio profile at 1.4~GHz has a narrow main pulse and a weak interpulse, which are separated by about $180^{\rm o}$. The PPA data are smooth, following a simple RVM (Desvignes 2009). The RVM Fitting at each epoch revealed that $\alpha\simeq 81^{\rm o}$ while $\beta$ varied from $6.8^{\rm o}$ to about $11^{\rm o}$ between 2005 and 2009 for the main pulse and from $11.2^{\rm o}$ to $7^{\rm o}$ for the interpulse (Desvignes et al. 2013).

The derived beams for both poles show asymmetric structures that are quite patchy.
In both of the beams, the intensity decrease as the LOS moves away from the magnetic pole (see the lower right panel of Fig. \ref{figure:obsbeam} for the main pulse). Since the LOSs in both poles are much further from the polar cap boundary ($\rho_{\rm pc}=3.3^{\rm o}$ for $P=0.144$~s), this limb-darkening feature is consistent with the radial intensity trend for the outer part of the fan beam. The pulse width is nearly constant for the main pulse, but not clear for the interpulse due to poor signal to noise ratios\footnote{Owing to decreasing flux density, it is difficult to measure the pulse width for the interpulse at the same level, e.g. 10\% of peak intensity, in all the epoches (Desvignes 2009, Kasian 2012).}. The near constant pulse width may be reflect the change of transverse intensity distribution at different radius, suggesting that the real physical process could be more complicated than our assumptions.

It is worth noting that absence of emission in the other parts along our LOS, e.g. the black point A in the lower right panel of Fig. \ref{figure:obsbeam}, which has the same radius as the white point B in the bright beam, is unlikely due to the limb-darkening effect. We suggest that perhaps only one flux tube is active in each pole for PSR J1906$+$0746, if there are more, they must be far away from the median and less active. As to PSR J1141$-$6545, two active flux tubes seem to be responsible to the asymmetric beam.

In order to compare with the observation, we simulate the model beams with Model B2 for PSR J1141$-$6545 and the main pulse of PSR J1906$+$0746, as shown by the upper panels in Fig. \ref{figure:obsbeam}. The observational beams of PSR J1141$-$6545 and PSR J1906$+$0746, shown in the lower panels, are taken from Manchester et al. (2010) and Desvignes et al. (2013), respectively.

For PSR J1141$-$6545, two flux tubes are used in the simulate. Both of them have $\chi_0=0.7$, $\Re_0=0.3$ and the same peak multiplicity factors, but with different central azimuths, viz. $\varphi_{0,{\rm a}}=0$ and $\varphi_{0,{\rm b}}=35^{\rm o}$, as indicated by a short and a long white arrows in the left upper panel. The multiplicity is assumed to follow the same two-dimensional Gaussian distribution across the cross section of both flux tubes. Another important difference is the radial limb-darkening index, which is assumed to be $\delta=-4$ for the flux tube at the meridian plane and $\delta=-2$ for the lateral one. Although the simulated beam is not a satisfactory reproduction to the observed beam, but the general intensity structure is similar.

For the main pulse beam of PSR J1906$+$0746, only one flux tube with $\chi_0=0.7$, $\Re_0=0.3$ and $\varphi_0=40^{\rm o}$ is used in the simulation. The arrows in the simulated and observed beams represent approximately the direction of radial intensity gradient. The simulations presented here is for the purpose of comparing the general intensity distribution pattern between the simple models and observations. Some complex local structures in the observed beams are not reproduced and require detailed models.

PSR B1913$+$16 has a precession rate of 1.2 deg$/$yr. The radio pulse profile at 1.4~GHz consists of double peaks and a shallow bridge. Although the peak separation gradually decreases with time, the pulse width measured at a level below 50\% peak intensity is roughly constant between 1981 and 2003 (Weisberg \& Taylor 2005, Clifton \& Weisberg 2008). In fact, the pulse width at very low intensity levels, e.g. below 10\%, after keeping constant for about 10 years, has undergone a subtle increasing since 1992. Through fitting the profiles from 1981 to 2001 with a geometrical model of precessing radio beam and comparing the slope rate of PPA curve expected by the RVM and the observed data, Weisberg \& Taylor (2002) determined that the impact angle varied from about $-3.6^{\rm o}$ in 1981 to $-6.6^{\rm o}$ in 2001. Combing with the best-fit inclination angle $\alpha\sim158^{\rm o}$, the negative $\beta$ angles mean that our LOS was sweeping across the beam between the magnetic pole and the equatorial plane and was moving further from the magnetic pole.

Obviously, a circular conal beam model can not account for the constant and subtle increasing trend of pulse width in such viewing geometry. The authors proposed that the beam should be elongated in the latitudinal direction and pinched in longitude near the center, forming a hourglass-shaped beam (Weisberg \& Taylor 2002, Weisberg \& Taylor 2005, Clifton \& Weisberg 2008), which works well in modeling the historical data. Although the subtle pulse broadening is not so significant as the above simple fan beam model predicts, it is possible to be explained by taking into account some modifications, e.g. the aberration and retardation effects, the rotating dipole field and intensity modulation that may depend on emission altitude and azimuth. The narrowing of peak separation and increasing bridge emission may also be explained by this modified version if an appropriate number of flux tubes, e.g. three (corresponding to the leading, bridge and trailing components), with different altitude-dependent intensity distributions are assumed. We notice that the current hour-glass model predicts decreasing pulse width between 2003 and 2020 (Clifton \& Weisberg 2008), then the data in near future is hopeful to provide clear test for the hour-glass beam model and our fan beam model. The other quick test is to examine the evolution of flux density, if they exist in previous observations. We didn't find useful information on this respect, because the multi-epoch profiles have been normalized before constructing 2-D beam maps in published papers.

PSR B1534$+$12, a 37.9$-$ms millisecond pulsar (MSP) with broad main pulse and inter-pulse, locates in a binary system orbiting with another neutron star (Wolszczan 1991). The most recently measured precession rate is $0^{\rm o}.59$/yr (Fonseca et al. 2014). The main pulse width of PSR B1534+12 at 3\% level of peak intensity is roughly increasing at a rate of $0^{\rm o}.5\sim1^{\rm o}$ per year as estimated from the profiles in Arzoumanian (1995) and Stairs et al. (2000). This tendency of profile broadening is confirmed by later observations (Stairs et al. 2004, Fonseca et al. 2014), accompanied with a secular decreasing trend in intensity of the central component with respect to the leading and trailing wings. The impact angle is constrained by fitting the observed PPA with RVM, which changed from about $-3^{\rm o}.3$ in 1993 to $-6^{\rm o}.8$ in 2009, indicating that the line of sight was moving away from the magnetic pole (Fonseca et al. 2014). The tendency that the main pulse width increases with increasing impact angle (absolute value) is generally consistent with our model. It would be interesting to check whether the flux density decreases with time, as the fan beam model predicts, but the flux density data were not presented in the above references due to normalization of multi-epoch profiles.

PSR J0737$-$3039A, a 22.7$-$ms MSP in the double pulsar system, has a predicted precession rate of $4^{\rm o}.8$/yr (Burgay et al. 2003), which is comparable to that of PSR J0737$-$3039B. However, the multi-epoch pulse profiles of A were found to be fairly stable (Manchester et al. 2005, Ferdman et al. 2013), which is explained as a consequence of very small misalignment between the pulsar spin axis and the orbital momentum. The broad main pulse and interpulse were modeled by circular cones, where the inclination angle was constrained to be nearly $90^{\rm o}$ and the two cones should be originated from two opposite poles (Ferdman et al. 2013, Perera et al. 2014). We would suggest that the profile can be alternatively modeled by the fan beam model. Since our line of sight is stable for this pulsar, it is unable to distinguish the beam models directly. Nevertheless, we notice that PSR J0737$-$3039A and PSR B1534$+$12 have some features similar to many other MSPs, e.g. broad pulse width and complex profile shape. In the coming paper II, we will discuss how the fan beam model, in the context of emission flux tubes, has advantages to explain these features for MSPs.

Through numerically modeling the distortion of PSR J0737-3039B's magnetosphere induced by PSR J0737-3039A, Lomiashvili \& Lyutikov (2013) determined that the emission beam is horse-shoe shaped, which is somewhat similar to the arc-like structure derived by Perera et al. (2010, 2012), and the emission region is located at about 3750 stellar radius ($\sim30$\% light cylinder radius), greater than the maximum altitude of 2500 stellar radius given by Perera et al. (2012). Considering that the orbital-phase dependent distortion of B's magnetosphere may influence the intrinsic emission beam structure, we choose to discard this pulsar.

\subsection{Statistical tests}

In Section 2 we have derived two major relationships for the fan beam model, i.e. the radial limb-darkening relationship and the $W-|\beta|$ relationship. A third relationship between the upper boundary of $|\beta|$ and pulsar distance $d$ will be derived in this section. In order to test these relationships, we collected a sample of pulsars with known pseudo radio luminosity $L_{\rm app}=F_\nu d^2$, pulse width $W_{10}$, $d$ and impact angle, where $F_\nu$ is the flux density at a particular frequency.

Unlike the other parameters, the impact angle can not be directly measured. In literature, at least four kinds of methods
have been proposed to constrain the impact and inclination angles. (1) When a pulsar presents a smooth S$-$shaped PPA curve, $\alpha$
and $\beta$ can be obtained by fitting the PPA data with the following relation given by the RVM,
%
\begin{eqnarray}
\tan(\Psi-\Psi_0)=\sin\alpha\sin(\Phi-\Phi_0)[\sin(\alpha+\beta)\cos\alpha \nonumber \\
-\cos(\alpha+\beta)\sin\alpha\cos(\Phi-\Phi_0)]^{-1},
\label{eq:RVM}
\end{eqnarray}
where $\Psi$ is the PPA, the subscript ``0'' denotes the values at the phase where the slope rate of PPA curve reaches its extremum. (2) The parameters can be constrained by fitting the pulse width and some properties of subpulse drifting for a few pulsars, e.g. the interval between successive sub-pulses in the same period. (3) For a few binary pulsars with considerable precession rate, the parameters can be derived by fitting the profiles and PPA variation in terms of a precession model. (4) For gamma-ray pulsars, the parameters can be constrained by fitting the radio and gamma-ray profiles with gamma-ray emission and radio conal beam models. In this paper, we prefer to the first three geometrical methods to avoid the dependence of emission models, and the related references can be found in Table 1.

We collected a sample of 64 normal pulsars\footnote{With a period of $59$~ms and the surface magnetic field of $2.28\times 10^{10}$~G, PSR B1913+16 locates between the majorities of normal pulsars and MSPs in the $P-\dot{P}$ diagram. In some literature, e.g. Kramer et al. (1998), it is classified as a MSP. In this paper, we adopt it in the sample because its intermediate position in the $P-\dot{P}$ diagram. Although we believe that the fan beam model should be applicable to MSPs, we limit the model testing for normal pulsars in this paper, because some factors in MSPs, e.g. more efficient aberration effect in compact magnetosphere and possibly complex magnetic field structure, can cause deviation from the current simple model, which should be studied elsewhere.} from literature, of which the inclination and impact angles are derived with method (1) for 62 pulsars (including the binary pulsar PSR J1906$+$0746) and with methods (2) or (3) for the other 2 pulsars. A handful of pulsars that have too large uncertainties for the impact angle are not included. 12 pulsars in the sample have interpulse emission, and hence contribute beam information from double poles. Table 1 gives the data for the total 76 beams. From the second column are the inclination angle $\alpha$, impact angle $\beta$, frequencies at which $\alpha$ and $\beta$ are derived (with marked numbers for references), 10\% peak pulse widths $W_{10}$ with uncertainties, pulse widths $W_{10}^1$ at lower frequencies and $W_{10}^2$ at higher frequencies, pairs of frequencies at which $W_{10}^1$ and $W_{10}^2$ are measured, number of profile components $N_{\rm c}$ that is identified by eye, pulsar period $P$, period derivative $\dot{P}$, pseudo luminosity at 400 MHz $L_{400}$ and at 1400 MHz $L_{1400}$, rotation energy loss rate $\lg\dot{E}$ and pulsar distance $d$.

For most pulsars, the uncertainty of pulse width induced by frequency dependence of profiles is larger than the observational error at a single frequency. To count in this major error source, we collected the pulse widths at two well separated frequencies ($W_{10}^1$ and $W_{10}^2$), mostly at 0.4~MHz and 1.6~MHz. The uncertainty is then figured out by $|W_{10}^2-W_{10}^1|/2$. For those pulsars with only one frequency observation in literature, we assume an uncertainty of 10\% for $W_{10}$. The data in the last six columns are taken from the ATNF Pulsar Catalogue (Manchester et al. 1995).

\subsubsection{Test of $W-|\beta|$ relationship}

According to Eq. (\ref{eq:w-b}), the pulse width depends on $\alpha$, $\beta$ and the azimuthal width of a flux tube $\Delta\varphi=2\varphi$. These parameters are different for pulsars, causing dispersion of pulse widths. In order to compare the model with observations, we perform a Monte Carlo simulation by randomly assigning $\alpha$, $\beta$ and $\Delta\varphi$ to a sample of $\sim$50,000 pulsars. It is assumed that the projection of magnetic pole is uniformly distributed in the celestial sphere, then the probability density function of inclination angle is $P(\alpha)=\sin\alpha/2$\footnote{$P(\alpha)=\int_{0}^{2\pi}(1/4\pi)\sin\alpha {\rm d}\Phi=\sin\alpha/2$}. The probability density function of viewing angle is also assumed to be $P(\zeta)=\sin\zeta/2$. $\Delta\varphi$ is assumed to be uniformly distributed between $\Delta\varphi_1$ and $\Delta\varphi_2$, where the boundaries are free parameters that can be estimated by comparing the simulated results with observational data.

Since we assume that the effective beam radius can be treated as $\rho_{\rm b}=1\sim2\rho_{\rm pc}$ when the LOS sweeps across the inner part of the fan beam, the pulsar period will also affect the pulse width through its influence on $\rho_{\rm pc}$. We assume a lognormal distribution for the pulse period, i.e.
\begin{equation}
p=\frac{1}{P\sigma\sqrt{2\pi}}\exp[-\frac{(\ln P-\mu)^2}{2\sigma^2}],
\end{equation}
where $p$ is the probability density function of pulsar period $P$. The best-fit parameters $\mu$ and $\sigma$ are obtained by fitting the ATNF data of pulsar period longer than 50~ms, which are $\mu=-0.48$ (corresponding to a peak period 0.62~s) and $\sigma=0.90$. The period is also randomly assigned to each pulsar together with the other three parameters.

Once a pulsar is assigned with a group of $\alpha$, $\zeta$, $P$ and $\Delta\varphi$, the pulse width is calculated separately for two circumstances: using Eq. (\ref{eq:w-b}) when $|\beta|=|\zeta-\alpha|>2\rho_{\rm pc}$ and fixing it as $W=2\Phi$ figured out by substituting the assigned $\alpha$, beam radius $\rho_{\rm b}=2\rho_{\rm pc}$ and $\zeta=\alpha$ into Eq. (\ref{eq:rho-cosine}) when $|\beta|\leq 2\rho_{\rm pc}$. Selecting $|\beta|=2\rho_{\rm pc}$ as the criterion to separate the two circumstances and assuming the beam radius as $2\rho_{\rm pc}$ are phenomenological choices to make the simulation generally coincides with the lower boundary of the observed $W_{10}$ for small impact angles (see Fig. \ref{figure:w-b}). But it can be reasonably explained in the context of fan beam model, as will be shown below.

Given the other parameters, smaller inclination angles tend to produce wider profiles. To avoid possible contamination of this effect, we divide the sample into three groups with different $\alpha$, i.e. group A with $\alpha\leq 30^{\rm o}$, group B with $30^{\rm o}<\alpha\leq 60^{\rm o}$ and group C with $60^{\rm o}<\alpha\leq 90^{\rm o}$. The inclination angles in Table 1 that are larger than 90$^{\rm o}$ are converted to $180^{\rm o}-\alpha$ before grouping, and the corresponding $\beta$ are converted to $-\beta$. Each group has more than a dozen of pulsars, which are shown by the black dots in Fig. \ref{figure:w-b}. The simulated pulsars are also divided into three groups and are plotted by the grey dots.

One can see clearly the twofold $W-|\beta|$ relationship from the observed data in Fig \ref{figure:w-b}: $W$ is almost independent to $|\beta|$ when $|\beta|< 6\sim 8^{\rm o}$, while it increases with increasing $|\beta|$ when $|\beta|> 6\sim8^{\rm o}$. When selecting $\Delta\varphi_1$ and $\Delta\varphi_2$ as 40$^{\rm o}$ and 160$^{\rm o}$, respectively, we find that the observed distribution is well reproduced. For comparison, the $W-|\beta|$ relationships of Eq. (\ref{eq:w-b}) under various groups of $\alpha$ and $\varphi$ are plotted as dashed and dotted curves, e.g. curve ``1'' for $\alpha=5^{\rm o}$ and $\varphi=80^{\rm o}$. Obviously, such a simple equation only accounts for $W-|\beta|$ relationship for the outer beam.

These curves are helpful to understand why we select $|\beta|=2\rho_{\rm pc}$ as the criterion to distinguish the outer beam and the inner beam and set the effective beam radius as $\rho_{\rm b}=2\rho_{\rm pc}$. We have tried some other criteria but the simulations do not fit the lower part of the observed data. For instance, when selecting $|\beta|=\rho_{\rm pc}$ and $\rho_{\rm b}=\rho_{\rm pc}$, many simulated data points extends along these curves down to the small pulse width region, which is well below the lower boundary of the observed data points. When selecting $|\beta|=3\rho_{\rm pc}$ and $\rho_{\rm b}=3\rho_{\rm pc}$, the lower boundary of simulated widths will be higher than that of the observed data. That is the reason why we made the final choices of $2\rho_{\rm pc}$. A possible interpretation is that, in the statistical sense, the lowest coherent emission altitude may be a few times of $R$ so that the transition radius of twofold intensity distribution is $\rho_{\rm t}\sim 2\rho_{\rm pc}$, which then leads to twofold $W-|\beta|$ relationships separated by $|\beta|\sim 2\rho_{\rm pc}$.

Two possible selection effects can be excluded for the increasing $W_{10}-|\beta|$ trend. First, we have checked the data of $W_{10}$ and pulsar period $P$. No trend is found between these two parameters for the sample, indicating that the increasing $W_{10}$ trend is not induced by possible dependence of pulsar period. Second, it is known that the inclination and impact angles are hardly constrained with the RVM fitting method for very narrow profiles, because the PPA curves for a number of $\alpha$-$\beta$ pairs that produce the same maximum slope rate $k=\sin\alpha/\sin\beta$ can be barely distinguished in a very narrow phase interval around the flection point of PPA curve. Then one may suspect whether this limitation can cause bias and hence the observed trend can not be regarded as a robust evidence for the fan beam model. Especially when one considers an opposite case that the pulse width decreases with increasing $|\beta|$ as predicted by the conal beam model, this limitation might cause the lack of samples with large impact angles (due to narrow pulse widths) and would be unfavorable to test the prediction of conal beam model. However, the limitation can be overcome by the large phase separation between the main pulse and interpulse when both of them are observed, even though the pulses are narrow. For the 25 main and interpulse beams in our sample (only the main pulse width is measurable for PSR J1932$+$1059), 11 cases have $|\beta|>8^{\rm o}$ (6 with $|\beta|\ge 12^{\rm o}$). With so many large-impact-angle samples, no violation to the increasing trend of $W_{10}$ is observed, therefore, this selection effect can be ruled out.

\begin{figure*}
\centering
\resizebox{13cm}{16cm}
{
\includegraphics{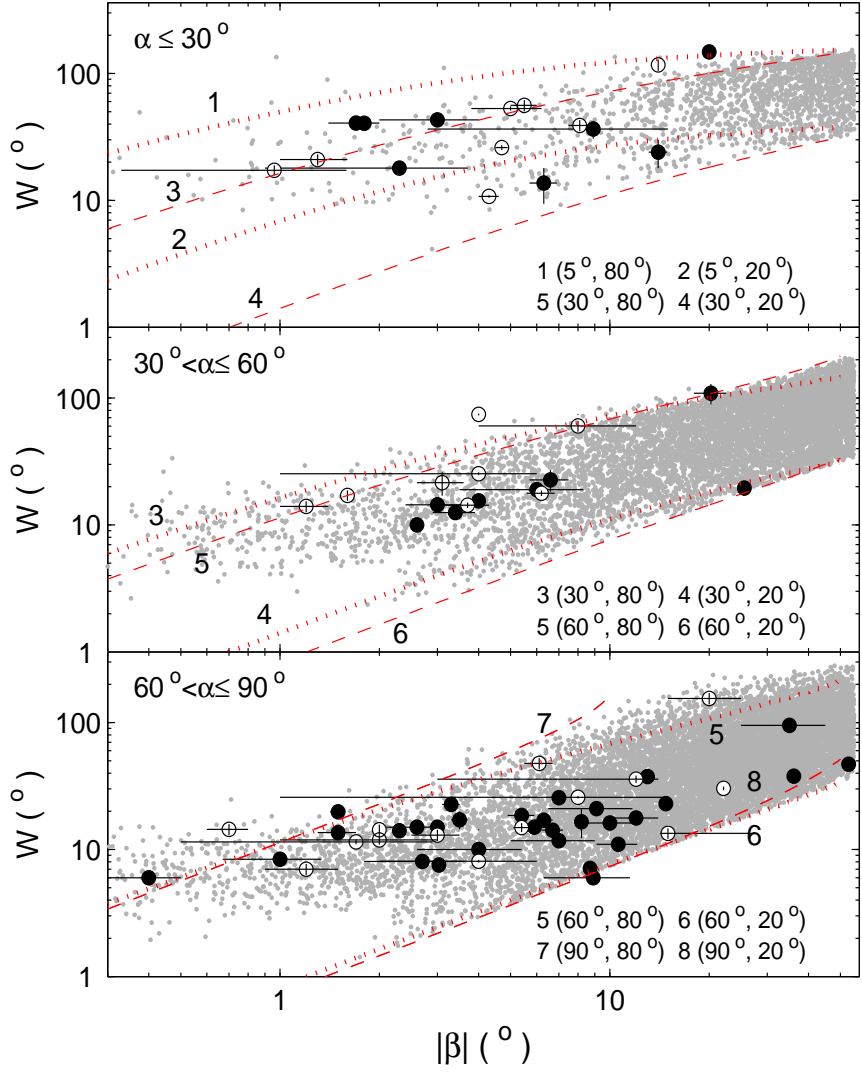}
}
\caption{The pulse width $W_{10}$ and $|\beta|$ for 76 beams from 64 pulsars (black dots) and the simulated data points for $\sim 50,000$ pulsars with the fan beam model (gray dots). The data are divided into three groups: $\alpha\le 30^{\rm o}$, $30^{\rm o}<\alpha\le 60^{\rm o}$ and $60^{\rm o}<\alpha\le 90^{\rm o}$, where the inclination angles larger than $90^{\rm o}$ in Table 1 are converted to $180^{\rm o}-\alpha$. Black dots and open circles stand for positive and negative impact angles, respectively. In the simulation, the half azimuthal widths of flux tubes are assumed to be uniformly distributed within $20^{\rm o}$ and $80^{\rm o}$. The pulse width $W$ is calculated with Eq. (\ref{eq:w-b}), except that in the case $|\beta|\leq 2\rho_{\rm pc}$, it is set as a constant as if the LOS sweeps across an effective beam with radius of $2\rho_{\rm pc}$(see Section 3.2.1 for details). The relationship of Eq. (\ref{eq:w-b}) is plotted by the dotted and dashed curves for several groups of inclination angle and half azimuth width $\varphi$. For instance, the curve 1 ($5^{\rm o}$, $80^{\rm o}$) stands for $\alpha=5^{\rm o}$ and $\varphi=80^{\rm o}$.}
\label{figure:w-b}
\end{figure*}

Finally, one may notice that the $W-|\beta|$ relationships predicted by Eqs. (\ref{eq:w-b}) and (\ref{eq:W-unif}) are actually different for positive and negative impact angles for a given $\alpha$ and other parameters and the difference grows with decreasing inclination angle, as shown by the solid and dashed curves in Fig. \ref{figure:wbmodel}. However, the predicted difference is not seen in the distribution of current data in Fig. \ref{figure:w-b}, where the positive and negative impact angles are represented by black dots and open circles, respectively. The reason is probably that the scatter due to other parameters, e.g. $\alpha$, $\varphi$ or $\delta$, overtakes the difference caused by the sign of impact angle (see the dashed curves in Fig. \ref{figure:w-b} for the scatter induced by $\alpha$ and $\varphi$). Perhaps this difference can only be tested for a large sample of pulsars.

\subsubsection{Test of intensity-radius relationship}
The observed mean flux density, $F=F_{\rm peak}W_{\rm e}/P$, is a quantity averaged over the whole pulse period, where $F_{\rm peak}$ is the flux density at the maximum peak of averaged pulse profile, $W_{\rm e}/P$ is the pulse duty cycle. Combining with the pulsar distance and other parameters, $F_{\rm peak}$ can be converted to the emission intensity $I_{\rm peak}$ at $\rho_{\rm peak}$ in the radio beam, where $\rho_{\rm peak}$ is the radius for the maximum pulse peak. The fan beam model predicts that such an intensity follows $I_{\rm peak}\propto \rho^{2q-6}$ in the outer beam. Therefore, the derived intensity can be used to test this intensity-radius relationship.

The peak flux at a frequency range $\nu\pm\Delta\nu/2$ can be converted to intensity by
$I_{\rm peak,\nu}=F_{\rm peak, \nu}d^2\Delta\nu$, namely $I_{\rm peak,\nu}=F_{\nu}(W_{\rm e}/P)^{-1}d^2\Delta\nu$. Using Eq. (\ref{eq:Lomg2}), we have
$$I_{\rm peak,\nu}=AP^{q-4}\dot{P}\cos^2\alpha \left(f_{1}^{q-3}-f_{2}^{q-3}\right)\rho_{\rm peak}^{2q-6}.$$
In order to display the dependence of $\rho_{\rm peak}$ clearly, we defined an intensity indicator parameter
\begin{equation}
Y=I_{\rm peak,\nu} A^{-1}P^{4-q}\dot{P}^{-1}\cos^{-2}\alpha(f_1^{q-3}-f_2^{q-3})^{-1},
\label{eq:Y}
\end{equation}
so that
\begin{equation}
Y=\rho_{\rm peak}^{2q-6}.
\label{eq:Y-b}
\end{equation}
Inserting $I_{\rm peak,\nu}$ into the above equation, there is $Y=F_\nu d^2(W_{\rm e}/P)^{-1}\Delta\nu A^{-1}P^{4-q}\dot{P}^{-1}\cos^{-2}\alpha(f_1^{q-3}-f_2^{q-3})^{-1}$, where $F_\nu d^2$ is the pseudo luminosity at a particular frequency $\nu$.

When trying to test the $Y-\rho_{\rm peak}$ relationship, it is better to use
\begin{equation}
Y=K\rho_{\rm peak}^{2q-6}.
\label{eq:Y-b-2}
\end{equation}
instead of Eq. (\ref{eq:Y-b}) to fit data, where $K$ is a coefficient to be constrained, because there are several free parameters in $Y$, which may cause uncertainties. These parameters include the boundaries of flux tube $f_1$ and $f_2$, index $q$, multiplicity $M$, stellar radius $R$ and emission power of a single particle $i_{\rm e0}$. In the following calculation we take $M=10^3$, $R=10^6$~cm, $i_{\rm e0}=10^{-15}$~erg/s, $f_1=1$ and $f_2=100$. In fact, choosing different values does not affect the power-law index and the $I_{\rm outer}-\rho_{\rm peak}$ relationships recovered later.

The pseudo luminosity $L_{400}=F_{400} d^2$ and $L_{1400}=F_{1400} d^2$ at 400~MHz and 1400~MHz are taken from the ATNF pulsar catalogue, the frequency range $\Delta \nu$ is set as 100~MHz and 500~MHz, respectively. $W_{\rm e}/P$ is simply replaced by $W_{10}/360^{\rm o}$, which may introduce an maximum uncertainty by a factor around 2 or 3 (see data in Gould \& Lyne 1998). However, this uncertainty is much smaller than the uncertainties caused by the free parameters. Notice that $\cos\alpha\sim 0$ when $\alpha\sim 90^{\rm o}$, which means that the net charge number density $n_{\rm gj0}\sim0$ and hence $n_0=Mn_{\rm gj0}\sim0$ on the polar cap surface. We believe that the real secondary particle density of orthogonal rotators should be much higher than this and may be comparable to other cases with moderate and small $\alpha$, therefore, we simply replace the correction term $\cos^{-2}\alpha$ by a factor of 1.

Finally, the free parameter $q$ needs special treatment, because it affects both sides of Eq. (\ref{eq:Y-b-2}). We let it vary from -3 to 3 by a very small step size\footnote{$q=3$ means a particular case that the emission power of single particle rises so effectively with increasing altitude that it just cancels the effect of density attenuation and eventually leads to constant intensity in the beam.}. Given a $q$ value, $Y$ is calculated for individual pulsars, and the least square fitting is applied to the $Y$ and $\rho_{\rm peak}$ data to find a best fit index $\delta$ for the relationship $Y\propto\rho_{\rm peak}^\delta$. If the relationship of Eq. (\ref{eq:Y-b-2}) does apply to the sample, there should be a $q$ satisfying $2q-6=\delta$. Therefore, the fitting process is repeated for the whole range of $q$ to search for the solution. Because Eq. (\ref{eq:Y-b-2}) is only valid for the outer beam, in the following test the pulsars with $|\beta|\leqslant 2\rho_{\rm pc}$ are excluded.

When calculating $\rho_{\rm peak}$ with the peak pulse phase $\Phi_{\rm peak}$, we have considered the effect of profile shape on $\Phi_{\rm peak}$. The peak phases are set as $\Phi_{\rm peak}\simeq W_{10}/4$, $3W_{10}/8$ and $5W_{10}/12$ for profiles with double, quadruple and sextuple components, where the profile center is always assumed to have $\Phi=0$. While the component number is odd, we simply assume that the maximum peak occurs at the profile center. Then $\rho_{\rm peak}$ is figured out with Eq. (\ref{eq:rho-cosine}) for each pulsar. The number of component, as listed in Table 1 by the column $N_{\rm c}$, is identified by eye for each pulsar according to its multi-frequency profile shapes whenever they are available in literature.

\begin{figure*}
\centering
\resizebox{16cm}{6cm}
{
\includegraphics{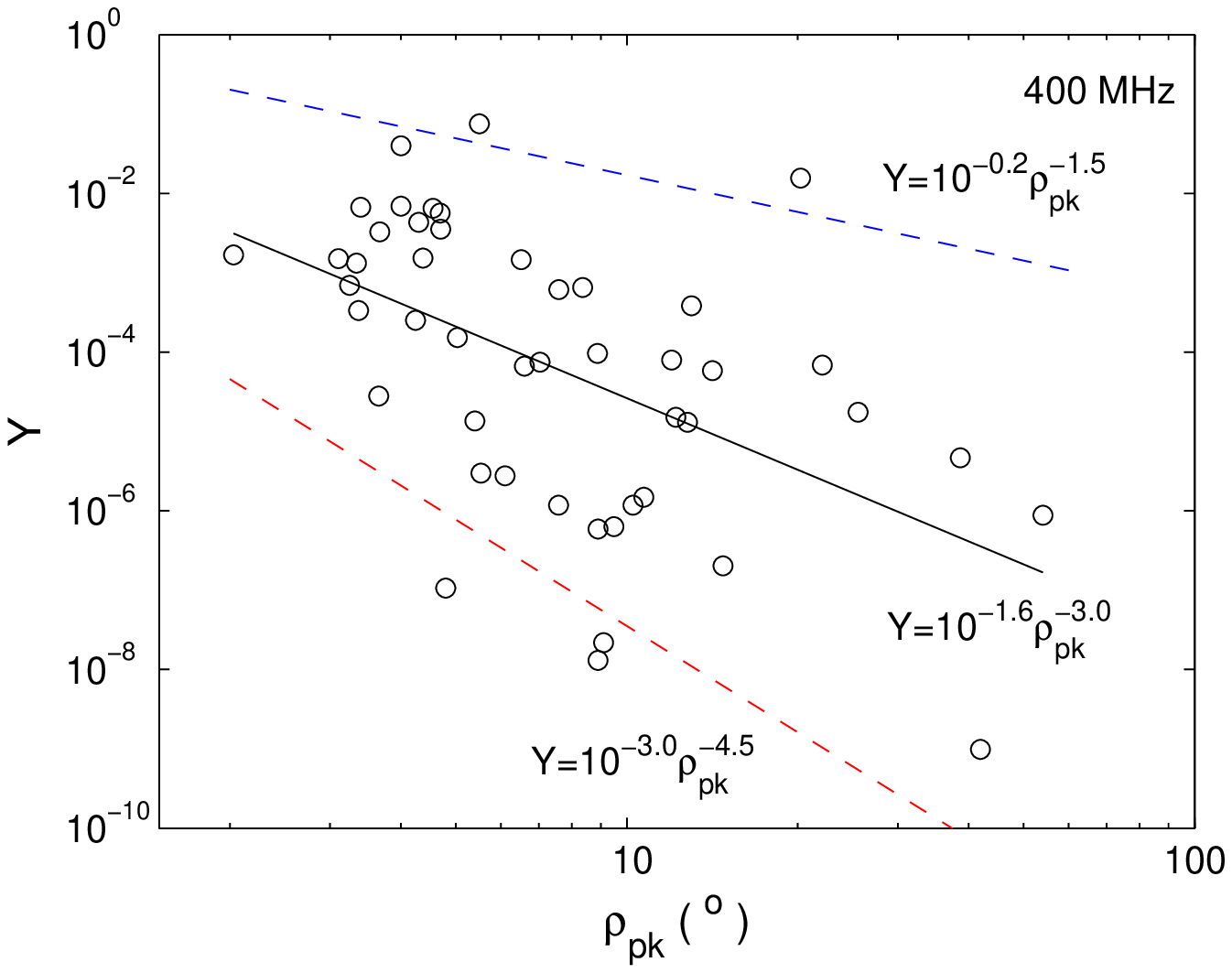}
\includegraphics{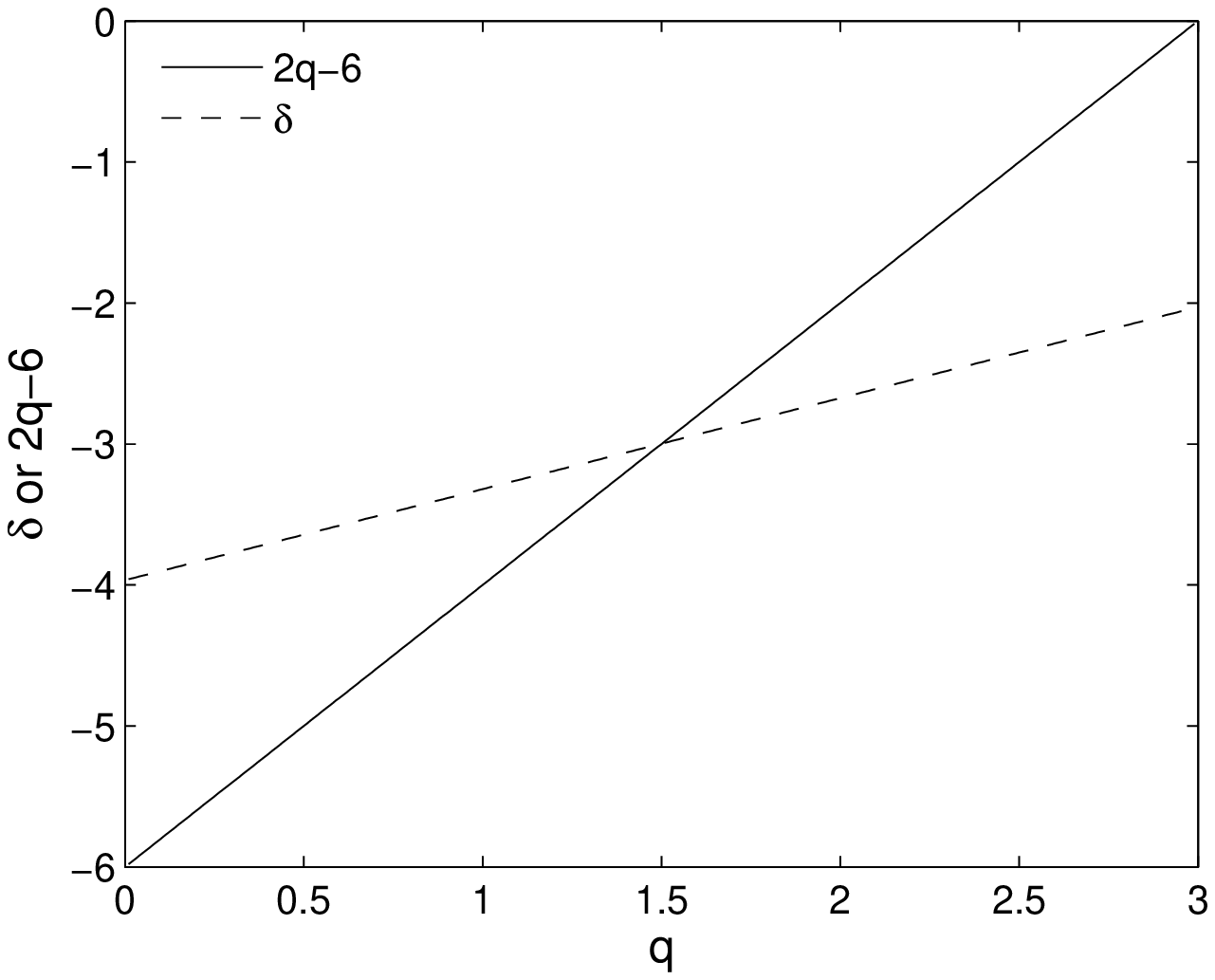}

}
\resizebox{16cm}{6cm}
{
\includegraphics{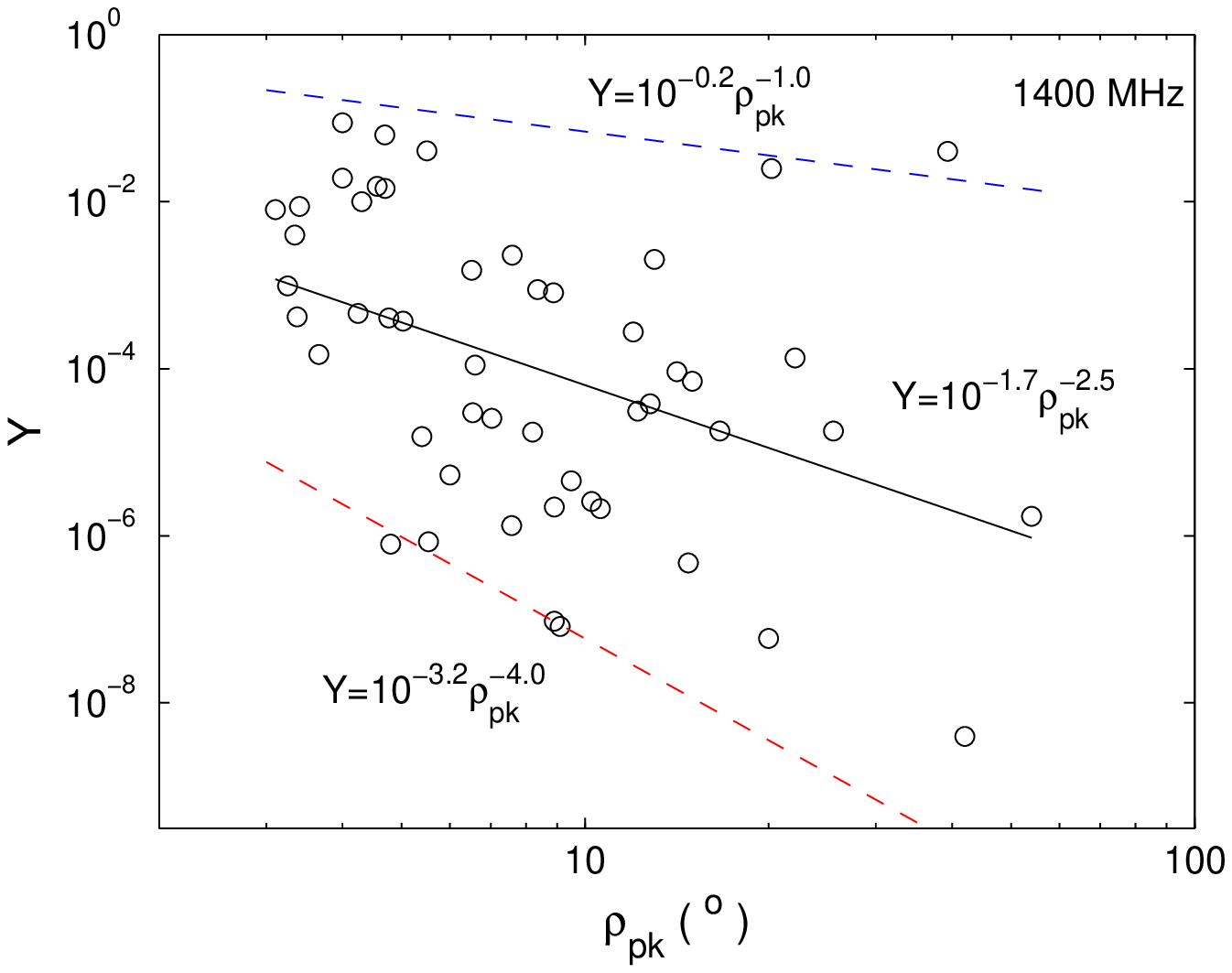}
\includegraphics{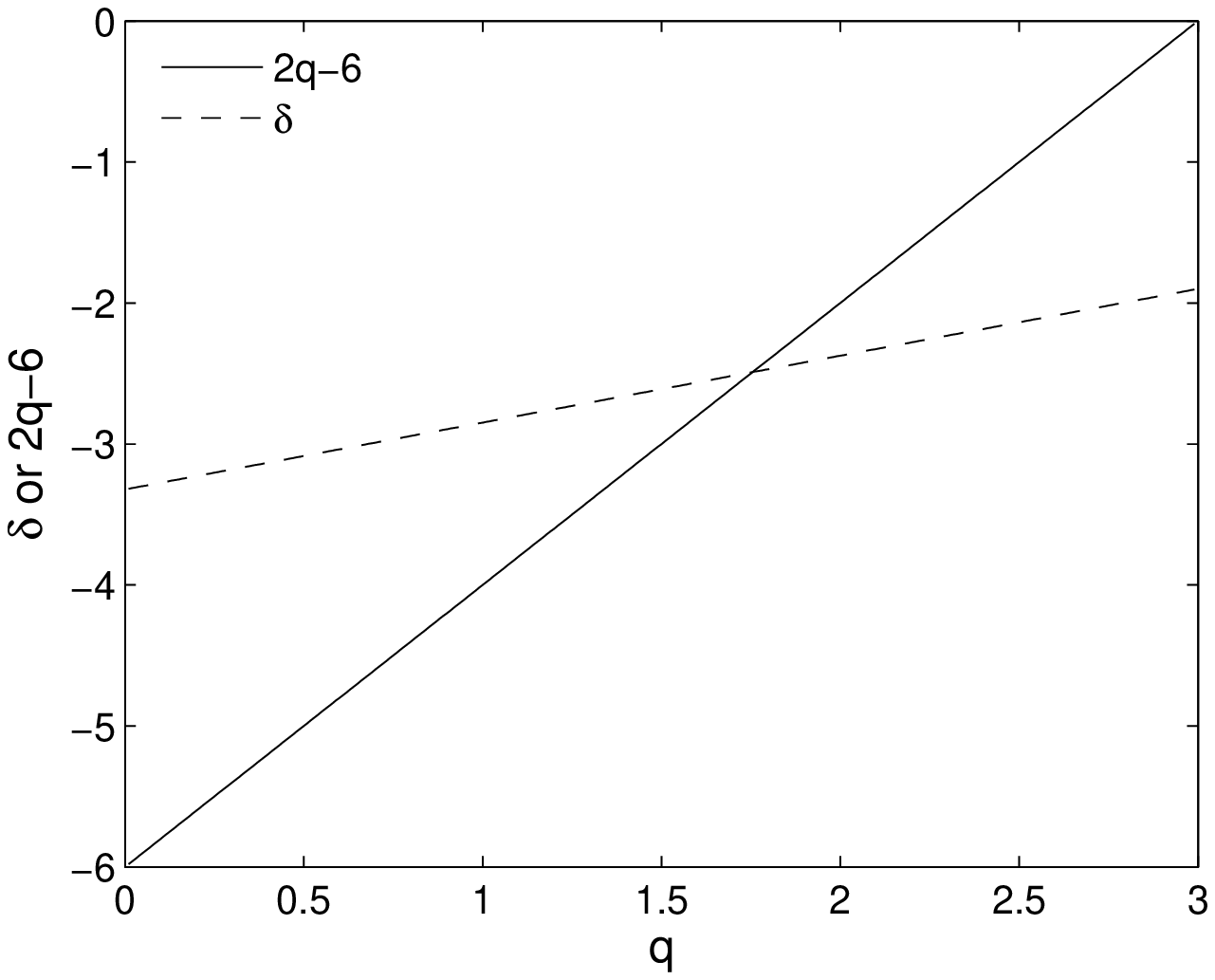}

}

\caption{Left: $Y-\rho_{\rm pk}$ diagram. The intensity indicator $Y$, obtained by correcting the ATNF data of $S_{400}d^2$ and $S_{1400}d^2$ with Eq. (\ref{eq:Y}), is expected to be a power-law of the radius $\rho_{\rm pk}$ that corresponds to the maximum pulse peak, i.e. $Y\propto \rho_{\rm pk}^\delta$. To avoid being contaminated by the data for the inner beam, which may follow a reverse intensity-radius relation, only the data with $|\beta|>2\rho_{\rm pc}$ are selected. The best fit relationship has an index $-3.0\pm1.5$ for 400~MHz data and $-2.5\pm1.5$ for 1400~MHz data at the 95\% confidence level. The solid and dashed lines represent the best-fit and the boundary relationships. Right: the best-fit index $\delta$ and modeled index $2q-6$ versus $q$, with the upper for 400~MHz and the lower for 1400~MHz, respectively. The intersections are the solutions of $q$. See Section 3.2.2 for details.}
\label{figure:l-b}
\end{figure*}

The right panels of Fig. \ref{figure:l-b} shows the modeled index $2q-6$ and the fitted index $\delta$ as a function of $q$ when the fitting process is applied to 400~MHz and 1400~MHz data, respectively. The intersections give the solution $q=1.5$ for 400~MHz and $q=1.75$ for 1400~MHz. The left panels of Fig. \ref{figure:l-b} present the $Y-\rho_{\rm peak}$ diagrams when the above solutions of $q$ are adopted. Despite the considerable scattering, the data show a trend that the intensity decrease when LOS becomes further from the beam center. The best fit limb-darkening relationship obtained with the Levenberg-Marquardt method is $Y=10^{-1.6\pm 1.4}\rho_{\rm peak}^{-3.0\pm 1.5}$ for 400~MHz and $Y=10^{-1.7\pm 1.5}\rho_{\rm peak}^{-2.5\pm 1.5}$ for 1400~MHz at the 95\% confidence level, respectively.
Combining with the results for 400~MHz and 1400~MHz, we have an approximate limb-darkening relationship
$I_{\rm out}\propto\rho^{-(2.8\pm 1.8)}$, with the index $q\simeq1.6\pm0.9$.

Using the best fit $Y-\rho_{\rm peak}$ relationships, we can recover the $I_{\rm outer}-\rho_{\rm peak}$ relationships with the following equation
$$I_{\rm outer}=YA(f_1^{q-3}-f_2^{q-3})\Delta\nu_{\rm MHz}^{-1}P^{q-4}\dot{P}_{-15}10^{-15},$$
where $I_{\rm outer}$ is in unit of erg$/$s$/$MHz$/$sr, $P$ is in unit of seconds, $\rho_{\rm peak}$ in unit of degrees,
$\dot{P}_{-15}=\dot{P}/(10^{-15}{\rm s/s})$, $\Delta\nu_{\rm MHz}=100$ and 500 for 400~MHz and 1400~MHz, respectively.
Substituting into the above parameters, we have
\begin{eqnarray}
I_{\rm outer}^{400{\rm MHz}} =  10^{27.2\pm1.4}P^{-2.50\pm0.75}\dot{P}_{-15}\rho_{\rm peak}^{-3.0\pm1.5} \nonumber \\
 {\rm erg/s/MHz/sr},
\label{eq:Ib_400M}
\end{eqnarray}
and
\begin{eqnarray}
I_{\rm outer}^{1400{\rm MHz}} =  10^{25.7\pm1.5}P^{-2.25\pm0.75}\dot{P}_{-15}\rho_{\rm peak}^{-2.5\pm1.5} \nonumber  \\
 {\rm erg/s/MHz/sr}.
\label{eq:Ib_1400M}
\end{eqnarray}

Although the above $I_{\rm outer}$ stands for the intensity at $\rho=\rho_{\rm peak}$ in the emission beam, it is reasonable to believe that these empirical relationships, in the statistical sense, can be used to describe the radial limb-darkening relationship for pulsar radio beams.

\subsubsection{Impact angle vs. pulsar distance}
Given an observing sensitivity, luminous pulsars have more chance to be detected at large distances. According to the radial limb-darkening relation in our model, a higher luminosity requires a smaller impact angle. Then, for a sample of pulsars, the upper limit of impact angle should decrease with pulsar distance. In the nearby region to the earth, less luminous pulsars may still reach the sensitivity, thus the impact angles are expected to be more scattered than in further regions.

To find out the relation between the distance $d$ and $|\beta|$, let us assume a fixed minimum detectable flux density $S_{\rm min}=C_{\rm s}\sqrt{W/(P-W)}$ (Lorimer \& Kramer 2005), where $W$ is the equivalent pulse width, $C_{\rm s}$ is a constant determined by some observational parameters, such as bandwidth, total integration time, etc. The corresponding peak flux density is $S_{\rm peak}=S_{\rm min}P/W\simeq C_{\rm s}\sqrt{P/W}$ when $P-W\simeq P$. In the fan beam model, an intensity $I_{\rm min}\propto|\beta|_{\rm max}^\delta$ would be detectable if
$S_{\rm peak}\sim I_{\rm min}/d^2$. This leads to
\begin{equation}
|\beta|\propto C_{\rm s}^{2/(1+2\delta)}d^{4/(1+2\delta)},
\label{eq:btd}
\end{equation}
if we use $W\propto|\beta|$ (a viable approximation to Eqs. (\ref{eq:w-b}) and (\ref{eq:W-unif}), see Fig. \ref{figure:wbmodel}).

\begin{figure}
\centering
\resizebox{8cm}{5.7cm}
{
\includegraphics{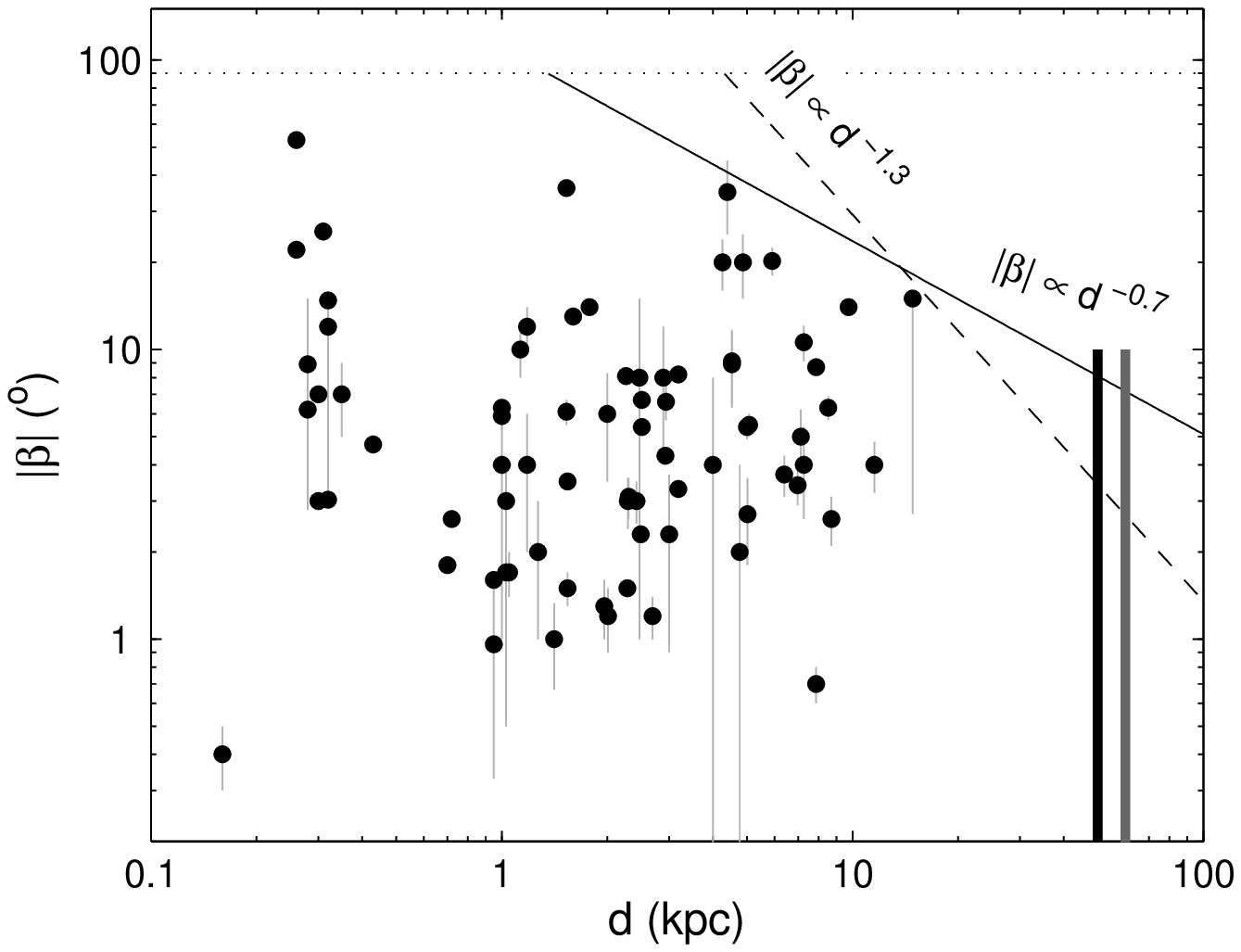}
}

\caption{Diagram of impact angle $|\beta|$ and pulsar distance $d$. The solid and dashed lines represent the power-law relationship $|\beta|\propto d^{4/(1+2\delta)}$ predicted by the limb-darkening relation $L\propto|\beta|^\delta$ when $\delta=-3.5$ and $-2.0$, respectively. $\beta$ is assumed not be greater than 90$^{\rm o}$ for each magnetic pole. The two vertical lines at the right hand stand for the expected $|\beta|$ range for the known pulsars in the Large (black line) and Small (gray line) Magellanic Clouds.  }
\label{figure:btd}
\end{figure}

Using the ATNF data of distance, we plot the $|\beta|-d$ diagram for the sample (Fig. \ref{figure:btd}). It does show that the impact angles are less scattered at larger distances and the upper boundary of $|\beta|$ decreases with distance until $\sim$15~kpc, where no more pulsars exist in the sample. The sparse data points at $d<1$~kpc is probably due to imcompleteness of the data set, but it does not affect the trend of upper limit. To compare the above relationship with the data, we plot two $|\beta|-d$ curves with $\delta=-2.0$ and $-3.5$, which corresponds to $|\beta|\propto d^{-1.3}$ (dashed) and $|\beta|\propto d^{-0.7}$ (solid) respectively. $|\beta|$ is restricted within 90$^{\rm o}$, because a $\beta$ larger than 90$^{\rm o}$ means viewing into the other pole and it can be replaced by its supplementary angle. Therefore the theoretical upper boundary at small distances are replaced by $|\beta|=90^{\rm o}$. As shown in the figure, the data points are all placed under these boundary lines, and the radial limb-darkening relationship with $\delta\sim -3$ generally matches the upper boundary of the data.

In order to examine if this consistency can extend to a larger distance, we have reviewed the currently known most distant pulsars in the Magellanic Clouds. Unfortunately, there is little polarization observation for these weak sources. But it was found that their pulse profiles are much narrower than those of Galactic pulsars (e.g. Manchester et al. 2006). This is a sign of small impact angle, if we believe that these two populations follow the same physics. We estimated $W_{10}$ for all the 21 normal pulsars in the Large Magellanic Cloud (LMC)\footnote{Two other pulsars with periods of 16ms and 50ms are not included.} and the 5 normal pulsars in the Small Magellanic Cloud (SMC) using the published profiles (McConnell et al. 1991, Crawford et al. 2001a, Manchester et al. 2006, Ridley et al. 2013). For a few pulsar with the signal to noise ratio lower than 10, the lowest level pulse width is measured. The $W_{10}$ of LMC pulsars varies from 6$^{\rm o}$ to $45^{\rm o}$, with only 4 larger than $22^{\rm o}$. The averaged $W_{10}$ is 15$^{\rm o}$ for the 17 small-width pulsars and 20$^{\rm o}$ for all the 21 pulsars. The $W_{10}$ of SMC pulsars varies from 11$^{\rm o}$ to $22^{\rm o}$, with an average value about 15$^{\rm o}$. These are considerably smaller than the averaged $W_{10}$ of 27$^{\rm o}$ of our sample. As shown by the data points in Fig. \ref{figure:w-b}, most pulsars with $W_{10}\leq20^{\rm o}$ have $|\beta|\leq 10^{\rm o}$, therefore, if there is no difference in the statistical property of emission beam between MC and Galactic populations, it can be inferred that most LMC and SMC pulsars have impact angles $|\beta|\leq 10^{\rm o}$. As to the 4 LMC pulsars with larger $W_{\rm 10}$, they may also have small impact angles if the inclination angle is not too large, say $\alpha<60^{\rm o}$ (see the upper two panels in Fig. \ref{figure:w-b}).

The expected range of impact angles are plotted in Fig. \ref{figure:btd} with two short lines at $d=50$kpc for LMC (Pietrzy\'{n}ski et al. 2013) and $d=60$kpc for SMC (Hilditch et al. 2005). Again, the combined upper boundary of Galactic and MC samples is consistent with the limb-darkening relation with the index around $-3$.

Some other effects may influence the $|\beta|-d$ distribution, but they can not account for the trend of the upper boundary lines. The following analysis will strengthen the conclusion that the impact-angle-distance upper boundary line is evidence to our model.

(1) Given the luminosity and pulsar distance, a narrow pulse profile is easier to be detected than a wide pulse profile. To estimate the impact of this selection effect quantitatively, one has to assume relationship between pulse width and impact angle. A possible estimation can be made as follows. Given a sensitivity $S_{\rm min}=C_{\rm s}\sqrt{W/(P-W)}$, a pulsar with luminosity $L$ will be detectable at $d$ when its flux density is comparable to $S_{\rm min}$, namely, $L/d^2\sim S_{\rm min}$. A simple derivation leads to $W\propto d^{-4}$. If still using the approximation in the fan beam model, $W\propto |\beta|$, we would have $|\beta|\propto d^{-4}$, which is too steep to account for the flat upper boundary line in the $|\beta|-d$ diagram. In order to account for the apparent power law index of the upper boundary line ($\sim-1$), one has to assume $W\propto |\beta|^{\sim4}$, which is not reliable.

(2) The disparity in sensitivity in different pulsar searching projects is unlikely to cause severe bias to the upper boundary line. For example, the limiting sensitivity is 0.14~mJy for Parkes multi-beam survey (PMS, Manchester et al. 2001) and 0.08~mJy for 5\% duty cycle for the survey of MC pulsars (Manchester et al. 2006). This difference can be denoted as $C_{\rm s, MC}/C_{\rm s, PMS}=4/7$. Using Eq. (\ref{eq:btd}), the sensitivity of 0.08~mJy causes an increment of $|\beta|$ by about 45\% (or 25\%) for MC pulsars compared with the $|\beta|$ estimated with the sensitivity of 0.14~mJy, if $\delta=-2$ (or -3) is used. Such an increment does not deviate much from the upper boundary line in Fig. \ref{figure:btd}.

(3) Although the pulsar period is not written in Eq. (\ref{eq:btd}), it does affect that relationship in our model. However, no statistically distinction is found for $P$ in all the distance ranges, including the MC pulsars. Therefore, the effect of $P$ should be trivial.

\section{Tests for the conal beam models}

The conal beam model has two categories: the canonical and patchy conal beam models. A canonical conal beam consists of one or more hollow cones and a core component. The shape of cones is circular or elliptical, and the cone is fully filled with emission. In the patchy conal beam models, cones are assumed to consist of some separated luminous spots, but the global conal structure is still recognizable (e.g. Rankin 1993, Karastergiou \& Johnston 2007). The global conal structure is obviously inconsistent with the very patchy beam patterns observed for PSR J1141$-$6546 and PSR J1906$+$0746. But are these two pulsars just minority outliers? Is the conal beam model still general for radio pulsars? Below we make statistical tests for the conal beam models based on the sample of 64 pulsars, similar to those have been done for the fan beam model.

Given a circular or an elliptical conal beam with a fixed beam width, the canonical conal beam model predicts that a larger impact angle leads to a narrower pulse width. Although pulsars may have different beam radii, but this trend should still exist for a sample of pulsars. In order to compare with the data and the model prediction, we also simulate a sample of $\sim 50,000$ pulsars to check the pulse width distribution as predicted by the conal beam model.

The procedure is very similar to the simulation in Section 3.2.1 except the treatment of pulse width. Firstly, we randomly assign the parameters $\alpha$, $\beta$ and the cone radius $\rho$ to each pulsar, and then calculate $\Phi$ with Eq. (\ref{eq:rho-cosine}). The the pulse width then reads $W=2\Phi$. Assuming a uniform distribution for $\rho$, we found that the range of $\rho$ needs to be within $3^{\rm o}$ and $50^{\rm o}$ to match the range of observed pulse widths.

\begin{figure*}
\centering
\resizebox{13cm}{15cm}
{
\includegraphics{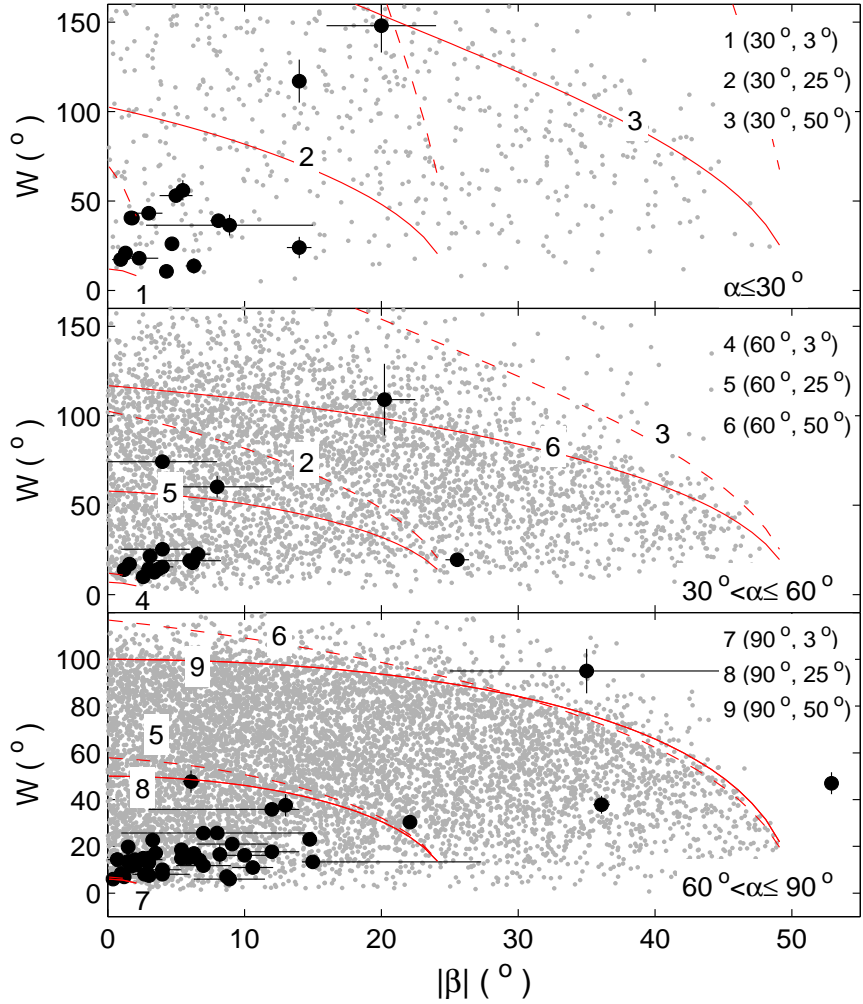}
}

\caption{The pulse width $W_{10}$ and $|\beta|$ for 76 beams from 64 pulsars (black dots) and the simulated data for $\sim 50,000$ pulsars with the canonical conal beam model (gray dots). In the simulation, the radii of circular beam of pulsars are assumed to be uniformly distributed within $3^{\rm o}$ and $50^{\rm o}$. The pulse width $W$ is calculated with Eq. (\ref{eq:rho-cosine}), where $W=2\Phi$. The relationship of Eq. (\ref{eq:rho-cosine}) is plotted by the dotted and dashed curves for several groups of inclination angle and beam radius. For instance, the curve 1 ($30^{\rm o}$, $3^{\rm o}$) stands for $\alpha=30^{\rm o}$ and $\rho=3^{\rm o}$. See Section 4 for details.}
\label{figure:w-b-cone}
\end{figure*}

Fig. \ref{figure:w-b-cone} shows the $W-|\beta|$ diagram for the simulated and observed data. The plot is made in linear scale for clarity. The modeled $W-|\beta|$ curves calculated with Eq. (\ref{eq:rho-cosine}) for groups of $\alpha$ and $\rho$ are also displayed. Obviously, the modeled curves show an opposite trend against the observed data. A striking discrepancy in the distributions of simulated and observed data can be seen in the left upper region of each panel. The selection effect of lack of data for small impact angles can be excluded, because in fact more data points in our sample are located in the small impact angle region than in the large impact angle region. Therefore, this discrepancy is definitely caused by the widening nature of pulse width at small impact angles in the conal beam model.

In the canonical conal beam model, the emission intensity could be either uniformly or randomly distributed in the cone, anyhow, it is independent to the impact angle. Since there is no predicted relationship between intensity and $\beta$ for conal beam models, we can not do further tests on this respect.

Besides the $W-\beta$ relationship, the upper boundary relationship between $|\beta|$ and $d$ can also be used to test the model. Since the pulse width is anticorrelated with $|\beta|$ in the conal beam model, we use an approximation $W\propto |\beta|^a$ for this relationship, where $a<0$. Given a limiting sensitivity $S_{\rm min}\propto\sqrt{W/(P-W)}$, following the derivation in the paragraph of point (1) in Section 3.2.3, one has a upper boundary line $|\beta|\propto d^{-4/a}$. Since $a<0$, the index will be $-4/a>0$. However, this is completely opposite to the apparent trend in Fig. \ref{figure:btd}.

To summarize, the canonical conal beam model are disfavored by the above two statistical tests and the patchy beams of two precessional pulsars. From geometrical view of point, in order to meet the observed $W-|\beta|$ trend, one may modify the beam shape by compressing the beam transversally at small impact angles while stretching it at large impact angles. A possible scenario will be presented in Section 5.

Since the global structure of patchy conal beam models is still conal, it will meet the same problems as the canonical
conal beam model when trying to interpret the statistical relationships. Therefore, we will not discuss this type of models further.

\section{On a patchy beam model with narrowband assumption}

We notice that an extreme case of patchy conal beam model associated with narrowband assumption, which has not been explored critically before, can generate a patchy beam pattern similar to those of binary pulsars. In this model, only one or a few flux tubes is active, where the emission is narrowband, i.e. a single frequency is mapped to a single altitude. A major distinction of this model from the other conal beam models lies that different parts of the patchy beam come from different layers of open field lines, and so does the leading or trailing beam boundary; while in traditional conal beam models, the boundary of the outermost conal beam is normally assumed to be confined by the last open field lines. This new model can account for the observed patchy beams and the statistically relationships, under a very particular assumption for the shape of flux tubes. However, it has big problems to explain the other phenomena, which will be discussed later in this section, therefore it is unlikely a general model for radio pulsars.

\begin{figure*}
\centering
\resizebox{15cm}{10cm}
{
\includegraphics{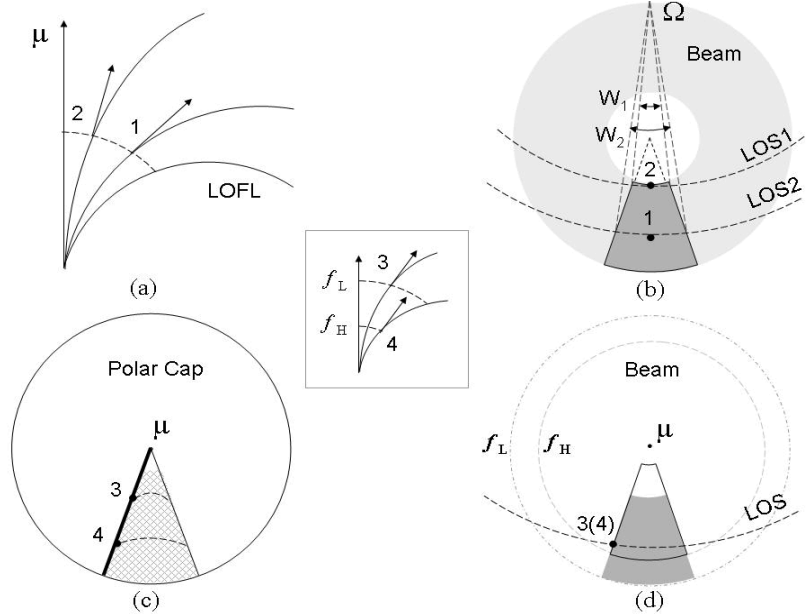}
}

\caption{The cartoon for a patchy beam model with the assumption of narrowband emission. (a) Open field lines in the meridian plane, where emissions at a single frequency come from the same altitude (dashed curve). (b) The patchy beam (dark) at a single frequency. The dashed curves stand for the lines of sight with different impact angles, LOS1 and LOS2, which see the emissions from outer field lines (e.g. 1 in panel (a)) and inner field lines (e.g. 2), respectively. The light grey annulus represents the global conal structure for a reference. With this presumed beam shape, the pulse width for LOS1, $W_1$, is smaller than that for LOS2, $W_2$. This is consistent with the statistical $W_{10}-|\beta|$ relationship. To form such a beam shape, the cross section of the flux tube needs to be sector-shaped, as shown by the hatched region in panel (c). Panels (c) and (d) illustrate the frequency evolution of the beam, where the low frequency $f_{\rm L}$ is assumed to be generated at a high altitude while the high frequency $f_{\rm H}$ at a low altitude (see the inset), and the LOS is fixed (panel (d)). The dashed curves in panel (c) stand for the foot points of those field lines viewed by the LOS at different frequency, with ``3'' for $f_{\rm L}$ and ``4'' for $f_{\rm H}$. The inset represents the field lines in the left-side lateral boundary of the flux tube (thick line in panel (c)). Points ``3'' and ``4'' have the emission direction, but they can only be observed by the LOS at different frequency. (d) The patchy beam at $f_{\rm L}$ (dark region) will shrink radially as frequency increases. The high frequency beam for $f_{\rm H}$ is presented by a region with solid boundary. However, because of the sector-shaped cross section of flux tube, the lateral boundaries of the patchy beam do not change with frequency. Therefore, the pulse width seen by the LOS keeps constant. As a comparison, in the framework of conal beam model, the conal beam boundary (dashed circles) shrinks with increasing frequency so that the LOS will see a pulse profile narrowing. }
\label{figure:patchycartoon}
\end{figure*}

In order to explain the observed beam pattern of PSR J1141$-$6545, one has to assume an extreme case that the active emission originates from one (or maybe two) flux tube, therefore only one (or maybe two) spot is bright in the beam. Under the assumption of narrowband emission, the emission in the inner part of the beam must come from more interior filed lines (Fig. \ref{figure:patchycartoon}(a) and (b)). To account for the elongated beam of PSR J1141$-$6545, the innermost emission at $\beta=0.9^{\rm o}$ should come from field lines very close to the magnetic pole, while the outermost emission at $\beta=3.7^{\rm o}$ should come from outer field lines, but not necessary the last open field lines, because the mildly intensity gradient implies that fainter emission probably exist out of the currently observed beam boundary, which should come from even outer open field lines. If we assume that the intensity is directly related to particle density, the observed intensity distribution requires a peak particle density at a particular layer of field lines close to the magnetic pole with a density gradient towards both the last open field lines and the magnetic pole. A similar scenario is required to account for the beam of PSR J1906$+$0746.

If the above scenario is a general model for pulsar radio beam, the cross section of the active flux tube on the polar cap should have a sector-like shape (similar to Model A in Section 2.4), namely, the azimuthal scale of any equi-colatitude layer of field lines is nearly constant, otherwise the positive correlation between the $W_{10}$ and $|\beta|$ can not be reproduced (see Fig. \ref{figure:patchycartoon}(b) and (c) for the illustration)\footnote{A more common assumption is that the cross section of a flux tube is circular. In the case of narrowband emission, a flux tube will form a quasi-circular beam. This picture can be regarded as an extreme case of patchy conal beam, namely, only one spotty structure is active in the cone. However, it has a problem to account for the $W_{10}-|\beta|$ distribution, because there will always be considerable chance for the LOS to sweep near the edge of the beam, which will results in very narrow pulse width, no matter how large the beam size and the impact angle are.}. However, such a model has to face the following challenges.

(1) The shape of the cross section of flux tube where pair cascades develop is not consistent with either the shape of cascade regions in the space-charge-limited-flow model (e.g. Figs. 2 and 5 in Harding \& Muslimov 2011) or the shape of sparks in the inner vacuum (or partially screened) gap model (eg. Fig. 1 in GS00). Note that the sector-shaped cross section is required to reproduce the $W-|\beta|$ relationship. As a comparison, in the fan beam model with broadband assumption, that relationship is a naturally result of the divergence of dipolar flux tube rather than an outcome of flux tube shape; so we do not need some unusual assumption on the shape of flux tubes.

(2) It can not explain the pulse width narrowing with increasing frequency by using the RFM, as done by conventional conal beam models. This is because the lateral boundary of the extreme patchy beam is no longer confined by a single layer of open field lines, such as the last open field lines as usually assumed by the canonical conal beam model, but is composed of a set of open field lines, from the last open to the allowed innermost ones (see Fig. \ref{figure:patchycartoon}(a)-(c)).

This point can be explained with Fig. \ref{figure:patchycartoon}(c), (d) and the inset. In the inset, the point 3, originally observed at a low frequency $f_{\rm L}$ and located at the field line 3 in panel (c) (represented by its footpoint on the polar cap), will be missed by the LOS at a high frequency $f_{\rm H}$, because the high frequency emission is assumed to be generated at a lower altitude. Instead, the point 4 from an outer field line at a lower altitude (shown in both the inset and panel (c)) will be viewed by the LOS at $f_{\rm H}$. Since the azimuth scale the cross section of flux tube is nearly constant due to its sector-like shape, the replacement of point 3 by point 4 does not change the beam width along the LOS (see panel (d)), thus the pulse width stays constant. Slight deviation from sector-like shape will lead to slightly frequency-dependent pulse width, but this scenario certainly can not account for the variety in the high-frequency pulse narrowing phenomenon, and not for the pulse broadening as well.

(3) Since the emission altitude at one frequency is fixed, the outer boundary of the patchy beam is confined by the outermost open field lines. This feature is common to the beam models with narrowband assumption. Therefore, such an extremely patchy beam model has the same limitation as the the conal beam models in explaining the phenomena of precursor, postcursor and off-pulse emission.

With the above arguments, we suggest that this patchy beam model, based on the assumption of narrowband emission from one or a few flux tubes, are unlikely to be a general model for radio pulsars.

%

\section{Conclusions and discussions}

We propose a new radio beam model based on the assumptions that the emission from secondary relativistic particles streaming along a flux tube is broadband and coherent. Under a set of basic assumptions on the dipolar magnetic field, the free flow of secondary particles and emission mechanisms/propagation effects, we demonstrate that the collective emission from a number of coherent volumes produces an emission beam pattern with the following features.
\begin{itemize}
  \item A flux tube forms a radially extended sub-beam. Starting from the innermost boundary of the sub-beam, the emission intensity increases with increasing beam radius until a transition radius near the polar cap boundary, and then keeps decreasing outwards.
  \item A Gaussian distribution of particle number density across the cross section of a flux tube leads to a transverse Gaussian distribution of emission intensity across the sub-beam.
  \item The transverse width of the sub-beam increases with radius due to the divergence nature of flux tube in the dipolar magnetic field. The whole radio beam may consist of a handful of sub-beams, forming a fan-like, limb-darkening beam pattern.
  \item When a very limited number of flux tubes, e.g. one or two, are active in emission, the beam becomes very patchy.
\end{itemize}

The fan beam model is different from the conal beam and patchy beam models in several respects, including the beam structure and predictions. Besides the shape of sub structures in the beams, another important difference is that the fan beam has no boundary while the conal and patchy beams have a circular or an elliptical boundary. The fan beam model predicts that the pulse width increases with increasing impact angle (the absolute value) in the outer part of radio beam (may differs in the inner part), while the conal beam models predict the opposite trend.

The observational evidence for the fan beam model comes from both individual sources and statistical relationships.
\begin{itemize}
  \item The limb-darkening patchy beams derived from long-term observations for two precessional binary pulsars, PSR J1141$-$6545 and J1906$+$0746, provide direct and strong evidence.
  \item We have compiled a sample of 64 pulsars with known impact angles, which are mostly constrained by the method of fitting the polarization position angle data with the RVM. Since 12 pulsars have interpulses, there are totally 76 groups of data. It is found that relationship between the 10\% peak pulse width $W_{10}$ and the absolute value of impact angle $|\beta|$ is twofold: a constant trend for small $|\beta|$ plus an increasing trend for large $|\beta|$, which is consistent with the $W-|\beta|$ relationship predicted by the model.
  \item  The derived emission intensity for the sample show an anti-correlation with $\rho_{\rm peak}$, the beam radius of the maximum pulse peak, which is consistent with the radial limb-darkening relationship, i.e. the decreasing intensity-radius relationship, in the outer part of fan beam.
  \item The upper boundary in the scatter plot of pulsar distance and $|\beta|$ shows that we tend to see pulsars with smaller impact angles at larger distances. This can be well reproduced by the radial limb-darkening relationship of the model.
\end{itemize}

We also tested conal beam models with the observational facts. Besides the difficulty in explaining the patchy beams of two binary pulsars, it is found that the statistical relationships disfavor both the canonical and patchy conal beam models. A patchy beam model based on the assumption of narrowband emission from a few flux tubes is also investigated and found to be unlikely a general model for radio pulsars.

Although the statistical relationships found for a sample of 64 pulsars disfavor conal beam models, we can not completely exclude the possibility that conal beam models, or in a broader sense, narrowband emission models may work for some pulsars. If they were mixed into our sample, their $|\beta|$ must happen to be considerably smaller than their beam radii, otherwise they would have violated the $W-|\beta|$ trend in Fig. \ref{figure:w-b}. From this point, we can infer that their percentage must be low. Future tests with larger samples will help to constrain the upper limit of the fraction.

We are aware that the current simple model has its limitation in several respects. (i) The distortion of static dipolar magnetic field, aberration and retardation effects are not taken into consideration. If they are included, the shape of sub-beam for a flux tube would be distorted at large radii. Numerical simulation will be helpful to quantitatively study the impacts of these effects. (ii) We notice that the beams of PSR J1141$-$6545 and J1906$+$0746 reveal some features that are not well reproduced by the simple version of fan-beam model, suggesting that the configuration of flux tubes and the distribution of limb-darkening index may be more complex than what we have assumed. More efforts are needed to model these individual beams. (iii) The empirical index $q$, as constrained to be within $0.7$ and $2.5$ by fitting the intensity data, requires physical explanation. These issues will be studied in subsequent papers. In the coming paper of this series, Paper II, we will focus on the configuration of flux tubes and the interpretation for a variety of observational properties of radio pulse profile.

Our model has implication to population studies on both radio and gamma-ray pulsars. In previous works, on one hand, the empirical relations of radio luminosity, e.g. $L=L_0 P^a \dot{P}^b$ (Arzoumanian et al. 2002, Gonthier et al. 2004, Story et al. 2007), or the luminosity function constrained from the observed flux densities and distances of radio pulsars (e.g. Chennamangalam et al. 2013), do not include the dependence of luminosity on the impact angle. On the other hand, the pulse width or beam radius is normally assumed to follow en empirical relation, $W=W_0P^{-1/2}$ or $\rho=\rho_0P^{-1/2}$ (Smits et al. 2009, Watters et al. 2009, Pierbattista et al. 2012, Perera et al. 2013), which is based on the concept of circular beam. When the above relationships are replaced by the radial limb-darkening intensity distribution and the $W-|\beta|$ relationship of the fan beam model, results will be different. We have presented the statistical relationship for the intensity (Eqs. (\ref{eq:Ib_400M}) and (\ref{eq:Ib_1400M})) and the approach to simulate the pulse width in our model. In general, the feature of extended beam will enhance the detection rate of radio pulsars, while the radial limb-darkening effect will reduce it. The final result depends on the competition of these two factors. Another related issue is the viewing geometry of radio-aloud gamma ray pulsars. Recent simulations with the conal radio beam model and the two-pole caustic (or outer gap) gamma-ray model show that radio-loud gamma-ray pulsars tend to have large inclination angles (Watters et al. 2009). However, with the radio fan beam model, gamma-ray pulsars with small inclination angles can still be radio-loud. This can be tested by future observations with high radio sensitivity.

In the fan beam model, we have suggested that remarkable disparity in secondary particle density probably exists among flux tubes. The density inhomogeneity may be helpful to understand the diversity of PPA swing for radio pulsars. The particle density plays an important role in propagation effects, which can dramatically affect the polarization behavior of pulsar emission. It was found that the PPA curve may maintain the RVM shape at low particle multiplicities (e.g. $M<100$) but deviate drastically from the RVM shape at much higher multiplicities (Wang et al. 2010, Andrianov \& Beskin 2010, Beskin \& Philippov 2012). In previous researches the particle density distribution is normally assumed to be symmetrical around the magnetic pole for the purpose of simplification. However, this is probably not true. As per the suggestion that the polarization is formed at a distance $r\sim 1000$ stellar radius for a pulsar period of 1~s (Wang et al. 2010, Beskin \& Philippov 2012), if only a few flux tubes near the meridian plane have high particle densities, the emission generated therein will propagate through other low-density flux tubes in the polarization formation region after the pulsar rotates for a long time $\Delta t=r/c$ (compared with the short duration of pulse window). Therefore, the simple RVM-type PPA curve may still survive. On the contrary, if a lateral flux tube met by the waves in the formation region has a high density, the resultant PPA curve may deviate from the RVM shape significantly. Although other parameters, e.g. the Lorentz factor of secondary particles and the initial emission altitude, also affect the polarization behavior, we believe that involving the inhomogeneity in flux tubes will be helpful to understand the polarization properties better. Anyway, the presumed lateral flux tubes, no matter with low or high particle density, need to be tested by future observations with high sensitivity.

\clearpage

%
\begin{deluxetable}{lllllllllllllll}
\tabletypesize{\scriptsize}
\rotate
\tablecaption{Parameters of pulsars}

\tablewidth{0pt}
\tablehead{
PSR J Name & $\alpha$ & $\beta$ & Freq. & $W_{10}$  & $W^1_{10}$ & $W^2_{10}$  & Freq. & $N_{\rm c}$ & $P$ & $\lg \dot{P}$ & $L_{400}$ & $L_{1400}$ & $\lg \dot{E}$ & $d$ \\
 & ($^{\rm o}$) & ($^{\rm o}$)  & (GHz) & ($^{\rm o}$)& ($^{\rm o}$) &($^{\rm o}$) & (GHz) & & (s) & (s/s) & (mJy$\cdot$ kpc$^2$) & (mJy$\cdot$ kpc$^2$) & (erg/s) & (kpc)\\
}

\startdata
0034$-$0721$^{\rm a}$	&	$3\pm1$	&	$3\pm1$	&	1.17 	$^1$	&	$43.2\pm2.3$	&	40.9 	$^2$	&	45.4 	$^2$	&	 0.4/1.4	&	2	&	0.943 	&	-15.39 	&	55.17 	&	11.67 	&	31.28 	&	1.03 	\\
0133$-$6957	&	$80\pm90$	&	$-3.0\pm0.5$	&	0.66 	$^3$	&	$13.0\pm1.3$	&	13.0 	$^3$	&			&	0.66 	&	 2	&	0.464 	&	-15.92 	&	29.28 	&		&	31.68 	&	2.42 	\\
0134$-$2937	&	$14\pm3$	&	$14.0\pm0.9$	&	0.44 	$^3$	&	$24.0\pm6.0$	&	30.0 	$^3$	&	17.9 	$^4$	&	 0.44/1.4	&	2	&	0.137 	&	-16.11 	&	28.52 	&	7.60 	&	33.08 	&	1.78 	\\
0248$+$6021	&	$46^{+64}_{-21}$	&	$6.0^{+2.3}_{-2.5}$	&	2.10 	$^5$	&	$19.0\pm2.0$	&			&	19.0 	$^5$	&	 2.10 	&	1	&	0.217 	&	-13.26 	&		&	54.80 	&	35.32 	&	2.00 	\\
0304$+$1932	&	$162.4\pm11.8$	&	$0.96\pm0.63$	&	1.42 	$^6$	&	$17.3\pm1.6$	&	18.9 	$^2$	&	15.7 	$^2$	&	 0.4/1.6	&	2	&	1.388 	&	-14.89 	&	24.37 	&	2.71 	&	31.28 	&	0.95 	\\
0332$+$5434	&	$51\pm28$	&	$-4^{+3}_{-2}$	&	0.41 	$^7$	&	$25.4\pm0.7$	&	26.0 	$^2$	&	24.7 	$^2$	&	 0.4/1.6	&	3	&	0.715 	&	-14.69 	&	1500.00 	&	203.00 	&	32.35 	&	1.00 	\\
0528$+$2200	&	$116.8\pm4.6$	&	$-1.50\pm0.08$	&	1.42 	$^6$	&	$19.8\pm1.0$	&	20.8 	$^2$	&	18.8 	$^2$	&	 0.4/1.6	&	2	&	3.746 	&	-13.40 	&	296.31 	&	46.79 	&	31.48 	&	2.28 	\\
0540$-$7125	&	$42\pm16$	&	$-1.2\pm0.2$	&	0.43 	$^3$	&	$14.0\pm1.4$	&	14.0 	$^3$	&	 		&	0.43 	&	 4	&	1.286 	&	-15.09 	&	36.18 	&		&	31.18 	&	2.69 	\\
0627$+$0706m	&	$86\pm0.2$	&	$8.7\pm0.3$	&	1.40 	$^8$	&	$7.1\pm0.7$	&	7.1 	$^8$	&			&	1.40 	&	2	 &	0.476 	&	-13.53 	&	234.46	&	130.92	&	34.04 	&	7.88 	\\
0627$+$0706i	&	$94\pm0.3$	&	$0.7\pm0.1$	&	1.40 	$^8$	&	$14.4\pm1.4$	&	14.4 	$^8$	&			&	1.40 	&	 2	&	0.476 	&	-13.53 	&	138.11	&	48.53	&	34.04 	&	7.88 	\\
0630$-$2834	&	$70\pm50$	&	$-12^{+9}_{-2}$	&	1.41 	$^9$	&	$35.8\pm2.4$	&	38.2 	$^2$	&	33.4 	$^2$	&	 0.4/1.6	&	1	&	1.244 	&	-14.15 	&	21.09 	&	2.36 	&	32.18 	&	0.32 	\\
0659$+$1414	&	$29\pm23$	&	$8.9\pm6.1$	&	1.42 	$^6$	&	$36.5\pm5.8$	&	42.2 	$^2$	&	30.5 	$^2$	&	 0.4/1.6	&	1	&	0.385 	&	-13.26 	&	0.51 	&	0.29 	&	34.58 	&	0.28 	\\
0738$-$4042	&	$62$	&	$13$	&	0.69 	$^{10}$	&	$37.6\pm4.7$	&	42.3 	$^{10}$	&	32.9 	$^{10}$	&	0.69/3.1	&	 3	&	0.375 	&	-14.79 	&	486.40 	&	204.80 	&	33.08 	&	1.60 	\\
0814$+$7429$^{\rm b}$	&	$8.7\pm0.2$	&	$-4.7\pm0.2$	&	0.33 	$^{11}$	&	$26.1\pm1.1$	&	25.0 	$^2$	&	27.2 	 $^2$	&	0.4/1.6	&	1	&	1.292 	&	-15.77 	&	14.74 	&	1.87 	&	30.49 	&	0.43 	\\
0826$+$2637m	&	$98.9\pm0.7$	&	$-3.03\pm0.01$	&	1.42 	$^6$	&	$7.6\pm0.9$	&	8.4 	$^2$	&	6.6 	$^2$	&	 0.4/1.6	&	1	&	0.531 	&	-14.77 	&	7.48	&	1.02	&	32.66 	&	0.32 	\\
0826$+$2637i	&	$81.1\pm0.7$	&	$14.77\pm0.01$	&	1.42 	$^6$	&	$23.0\pm2.3$	&	 		&	23.0 	$^6$	&	 1.4	&	1	&	0.531 	&	-14.77 	&	0.07	&	0.01	&	32.66 	&	0.32 	\\
0835$-$4510	&	$60$	&	$-6.2\pm0.6$	&	2.30 	$^{12}$	&	$17.8\pm0.9$	&	18.7 	$^{13}$	&	16.9 	$^{12}$	&	 0.4/2.3	&	2	&	0.089 	&	-12.90 	&	392.00 	&	86.24 	&	36.84 	&	0.28 	\\
0837$+$0610	&	$50$	&	$2.6$	&	0.33 	$^{14}$	&	$10.0\pm0.5$	&	9.6 	$^2$	&	10.5 	$^2$	&	0.4/1.4	&	2	 &	1.274 	&	-14.17 	&	46.14 	&	2.07 	&	32.11 	&	0.72 	\\
0908$-$4913m	&	$96.1\pm0.4$	&	$-5.9\pm0.6$	&	1.40 	$^{15}$	&	$15.0\pm1.5$	&			&	15.0 	$^{15}$	&	 1.40 	&	2	&	0.107 	&	-13.82 	&	20.14	&	7.19	&	35.69 	&	1.00 	\\
0908$-$4913i	&	$83.9\pm0.2$	&	$6.3\pm0.4$	&	1.40 	$^{15}$	&	$17.0\pm1.7$	&			&	17.0 	$^{15}$	&	1.40 	 &	2	&	0.107 	&	-13.82 	&	7.86	&	2.81	&	35.69 	&	1.00 	\\
0934$-$5249	&	$13\pm3$	&	$-4.3\pm0.3$	&	0.66 	$^3$	&	$10.8\pm0.3$	&	11.0 	$^3$	&	10.5 	$^{16}$	&	 0.66/1.4	&	1	&	1.445 	&	-14.33 	&	154.53 	&	10.30 	&	31.78 	&	2.93 	\\
0942$-$5657	&	$90$	&	$2.7\pm0.9$	&	0.66 	$^3$	&	$8.1\pm0.9$	&	9.0 	$^3$	&	7.1 	$^{16}$	&	0.66/1.4	&	 2	&	0.808 	&	-13.40 	&	327.61 	&	18.14 	&	33.47 	&	5.02 	\\
0953$+$0755m	&	$105.4\pm0.5$	&	$22.1\pm0.1$	&	1.42 	$^6$	&	$30.4\pm1.0$	&	29.4 	$^2$	&	31.4 	$^2$	 &	0.4/1.6	&	1	&	0.253 	&	-15.64 	&	26.72 	&	5.61 	&	32.75 	&	0.26 	\\
0953$+$0755i	&	$74.6\pm0.5$	&	$52.9\pm0.1$	&	1.42 	$^6$	&	$47.0\pm4.7$	&			&	47.0 	$^6$	&	 1.42 	&	2	&	0.253 	&	-15.64 	&	0.53 	&	0.11 	&	32.75 	&	0.26 	\\
1015$-$5719	&	$101\pm5$	&	$20\pm5$	&	1.37 	$^{17}$	&	$155\pm16$	&			&	155.0 	$^{17}$	&	1.37 	&	3	&	 0.140 	&	-13.24 	&		&	21.35 	&	35.92 	&	4.87 	\\
1036$-$4926	&	$70\pm50$	&	$2.6\pm0.5$	&	0.66 	$^3$	&	$15.0\pm1.5$	&	15.0 	$^3$	&			&	0.66 	&	3	 &	0.510 	&	-14.78 	&	682.78 	&		&	32.69 	&	8.71 	\\
1042$-$5521	&	$32\pm9$	&	$3.4\pm0.5$	&	0.66 	$^3$	&	$12.5\pm0.5$	&	13.0 	$^3$	&	12.0 	$^{16}$	&	 0.66/1.4	&	1	&	1.171 	&	-14.17 	&	682.09 	&	30.21 	&	32.22 	&	6.98 	\\
1057$-$5226m	&	$75.2\pm0.4$	&	$36.1\pm0.6$	&	1.37 	$^8$	&	$37.8\pm3.8$	&			&	37.8 	$^8$	&	 1.37 	&	4	&	0.197 	&	-14.23 	&	107.02	&		&	34.48 	&	1.53 	\\
1057$-$5226i	&	$104.8\pm0.4$	&	$6.1\pm0.6$	&	1.37 	$^8$	&	$47.7\pm4.8$	&			&	47.7 	$^8$	&	1.37 	 &	3	&	0.197 	&	-14.23 	&	80.25	&		&	34.48 	&	1.53 	\\
1136$+$1551	&	$88\pm5$	&	$7\pm2$	&	1.40 	$^{18}$	&	$11.8\pm1.3$	&	13.0 	$^2$	&	10.5 	$^2$	&	0.4/1.6	&	 2	&	1.188 	&	-14.43 	&	31.48 	&	3.92 	&	31.94 	&	0.35 	\\
1141$-$6545$^{\rm c}$	&	$20^{+16}_{-8}$	&	$2.3\pm1.4$	&	1.40 	$^{19}$	&	$18.0\pm1.8$	&			&	18.0 	$^{19}$	&	 1.4	&	2	&	0.394 	&	-14.37 	&		&	29.70 	&	33.45 	&	3.00 	\\
1253$-$5820	&	$35\pm10$	&	$6.6\pm0.9$	&	0.44 	$^3$	&	$22.7\pm4.4$	&	27.0 	$^3$	&	18.3 	$^{16}$	&	 0.44/1.4	&	3	&	0.255 	&	-14.68 	&	172.87 	&	30.25 	&	33.70 	&	2.94 	\\
1328$-$4357	&	$55\pm25$	&	$3.0\pm0.6$	&	0.44 	$^3$	&	$14.4\pm1.6$	&	16.0 	$^3$	&	12.8 	$^{20}$	&	 0.44/1.6	&	2	&	0.533 	&	-14.52 	&	94.39 	&	10.49 	&	32.90 	&	2.29 	\\
1513$-$5908	&	$80$	&	$35\pm10$	&	1.35 	$^{21}$	&	$95.2\pm9.5$	&			&	95.2 	$^{21}$	&	1.35 	&	2	&	 0.151 	&	-11.82 	&	29.04 	&	18.20 	&	37.23 	&	4.40 	\\
1527$-$3931	&	$70\pm55$	&	$-1.2\pm0.3$	&	0.44/0.66	$^3$	&	$7.0\pm0.7$	&	7.0 	$^3$	&			&	0.66	&	 2	&	2.418 	&	-13.72 	&	44.44 	&		&	31.73 	&	2.01 	\\
1543$+$0929	&	$131.0\pm5.7$	&	$-20.2\pm2.3$	&	1.42 	$^6$	&	$109\pm20$	&	88.5 	$^2$	&	129.2 	$^2$	&	 0.4/1.4	&	3	&	0.748 	&	-15.36 	&	2715.18 	&	205.38 	&	31.61 	&	5.90 	\\
1549$-$4848m	&	$92.5\pm0.2$	&	$-3.5\pm0.2$	&	1.40 	$^8$	&	$17.2\pm2.8$	&	19.9 	$^3$	&	14.4 	$^8$	 &	0.66/1.4	&	2	&	0.288 	&	-13.85 	&	28.99	&	0.8	&	34.37 	&	1.54 	\\
1549$-$4848i	&	$87.5\pm0.5$	&	$1.5\pm0.2$	&	1.40 	$^8$	&	$13.6\pm1.4$	&			&	13.6 	$^8$	&	1.40 	 &	3	&	0.288 	&	-13.85 	&	11.31	&	0.31	&	34.37 	&	1.54 	\\
1550$-$5418	&	$160$	&	$14$	&	8.36 	$^{22}$	&	$117\pm12$	&			&	117.0 	$^{22}$	&	8.36	&	6	&	2.070 	 &	-10.63 	&		&	313.06 	&	35.00 	&	9.74 	\\
1559$-$4438	&	$37\pm12$	&	$-3.1\pm0.5$	&	0.66/1.5	$^3$	&	$21.5\pm4.5$	&	17.0 	$^3$	&	26.0 	$^3$	&	 0.66/1.5	&	2	&	0.257 	&	-14.99 	&	581.90 	&	211.60 	&	33.37 	&	2.30 	\\
1603$-$5657	&	$26\pm6$	&	$6.3\pm0.6$	&	0.66 	$^3$	&	$13.7\pm4.3$	&	18.0 	$^3$	&	9.4 	$^{16}$	&	 0.66/1.4	&	2	&	0.496 	&	-14.55 	&	580.72 	&	38.47 	&	32.96 	&	8.52 	\\
1622$-$3751	&	$40$	&	$4.0$	&	1.37 	$^{23}$	&	$15.5\pm1.6$	&			&	15.5 	$^{23}$	&	1.37	&	2	&	 0.731	&	-14.59 	&		&	3.04 	&	32.42 	&	11.55 	\\
1651$-$5222	&	$50\pm20$	&	$-3.7\pm0.6$	&	0.66 	$^3$	&	$14.3\pm0.7$	&	15.0 	$^3$	&	13.6 	$^{16}$	&	 0.66/1.4	&	2	&	0.635 	&	-14.74 	&	939.14 	&	118.41 	&	32.45 	&	6.39 	\\
1705$-$1906m	&	$94\pm2$	&	$-12\pm2$	&	4.85 	$^{24}$	&	$17.7\pm0.5$	&	17.3 	$^2$	&	18.2 	$^2$	&	 0.4/1.6	&	2	&	0.299 	&	-14.38 	&	35.13	&	9.65	&	33.79 	&	1.18 	\\
1705$-$1906i	&	$86\pm2$	&	$-4\pm2$	&	4.85 	$^{24}$	&	$8.1\pm0.2$	&	7.9 	$^2$	&	8.3 	$^2$	&	 0.6/1.6	&	1	&	0.299 	&	-14.38 	&	5.27	&	1.45	&	33.79 	&	1.18 	\\
1709$-$1640	&	$90\pm65$	&	$-2\pm1$	&	4.85 	$^{24}$	&	$11.9\pm1.0$	&	12.8 	$^2$	&	10.9 	$^2$	&	 0.4/1.6	&	1	&	0.653 	&	-14.20 	&	75.81 	&	6.45 	&	32.95 	&	1.27 	\\
1710$-$2616	&	$30$	&	$20$	&	1.37 	$^{23}$	&	$148\pm15$	&			&	148.0 	$^{23}$	&	1.37	&	4	&	0.954	 &	-16.70 	&		&	9.46 	&	29.95 	&	4.26 	\\
1722$-$3712m	&	$90.7\pm0.1$	&	$5.4\pm0.3$	&	1.40 	$^8$	&	$14.9\pm1.2$	&	16.0 	$^3$	&	13.7 	$^{16}$	&	 0.66/1.4	&	1	&	0.236 	&	-13.96 	&	145.46	&	18.63	&	34.51 	&	2.51 	\\
1722$-$3712i	&	$89.3\pm0.1$	&	$6.7\pm0.5$	&	1.40 	$^8$	&	$14.2\pm1$	&			&	14.2 	$^8$	&	1.40 	&	 2	&	0.236 	&	-13.96 	&	12.04	&	1.53	&	34.51 	&	2.51 	\\
1739$-$2903m	&	$84.2\pm0.3$	&	$3.3\pm0.2$	&	1.40 	$^8$	&	$22.7\pm1.2$	&	23.8 	$^2$	&	21.5 	$^8$	&	 0.6/1.6	&	2	&	0.323 	&	-14.10 	&		&	14.23	&	33.97 	&	3.19 	\\
1739$-$2903i	&	$95.8\pm0.4$	&	$-8.2\pm0.4$	&	1.40 	$^8$	&	$16.6\pm4.3$	&	20.8 	$^3$	&	12.3 	$^8$	 &	0.66/1.4	&	1	&	0.323 	&	-14.10 	&		&	6.12	&	33.97 	&	3.19 	\\
1740$+$1311	&	$90\pm86$	&	$-2\pm2$	&	4.85 	$^{24}$	&	$14.3\pm1.4$	&			&	14.3 	$^2$	&	1.4	&	3	&	 0.803 	&	-14.84 	&	546.07 	&	88.74 	&	32.05 	&	4.77 	\\
1750$-$3503	&	$11\pm2$	&	$-5.5\pm0.5$	&	0.66 	$^3$	&	$56\pm5.6$	&	56.0 	$^3$	&			&	0.66 	&	1	 &	0.684 	&	-16.42 	&	745.44 	&	20.31 	&	30.67 	&	5.07 	\\
1751$-$4657	&	$90$	&	$-1.7\pm1.2$	&	0.43 	$^3$	&	$11.5\pm0.5$	&	12.0 	$^3$	&	11.0 	$^{25}$	&	 0.43/0.95	&	2	&	0.742 	&	-14.89 	&	74.26 	&	10.61 	&	32.10 	&	1.03 	\\
1825$-$0935m	&	$95$	&	$-7$	&	0.69 	$^{10}$	&	$25.7\pm1.1$	&	26.7 	$^{10}$	&	24.6 	$^{10}$	&	0.69/3.1	 &	2	&	0.769 	&	-13.28 	&	2.95	&	0.98	&	33.66 	&	0.30 	\\
1825$-$0935i	&	$85$	&	$3$	&	0.69 	$^{10}$	&	$15.0\pm1.0$	&	16.0 	$^2$	&	14.0 	$^2$	&	0.6/1.6	&	2	 &	0.769 	&	-13.28 	&	0.29	&	0.1	&	33.66 	&	0.30 	\\
1828$-$1101m	&	$97.3\pm0.6$	&	$-10.6\pm1.5$	&	3.10 	$^8$	&	$11.0\pm1.1$	&			&	11.0 	$^8$	&	 3.10 	&	1	&	0.072 	&	-13.83 	&		&	121.32	&	36.19 	&	7.26 	\\
1828$-$1101i	&	$82.7\pm0.6$	&	$4.0\pm1.4$	&	3.10 	$^8$	&	$10.0\pm1.0$	&			&	14.0 	$^8$	&	3.10 	 &	1	&	0.072 	&	-13.83 	&		&	31.53	&	36.19 	&	7.26 	\\
1841$+$0912	&	$86.1\pm11.4$	&	$2.30\pm0.04$	&	1.42 	$^6$	&	$14.1\pm1.5$	&	12.6 	$^2$	&	15.5 	$^2$	&	 0.4/1.4	&	3	&	0.381 	&	-14.96 	&	124.00 	&	10.54 	&	32.89 	&	2.49 	\\
1852$-$2610	&	$17\pm3$	&	$-8.1\pm0.6$	&	0.43 	$^3$	&	$39.0\pm3.9$	&	39.0 	$^3$	&			&	0.43 	&	 2	&	0.336 	&	-16.06 	&	61.29 	&	7.15 	&	31.96 	&	2.26 	\\
1900$-$2600	&	$25$	&	$1.8$	&	0.33 	$^{26}$	&	$40.5\pm1.3$	&	41.8 	$^2$	&	39.2 	$^2$	&	0.4/1.6	&	5	 &	0.612 	&	-15.69 	&	64.19 	&	6.37 	&	31.54 	&	0.70 	\\
1906$+$0746m	&	$81^{+1}_{-66}$	&	$8.9\pm2.6$	&	1.4	$^{27}$	&	$6.0\pm0.6$	&			&	6.0 	$^{28}$	&	1.4	&	1	&	 0.144	&	-13.69	&	16.35	&	9.99	&	35.43	&	4.53 	\\
1906$+$0746i	&	$99^{+66}_{-1}$	&	$-9.1\pm2.6$	&	1.4	$^{27}$	&	$21.0\pm2.1$	&			&	21.0 	$^{28}$	&	1.4	&	 1	&	0.144	&	-13.69	&	2.12	&	1.3	&	35.43	&	4.53 	\\
J1915$+$1606	&	$157\pm5$	&	$5.0\pm1.2$	&	1.4	$^{29}$	&	$53.0\pm2.0$	&	55.0 	$^{30}$	&	51.0 	$^{29}$	&	 0.43/1.4	&	2	&	0.059	&	-17.06	&	203.35	&	45.75	&	33.23	&	7.13 	\\
1917$+$1353	&	$73\pm19.4$	&	$5.4\pm0.5$	&	1.42 	$^6$	&	$18.6\pm1.1$	&	19.7 	$^2$	&	17.5 	$^2$	&	 0.4/1.4	&	2	&	0.195 	&	-14.14 	&	1075.00 	&	47.50 	&	34.59 	&	5.00 	\\
1918$+$1444	&	$118.0\pm33.4$	&	$-1.00\pm0.33$	&	1.42 	$^6$	&	$8.4\pm0.8$	&	9.1 	$^2$	&	7.6 	$^2$	&	 0.4/1.6	&	3	&	1.181 	&	-12.67 	&	3.18 	&	1.99 	&	33.71 	&	1.41 	\\
1932$+$1059m	&	$35.97\pm0.95$	&	$25.55\pm0.87$	&	1.42 	$^6$	&	$19.5\pm1.4$	&	20.9 	$^2$	&	18.1 	$^2$	 &	0.4/1.6	&	3	&	0.227 	&	-14.94 	&	29.12 	&	3.46 	&	33.59	&	0.31 	\\
1941$+$0121	&	$138\pm32$	&	$8\pm4$	&	0.82 	$^{31}$	&	$60.3\pm6.0$	&	60.3 	$^{31}$	&			&	0.82 	&	2	&	 0.217 	&	-15.72 	&	14.30 	&	3.50 	&	32.87 	&	2.89 	\\
2043$+$2740	&	$67\pm15$	&	$10\pm2$	&	1.36 	$^{32}$	&	$16.2\pm1.6$	&			&	16.2 	$^{32}$	&	1.36 	&	2	 &	0.096 	&	-14.90 	&	19.15 	&		&	34.75 	&	1.13 	\\
2048$-$1616	&	$34$	&	$-1.6$	&	0.69 	$^{10}$	&	$17.1\pm1.6$	&	18.6 	$^2$	&	15.5 	$^2$	&	0.4/1.6	&	3	 &	1.962 	&	-13.96 	&	104.69 	&	11.73 	&	31.76 	&	0.95 	\\
2053$-$7200	&	$27\pm8$	&	$1.7\pm0.3$	&	0.66 	$^3$	&	$40.6\pm1.6$	&	39.0 	$^3$	&	42.2 	$^{33}$	&	 0.66/1.4	&	2	&	0.341 	&	-15.71 	&	31.97 	&	6.62 	&	32.29 	&	1.05 	\\
2113$+$4644	&	$38\pm37$	&	$-4\pm4$	&	4.85 	$^{24}$	&	$74.3\pm1.2$	&	73.1 	$^2$	&	75.5 	$^2$	&	 0.6/0.9	&	3	&	1.015 	&	-15.15 	&	3680.00 	&	304.00 	&	31.43 	&	4.00 	\\
2144$-$3933	&	$90$	&	$0.4\pm0.1$	&	0.66 	$^3$	&	$6.0\pm0.6$	&	6.0 	$^3$	&			&	0.66 	&	1	&	 8.510 	&	-15.30 	&	0.41 	&	0.02 	&	28.50 	&	0.16 	\\
2240$+$5832	&	$108^{+42}_{-98}$	&	$15^{+0.8}_{-12.3}$	&	1.41 	$^5$	&	$13.4\pm1.3$	&			&	13.4 	$^5$	&	 1.41 	&	3	&	0.140 	&	-13.81 	&		&	595.41 	&	35.34 	&	14.85 	\\
2337$+$6151	&	$90\pm88$	&	$-8\pm7$	&	4.85 	$^{24}$	&	$25.8\pm0.6$	&	25.2 	$^2$	&	26.3 	$^2$	&	 0.4/1.4	&	2	&	0.495 	&	-12.71 	&	61.01 	&	8.54 	&	34.80 	&	2.47 	\\
2346$-$0609	&	$28\pm9$	&	$-1.3\pm0.3$	&	0.44 	$^3$	&	$21.0\pm2.1$	&	21.0 	$^3$	&			&	0.44 	&	 3	&	1.181 	&	-14.87 	&	42.26 	&	7.68 	&	31.51 	&	1.96 	\\


\enddata
\tablenotetext{\space}{ $^{\rm a}$ $\alpha$ and $\beta$ were constrained by fitting the multi-frequency pulse widths of average profiles, the width of the average drift profile, the fractional drift intensity, the interval between successive sub-pulses in the same period and parts of PPA (without the orthogonal mode jumps).}
\tablenotetext{\space}{ $^{\rm b}$ $\alpha$ and $\beta$ were constrained by fitting the PPA sweep rate and multi-frequency pulse widths. }
\tablenotetext{\space}{ $^{\rm c}$ $\alpha$ and $\beta$ were constrained by fitting multi-epoch pulse profiles and PPA variations with a precessional model. }

\tablenotetext{\space}{ The symbol ``m'' stands for main pulses, while ``i'' means interpulses.}
\tablenotetext{\space}{ References: 1 $-$ Smits et al. (2007), 2 $-$ Gould \& Lyne (1998), 3 $-$ Manchester et al. (1998),
4 $-$ Johnston et al. (2008), 5 $-$ Theureau et al. (2011), 6 $-$ Everett \& Weisberg (2001), 7 $-$ Gil \& Lyne
(1995), 8 $-$ Keith et al. (2010), 9 $-$ Becker et al. (2005), 10 $-$ Johnston et al. (2007), 11 $-$ Rankin et al.
(2006), 12 $-$ Krishnamohan \& Downs (1983), 13 $-$ Taylor et al. (1993), 14 $-$ Rankin \& Wright (2007),
15 $-$ Kramer \& Johnston (2008), 16 $-$ Hobbs et al. (2004), 17 $-$ Johnston \& Weisberg (2006), 18 $-$ Gangadhara
et al. (1999), 19 $-$ Manchester et al. (2010), 20 $-$ Wu et al. (1993), 21 $-$ Crawford et al. (2001b), 22 $-$ Camilo et al. (2008), 23 $-$ Tiburzi et al. (2013), 24 $-$ von Hoensbroech et al. (1998), 25 $-$ van Ommen et al. (1997), 26 $-$ Mitra \& Rankin (2008), 27 $-$ Desvignes et al. (2013), 28 $-$ Kasian (2012), 29 $-$ Weisberg \& Taylor (2002), 30 $-$ Taylor \& Weisberg (1982), 31 $-$ Boyles et al. (2013), 32 $-$ Noutsos et al. (2011), 33 $-$ Qiao et al. (1995).}
\label{table:1}
\end{deluxetable}

\acknowledgments

We appreciate many useful discussions with Guo-Jun Qiao, Ren-Xin Xu, Bing Zhang, Dick Manchester, Simon Johnston, Axel Jessner and Gorge Hobbs. H.G.W. acknowledges the funding for visiting CSIRO by the Chinese Scholarship Council (201308440093). This work is supported by NSFC key project 11178001.

\appendix
\section{Approximate relations for dipolar magnetic field}
\label{appendix:A}
Below we list the basic approximate relationships for dipolar magnetic field used in our derivation.
The dipolar magnetic field strength at $r$ (distance to the stellar center) can be written as
\begin{equation}
B\simeq B_{\rm s}\left(\frac{R}{r}\right)^3\left(\frac{1+9\cos^2\theta_{\rm pc}}{1+9\cos^2\theta}\right)^{1/2}\simeq B_{\rm s}\left(\frac{R}{r}\right)^3\left(1+\frac{9}{20}\sin^2\theta\right),
\label{eq:B_approx}
\end{equation}
where $B_{\rm s}$ is the surface magnetic field strength, $\theta$ is the polar angle of the radius vector $\mathbf{r}$ with respect to the magnetic pole, and $\theta_{\rm pc}$, usually very small, is the polar angle of the foot point of last open field line on the polar cap. Since the $\sin^2\theta$ term is minor, it is safe to use $B\simeq B_{\rm s}(R/r)^3$ when deriving the density of particle number density at $r$ (Eq. (\ref{eq:n})).

The beam radius $\rho$, namely the opening angle between the tangent of field line and magnetic pole, is related to
$\theta$ by $\tan\rho\simeq3\tan\theta/(2-\tan^2\theta)$. When $\theta$ is small, there is
\begin{equation}
\rho\simeq \frac{3}{2}\theta.
\label{eq:rho}
\end{equation}
With this approximation, the altitude will be
\begin{equation}
r=R_{\rm e}\sin^2\theta\simeq {4 \over 9} f R_{\rm c}\rho^2,
\label{eq:r}
\end{equation}
where $R_{\rm e}\simeq f R_{\rm c}$.

Then, the other variables that depends on $r$, e.g. the particle number density $n$ in Eq. (\ref{eq:n}), three-dimensional sizes of a sub emission region ${\rm d}h$, ${\rm d}w$ and ${\rm d}s$ in Appendix B and eventually the emission intensity $I$ in Eqs. (\ref{eq:dPf})-(\ref{eq:Lomg1}), can be expressed as a function of $\rho$.

\section{Derivation of emission intensity}
\label{appendix:B}

We first derive the volume ${\rm d}V_{\rm f}$ of an elementary region between the filed lines $f$ and $f+{\rm d}f$ in a subregion as shown in Fig. \ref{figure:region_geometry}(a), and then the emission intensity $I$ from the whole subregion. The emission direction is assumed to be along the tangent of magnetic field line. The following approximations are applicable to low emission altitudes ($r<<R_{\rm c}$), where the aberration and retardation effects are neglected.

\begin{figure*}
\centering
\resizebox{8cm}{5.7cm}
{
\includegraphics{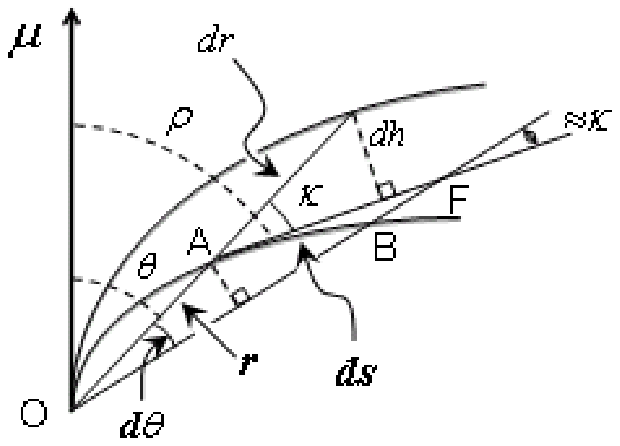}
}

\caption{Diagram for the poloidal cross section of an elementary region between field lines with $f$ (the lower thick curve) and $f+{\rm d}f$ (the upper solid curve). ${\rm d}h$ and the length AF are used as approximations for the height of this region and the arc length ${\rm d}s$ (AB), respectively.}
\label{figure:poloidal2}
\end{figure*}

The volume ${\rm d}V_{\rm f}$ can be determined by the arc length ${\rm d}s$, the transverse width ${\rm d}w$ and the height between the field lines $f$ and $f+{\rm d}f$. Fig. \ref{figure:poloidal2} gives the quantities needed to calculate the height ${\rm d}h$ and the arc length ${\rm d}s$, where ${\rm d}h$ is ${\rm d}h\simeq{\rm d}r\sin\kappa$. The arc length between A and B can be approximated by the length AF along the tangent of the field line, so we have ${\rm d}s\simeq r{\rm d}\theta/\sin\kappa$. When $r<<R_{\rm e}$ so that the polar angle $\theta=\arcsin (r/R_{\rm e})^{1/2}$ is small, it is convenient to use the
approximation $\rho\simeq 3/2\theta$, therefore, $\kappa=\rho-\theta\simeq \theta/2$. With $r=fR_{\rm c}\sin^2\theta$, one has
${\rm d}r=R_{\rm c}\sin^2\theta {\rm d}f$, then the height and arc length read
\begin{equation}
{\rm d}h\simeq\frac{1}{2}\theta^3 R_{\rm c}{\rm d}f,
\end{equation}
and
\begin{equation}
{\rm d}s\simeq2fR_{\rm c}\theta{\rm d}\theta.
\end{equation}
%

The transverse width ${\rm d}w$ is estimated as the distance between the two vertices that have the same altitudes $r$ and polar angles $\theta$, as marked by A and E in Fig. \ref{figure:region_geometry}(d). It is easy to find ${\rm d}w\simeq r{\rm d}\xi$. The angle ${\rm d}\xi$ between OA and OE can be figured out by using the law of cosines of spherical geometry
$\cos{\rm d}\xi=\cos^2\theta+\sin^2\theta\cos{\rm d}\varphi$,
where ${\rm d}\varphi$ is the azimuthal distance between OA and OE. Using small angle approximation, one has ${\rm d}\xi\simeq\sin\theta{\rm d}\varphi$, therefore,
\begin{equation}
{\rm d}w=fR_{\rm c}\theta^3{\rm d}\varphi.
\end{equation}
The volume of such a subregion is (Eq. \ref{eq:dv})
\begin{equation}
{\rm d}V_{\rm f}={\rm d}s{\rm d}w{\rm d}h\simeq R_{\rm c}^3\theta^7f^2{\rm d}f{\rm d}\theta{\rm d}\varphi.
\label{eq:dv-App}
\end{equation}
Substituting Eqs. (\ref{eq:sd})$-$(\ref{eq:dP}) into Eq. (\ref{eq:Lomg1}), one has the emission intensity
\begin{equation}
I=\frac{\int_{f_1}^{f_2}{\rm d}P_{\rm f}}{{\rm d}\Omega}=\frac{4}{9}\left(\frac{B_{\rm s}\cos\alpha}{Pce}\right)^2 M^2 \lambda^3 i_{\rm e0}R_{\rm c}^3\theta^6\int_{f_1}^{f_2}\left({R\over r}\right)^{6-q}f^2{\rm d}f.
\label{eq:Lomg-app}
\end{equation}
With Eqs. (\ref{eq:rho}) and (\ref{eq:r}), it can be reduced to Eqs. (\ref{eq:Lomg2}) and  (\ref{eq:A}).

\section{Derivation of the transverse intensity distribution for a sub beam in Model B}
\label{appendix:C}

\begin{figure*}
\centering
\resizebox{7.5cm}{8cm}
{
\includegraphics{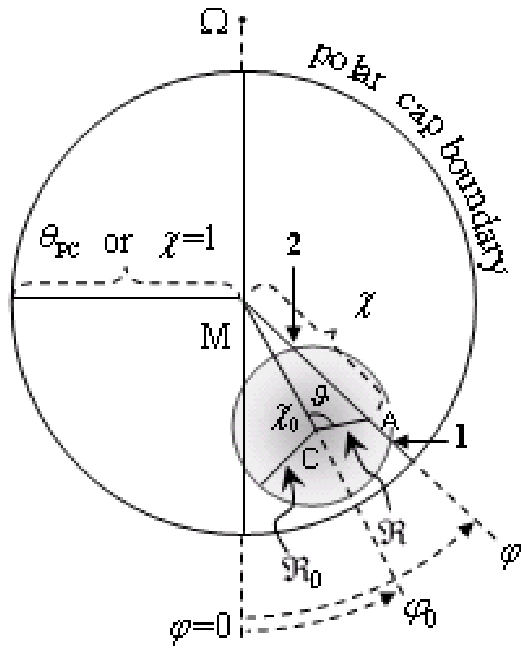}
}

\caption{Diagram for a circular discharging area on the polar cap. The symbol ``M'' and ``$\Omega$'' stand for the magnetic and rotation poles. The grey scale in the circle represent the Gaussian distribution of the multiplicity across the area. ``1''and ``2'' stand for the two intersections between the line MA and the circle. The line through $\Omega$ and M is the projection of the meridian plane, where its lower part is close to the equatorial side.}
\label{figure:PCpar}
\end{figure*}

In the following we use two coordinates, the magnetic azimuth $\varphi$ and the dimensionless colatitude $\chi$, to denote a foot point of an open field line on the polar cap (see Fig. \ref{figure:PCpar}), where $\varphi$ is counted anticlockwise from the equatorial side of meridian plane, and $\chi$ is defined as the ratio between the polar angles of the foot point and the polar cap boundary, namely $\chi\equiv\theta/\theta_{\rm pc}$. Given $\chi$, the parameter $f$ of the field line will be $f=(9/4)(R/R_{\rm c})(\rho_{\rm pc}\chi)^{-2}$, which varies from 1 (the polar cap boundary where $\chi=1$) to infinite (the magnetic pole where $\chi=0$).

Let us suppose that there is a circular discharging area centered at ($\chi_0$, $\varphi_0$) with a radius $\Re_0$ on the polar cap, in which the multiplicity factor follows a gaussian distribution around the center ``C'', i.e., $M=M_0\exp[-\Re^2/(2\sigma^2)]$, where $\Re\equiv\Delta\theta/\theta_{\rm pc}$ is the dimensionless angular distance of a field-line foot point from the center. In order to locate the area between the magnetic pole and the polar cap boundary, we assume $\Re_0\le \min[\chi_0,(1-\chi_0)]$.

Below we derive the relationship between the emission intensity $I$ and $\varphi$. It is noticed that the multiplicity $M$ along a
path with a fixed $\varphi$ is not uniform, so we need to do integration over the path between ``1'' and ``2''
along the line MA with $\varphi$ in Fig. \ref{figure:PCpar}. Substituting
$\Re=\sqrt{\chi_0^2+\chi^2-2\chi_0\chi\cos(\varphi-\varphi_0)}$ into Eq. (\ref{eq:Lomg-app}), one has
\begin{equation}
I_{\rm outer}(\varphi)=A_2P^{q-4}\dot{P}\cos^2\alpha\rho^{2q-6} \int_{f_1}^{f_2}{\rm e}^{-[\chi_0^2+\chi^2-2\chi_0\chi\cos(\varphi-\varphi_0)]/\sigma^2}f^{q-4}{\rm d}f,
\label{eq:Lomg_gauss_out}
\end{equation}
where
\begin{equation}
\chi={3\over 2}\left(\frac{R}{fR_{\rm c}}\right)^{1/2}\rho_{\rm pc}^{-1},
\label{eq:chi}
\end{equation}
and
\begin{equation}
A_2=\left({{6.4\times10^{19}} \over {3ce}}\right)^2 \left( {{2c}\over{9\pi}}\right)^{q-3} R^{6-q}\lambda^3 i_{\rm e0} M_0^2
\label{eq:A_2_AC}
\end{equation}
for the outer beam ($\rho>(3/2)\sqrt{R/(R_{\rm c}f_1)}$).

Strictly to say, one should use spherical geometry relations for the quantities
in the triangle $\bigtriangleup$MCA, including $\Re$, however, since $\theta$ and $\theta_{\rm pc}$ are very small for normal pulsars, plane geometry relations are good approximations. The lower and upper limits $f_1$ and $f_2$ in Eq. (\ref{eq:Lomg_gauss_out}), which corresponds to the intersections ``1'' and ``2'', can be found out through the following steps.
In $\bigtriangleup$MCA, one has
\begin{equation}
\frac{\Re}{\sin(\varphi-\varphi_0)}=\frac{\chi_0}{\sin(\varphi-\varphi_0+\vartheta)}=\frac{\chi}{\sin\vartheta}.
\label{eq:Resin}
\end{equation}
With these relations, the $\vartheta$ angles for ``1'' and ``2'' are $\vartheta_{1,2}=\arccos[\sin(\varphi-\varphi_0)^2 \chi_0/\Re_0\mp\cos(\varphi-\varphi_0)\sqrt{1-(\chi_0/\Re_0)^2\sin(\varphi-\varphi_0)^2}]$, with the sign ``-'' for ``1'' and ``+'' for ``2''. Therefore, the $\chi$ parameters of the two intersections are $\chi_{1,2}=\Re_0\sin\vartheta_{1,2}/\sin(\varphi-\varphi_0)$, and hence $f_{1,2}=(9/4)(R/R_{\rm c})(\rho_{\rm pc}\chi_{1,2})^{-2}$.

In the inner beam, the LOS may miss some outmost part of the area if $\rho\leq (3/2)\sqrt{R/(R_{\rm c}f_1)}$. At the azimuth $\varphi$, the outmost visible field line for an arbitrary $\rho$ has
$f_{\rm o}=9R/(4R_{\rm c}\rho^{2})$. Therefore, the lower boundary of the integration should be determined by $f_1^{\prime}=\min[f_{\rm o}, f_1]$. Finally,
the intensity distribution of the inner beam reads
\begin{equation}
I_{\rm inner}(\varphi)=A_2P^{q-4}\dot{P}\cos^2\alpha\rho^{2q-6} \int_{f_1^\prime}^{f_2}{\rm e}^{-[\chi_0^2+\chi^2-2\chi_0\chi\cos(\varphi-\varphi_0)]/\sigma^2}f^{q-4}{\rm d}f.
\label{eq:Lomg_gauss_in}
\end{equation}
Eqs. (\ref{eq:Lomg_gauss_out}) (\ref{eq:A_2_AC}) and (\ref{eq:Lomg_gauss_in}) are exactly Eqs. (\ref{eq:Lout_caseB}), (\ref{eq:A_2}) and (\ref{eq:Lin_caseB}) in Section 2.3, respectively.

\end{document}